\documentclass[letterpaper,11pt]{article}

\usepackage{ucs}
\usepackage[utf8x]{inputenc}
\usepackage{graphicx}
\usepackage{amsfonts}
\usepackage{dsfont}
\usepackage{amssymb}
\usepackage{amsmath}
\usepackage{amsthm}
\usepackage{enumerate}
\usepackage{stmaryrd}
\usepackage{fullpage}
\usepackage{ifthen}
\usepackage{subfigure}
\usepackage{epic}
\usepackage{authblk}
\usepackage{textcomp}
\usepackage[small]{caption}

\usepackage[hypertexnames=false,colorlinks=true,linkcolor=blue,citecolor=blue]{hyperref}
\usepackage[numbers,comma,square,sort&compress]{natbib}

\usepackage{color}
\usepackage{titlesec}

\newcommand{\bqq}{\begin{equation}}
\newcommand{\eqq}{\end{equation}}
\newcommand{\bqs}{\begin{equation*}}
\newcommand{\eqs}{\end{equation*}}

\newcommand{\R}{\mathbb{R}} 
\newcommand{\C}{\mathbb{C}} 
\newcommand{\Z}{\mathbb{Z}}

\newcommand{\Rea}{\mathrm{Re}\,}

\newcommand{\A}{\mathcal{A}}
\newcommand{\B}{\mathcal{B}}
\newcommand{\K}{\mathcal{K}}
\newcommand{\cO}{\mathcal{O}}

\newcommand{\e}{\epsilon}
\renewcommand{\Re}{\mathrm{Re}\,}
\renewcommand{\Im}{\mathrm{Im}\,}

\newcommand{\rmd}{\mathrm{d}}

\newcommand{\rme}{\mathrm{e}}
\newcommand{\rmi}{\mathrm{i}}

\newtheorem{lem}{Lemma}[section]
\newtheorem{thm}{Theorem}
\newtheorem{prop}[lem]{Proposition}

\newtheorem{rmk}[lem]{Remark}
\newtheorem{defi}[lem]{Definition}

\newenvironment{Proof}%
 {\begin{trivlist} \item[]{\bf Proof. }}%
{\hspace*{\fill}$\rule{.4\baselineskip}{.4\baselineskip}$\end{trivlist}}

\newenvironment{Hypothesis}[1]%
  {\begin{trivlist}\item[]{\bf Hypothesis #1 }\em}{\end{trivlist}}

\numberwithin{equation}{section}

\title{Linear spreading speeds from nonlinear resonant interaction\footnote{MH was partially supported by the National Science Foundation through grant NSF-DMS-1516155. AS was partially supported by the National Science Foundation through grant NSF-DMS-1311740.}}

\author[1]{Gr\'egory Faye\footnote{Corresponding Author, gregory.faye@math.univ-toulouse.fr}}
\affil[1]{\small CNRS, UMR 5219, Institut de Math\'ematiques de Toulouse, 31062 Toulouse Cedex, France}
\author[2]{Matt Holzer}
\affil[2]{\small George Mason University, Department of Mathematical Sciences, Fairfax, VA 22030, USA}
\author[3]{Arnd Scheel}
\affil[3]{\small University of Minnesota, School of Mathematics, 206 Church Street S.E., Minneapolis, MN 55455, USA}

\begin{document}
\maketitle

\begin{abstract}
We identify a new mechanism for propagation into unstable states in spatially extended systems, that is based on resonant interaction in the leading edge of invasion fronts. Such resonant invasion speeds can be determined solely based on the complex linear dispersion relation at the unstable equilibrium, but rely on the presence of a nonlinear term that facilitates the resonant coupling. We prove that these resonant speeds give the correct invasion speed in a simple example, we show that fronts with speeds slower than the resonant speed are unstable, and corroborate our speed criterion numerically in a variety of model equations, including a nonlocal scalar neural field model.
\end{abstract}

{\noindent \bf Keywords:} traveling fronts; spreading speeds; resonances; amplitude equations; neural fields.\\

\section{Introduction}

We are interested in spreading speeds in spatially extended systems when more than one scalar mode participates in the instability. As a particular example, we are interested in systems possessing a homogeneous steady state that is unstable with respect to both homogeneous perturbations and perturbations near a fixed nonzero wavelength (a homogeneous-Turing instability). Dynamics of such systems can be captured well by amplitude equations for weak instabilities, a real scalar amplitude equation for the homogeneous mode, and a complex scalar equation for the Turing mode. Spreading speeds for scalar equations are reasonably well understood, in particular in the case when speeds are linearly determined. Criteria for the speed can be readily calculated and proofs for the invasion speed can be obtained using comparison principles. For systems, some results are available for particular structures, such as competitive or cooperative systems, which again allow for the use of comparison principles. Leaving this restrictive class, we aim at predictions for spreading speeds based on general properties of the linearization, in particular the linear dispersion relation. Part of our motivation stems from the effort of describing  pattern-forming fronts, when spreading of the instability leaves a pattern in the wake. Predictions of the spreading speed often come with predictions for an invasion frequency, which ultimately allows one to predict the pattern formed in the wake of the invasion process.

To motivate the effect that leads us to revisit criteria for spreading speeds that are well known and recognized in the literature, consider the pair of amplitude equations that one would derive near a homogeneous-Turing instability \cite{cross93},
\begin{subequations}
\begin{align}
U_T&= d_UU_{XX}+(a_1+a_2|A|^2)U+a_3 U^3+a_4|A|^2 \\
A_T&= d_AA_{XX}+(b_1+b_2U+b_3U^2)A+b_4A|A|^2,
\end{align}
\label{eq:amp}
\end{subequations}
where $U(T,X)\in\mathbb{R}$ represents the amplitude of the homogeneous perturbation, $A(T,X)\in\mathbb{C}$ represents the amplitude of the Turing mode and the real coefficients  $a_j$ and $b_j$ are determined from the particular system being studied. In the following, we restrict our considerations to the simplest case where $A\in\R$.   In (\ref{eq:amp}), the zero solution is unstable for $a_1,b_1>0$ and the linearization is diagonal, reducing to two uncoupled scalar equations. Each of those scalar instabilities corresponds to a linear spreading speed $s_U$ and $s_A$. The larger of these two spreading speeds therefore is a natural candidate for the spreading speeds in the system. Of course, one can imagine situations where the nonlinearity significantly amplifies growth and leads to spreading faster than the linear spreading speed, a situation which mostly is observed in subcritical instabilities and referred to as the ``pushed'', nonlinear invasion, rather than ``pulled'', linear invasion. Ignoring this possibility, which also does not occur in the parameter regimes we study here, we are interested in cases where the nonlinear interaction of the two modes can create a faster spreading speed $s_{AU}>\max\{s_A,s_U\}$. Our main findings point to precise parameter regions where this acceleration through interaction is possible. Interestingly, this accelerated spreading speed $s_{AU}$ is \emph{independent of the strength of the nonlinearity}, but rather reliant only on the mere presence of a quadratic coupling term, $a_4\neq 0$. 

The principle objective of this article is to derive and corroborate a criterion for spreading speeds that incorporates the possibility of nonlinear mode interaction. In somewhat more detail, our contributions are as follows. 

\paragraph{Quadratic resonance speeds --- criteria.}

After reviewing more classical, simple-mode spreading criteria we introduce our new \emph{quadratic resonance speed} $s_\mathrm{quad}$, enabled by quadratic nonlinearities. Given a dispersion relation $\lambda=\lambda(\nu)$  for spatio-temporal modes $\rme^{\lambda t + \nu x}$, we find spreading speeds as critical points of the envelope speed $\Re(\lambda(\nu_1))/\Re(\nu_1)$, subject to a spatio-temporal resonance condition
\[
s_\mathrm{quad}=\min_{\Re\nu_1}\max_{\Im\nu_1} s_\mathrm{env}(\nu_1),\qquad \lambda(\nu_1)=\lambda(\nu_2)+\lambda(\nu_3),\quad \nu_1=\nu_2+\nu_3.
\]
The spatio-temporal resonance also needs to be supplemented with a pinching condition. We predict that such spreading speeds will be observed whenever a quadratic term in the equation actually couples modes $\nu_2$ and $\nu_3$ to form mode $\nu_1$, and whenever this speed exceeds other spreading speeds in the equation. The definition is motivated and stated precisely in several forms in Section \ref{sec:criterion}. 

\paragraph{Quadratic resonance speeds in a simple unidirectional amplitude equation --- proofs.}

We prove  that quadratic resonance speeds give  the correct spreading speed for typical initial data in a simple case of uni-directionally coupled amplitude equations; $b_2=b_3=0$ in \eqref{eq:amp}.  We first motivate the equations and compute quadratic resonance speeds, then show that the quadratic resonance speeds give the invasion speed in the parameter region predicted; see Section \ref{sec:uni}.
The proof exploits the unidirectional nature of the coupling and is based on comparison principles. The situation  is reminiscent of recent work on anomalous spreading in systems of coupled Fisher-KPP equations, see \cite{weinberger07,anomalousI,anomalousII}.

\paragraph{Linear instability of fronts with speed $s<s_\mathrm{quad}$ --- proofs.}

Beyond the simple uni-directional coupling, we corroborate our criterion by showing instability of fronts that propagate with speeds less than the quadratic resonance speed $s_\mathrm{quad}$. We demonstrate this instability in simple examples and explain how the mechanism carries over to more general situations. We view this criterion as an indicator that spreading speeds are necessarily at least $s_\mathrm{quad}$, although actual invasion does not take the form of a single invasion front, stationary or time-periodic in a comoving frame. Technically, we exhibit a necessary singularity in the Greens function that implies pointwise exponential growth of perturbations; see Section \ref{sec:linunstable}.

\paragraph{Validation of quadratic resonance speeds --- numerics.}

We demonstrate numerically the validity of our criterion in several contexts. We first compare numerical simulations with the theoretical results on amplitude equations from Section \ref{sec:uni}. We then explore bidirectionally coupled amplitude equations, systems of equations where a Swift-Hohenberg equation is coupled to a reaction-diffusion equation, and a nonlocal neural field equation. In each case, we compute the quadratic resonance speed from the dispersion relation and compare with direct simulations; see Section \ref{sec:num}. 

We conclude the paper with a discussion of our results and an outlook towards generalizations and related problems. We note that Sections \ref{sec:uni}, \ref{sec:linunstable}, and \ref{sec:num} can be viewed as independent justifications for our main criterion in Section \ref{sec:criterion}. In particular, any of them could be skipped at first reading.

\section{Quadratic resonance speeds --- the linear criterion}\label{sec:criterion}

We briefly review invasion speed theory, in particular linear criteria for the speed, Section \ref{s:rev}, min-max characterizations, Section \ref{s:min},  and then introduce our new quadratic resonance speed $s_\mathrm{quad}$, Section \ref{s:def}. We finally derive a complex space-time interaction criterion, equivalent to the criticality of the envelope speed, Section \ref{s:int}, and finally generalize to systems, Section \ref{s:sys}.

\subsection{Review of invasion speed theory}\label{s:rev}

We discuss some aspects of wave speed selection.  Our focus here is rather narrow and we emphasize those features pertinent for the results obtained in the remainder of the paper.  We point the interested reader to \cite{vansaarloos03} for a more in depth review and general treatment.

Localized perturbations of an unstable steady state grow in time and spread spatially. The spreading process is mediated by invasion fronts that propagate into the unstable state and select a secondary state in their wake.  A defining feature of these fronts is the speed at which they propagate.  Invasion fronts can be loosely characterized as either pulled or pushed.  Pulled fronts are driven by the instability of the unstable state ahead of the front interface and their speed can be calculated from the linearization of the system about this state.  On the other hand, the growth of perturbations can sometimes be enhanced by nonlinear effects such that the speed is determined by the nonlinearity.  Fronts of this variety  are commonly referred to as pushed.  

Determining the speed of pulled fronts involves calculating the {\em linear spreading speed}.  In words, the linear spreading speed is the critical speed at which a moving observer witnesses a transition from pointwise exponential stability to pointwise exponential instability.  At any speed faster than the linear spreading speed the observer outruns the instability while at slower speeds the instability outruns the observer.  In this way, linear spreading speeds are  related to the notion of absolute and convective instabilities, see for example \cite{briggs,bers84,brevdo88,chomaz,dee83,holzsch,huerre90,sandstede00}.  Mathematically, the linear spreading speed associated to an unstable state can be determined by locating pinched double roots of the dispersion relation.  Simple roots of the dispersion relation $D(\lambda,\nu)$ correspond to spatial modes $e^{\nu x}$ with temporal behavior $e^{\lambda t}$, where $\nu,\lambda\in\mathbb{C}$.  Double roots $(\lambda,\nu)$ correspond to a ``double'' mode with spatio-temporal behavior $e^{\lambda t + \nu x}$. Such double roots, together with a pinching condition, give rise to a singularity of the Green's function and therefore  induce spatio-temporal behavior  $e^{\lambda t + \nu x}$ for localized initial conditions, locally in space. Therefore, pointwise linear stability is equivalent to requiring that $\Re\lambda<0$ for all pinched double roots. Transforming to a frame of reference moving with speed $s$, the location of these pinched double roots varies with $s$ and marginal stability is achieved when the pinched double root satisfies $\lambda^*\in i\mathbb{R}$ for some value of $s=s_{lin}$; see for instance \cite{holzsch} for a recent and general account of the linear theory. 

A more subtle analysis of the singularity of the Green's function predicts a slow convergence to the front, with relaxation of the speed $s\sim s_\infty -\frac{3}{2t}$; see \cite{bramson} for a first proof of expansions for the speed in the case of the scalar KPP equation using probabilistic methods, \cite{saarloosebert} for an analysis in a more general context based on the Green's function, and the more recent \cite{hamel13} that partially recovers Bramson's result using PDE comparison techniques. Our emphasis here goes in a different direction, aimed at ``zeroth order'' speed selection in more complicated equations rather than higher order approximations in simple systems.

The possibly simplest example where complications arise is when the linearization has a skew-product structure that is, a subset of variables decouples from the others. It is then possible for multiple linear spreading speeds to exist within different components. To give an example, consider the system of equations studied in \cite{anomalousI,anomalousII}
\begin{eqnarray*}
u_t&=& du_{xx}+\alpha u-u^2+\beta v \\
v_t&=& v_{xx}+v-v^2.
\end{eqnarray*}
Linearizing the system about the unstable state, the $v$ component decouples and feeds into the $u$ component as a source term.  The dispersion relation for the full system is the product of the dispersion relations for the sub-systems, i.e. $D(\lambda,\nu)=D_u(\lambda,\nu)D_v(\lambda,\nu)$.  The linear spreading speed for the $v$ component is the Fisher-KPP speed of two and the solution converges to a traveling front with decay rate $xe^{-x}$.  However, ahead of the front steeper decay rates are observed.  These steep modes feed into the $u$ component as a source term and, depending on the values of $d$ and $\alpha$, lead to faster spreading speeds.  To determine the selected mode $\nu^*$, one imposes that $D_u(\lambda^*,\nu^*)=D_v(\lambda^*,\nu^*)$.  This \emph{resonance condition}, together with a pinching requirement, implies that $(\lambda^*,\nu^*)$ is a pinched double root of the full dispersion relation $D(\lambda,\nu)$ and determines the associated linear spreading speed. 

This paper is based upon the observation that the pinched double root criterion may be insufficient in cases where there exist multiple bands of unstable modes. The previous example illustrates that spreading can be thought of as being enabled by ``$1:1$--resonant  coupling'' between modes.  A crucial factor is the presence of the term $\beta v$ which enables the resonant coupling --- spreading speeds are slower when $\beta=0$; see also \cite{freidlin} for bidirectional coupling and associated discontinuity of spreading speeds. This point of view suggests that higher resonances may induce associated spreading speeds provided that nonlinear coupling terms are present in the system. The present work can be viewed as a case study for spreading speeds induced by \emph{``2:1--resonant coupling''}. In other words, we suggest that spreading of localized disturbances into an unstable medium can be studied in a similar fashion to instabilities in bounded domains, that is, deriving amplitude equations that take into account the crucial effect of nonlinear interaction. A key-difference is that such considerations here appear to be relevant even far from onset of instability, since speeds are determined in the leading edge of the front, at small amplitude, even when final states in the system have large finite amplitude. 

Both resonant interaction of modes as well as the transition between absolute and convective instabilities are well known and studied in the context of fluid and plasma instabilities \cite{bers84,chomaz,hasselmann,shatah10}. We are, however, not aware of a general criterion that would combine both concepts to predict spreading speeds.

We explore this phenomenon in the context of a differential equation with weakly unstable modes clustered around wavenumber $\nu=0$ and $\nu=\pm i \ell_c$ for some $\ell_c\neq 0$.  On the linear level all of these modes are uncoupled and linear spreading speeds can be computed using the pinched double root criterion as before, yielding spreading speeds associated with homogeneous modes $\nu\sim 0$ and patterned modes $\nu\sim i\ell_c$. However, nonlinearities with quadratic terms will couple the modes $\pm i \ell_c$ to the zero mode, thus acting as a forcing term for the zero mode.  By the same mechanism at play in the coupled Fisher-KPP equations above, this could lead to faster spreading speeds.

Our goal in this paper is to provide the correct prediction for spreading speeds induced by this phenomenon.  We reemphasize  that this speed is a "linear" spreading speed, since it can be determined from the linearization of the unstable state, alone,  assuming only generic nonlinearities that couple the relevant modes. Different from the speeds of pushed fronts, speeds here are independent of the exact strength of the nonlinear interaction. We also emphasize that such resonant speeds do not depend on a skew-product structure or some kind of degenerate decoupling mechanism, but occur in open classes of equations.

\subsection{Linear speeds from pointwise stability --- min-max characterizations}\label{s:min}

Consider a scalar partial differential equation  
\[ u_t=\mathcal{L} u+N(u),\quad N(u)=N_2[u,u]+\cO(|u|^3),\quad \widehat{\mathcal{L} u}(\ell)=A(i \ell)\hat{u}(\ell),\]
with $\Re\max_\ell{A}(i \ell)>0$, that is, $u\equiv 0$ is unstable. Associated to this unstable state is a dispersion relation, $D(\lambda,\nu)=A(\nu)-\lambda$, whose roots dictate the temporal evolution $e^{\lambda t}$, $\lambda\in\mathbb{C}$ of modes $e^{\nu x}$, with $\nu\in\mathbb{C}$. 

To motivate the following definition of a resonant spreading speed, recall the criterion for linear spreading speeds in scalar equations. With $\lambda=\lambda(\nu)$ from the dispersion relation in a steady frame\footnote{For systems, we take $\lambda(\nu)$ to be the root with largest real part}, we can define an envelope velocity $s_\mathrm{env}(\nu)=-\frac{\Re(\lambda(\nu))}{\Re\nu}$. In the simplest case of order preserving systems, assuming that perturbations travel at most as fast as linear perturbations, the linear (or $1: 1$-resonant) spreading speed can be obtained as a minimum of the envelope velocity, 
\[
s_\mathrm{lin}=\min_{\nu\in\R} (s_\mathrm{env}(\nu)).
\]
We find the extremality condition by differentiating, \[
0=-\Re(\lambda'(\nu*))+\frac{\Re(\lambda(\nu_*))}{\Re(\nu_*)}=:\Re(s_\mathrm{g}(\nu_*))-s_\mathrm{env}(\nu_*),
\]
where we wrote $s_\mathrm{g}(\nu):=-\lambda'(\nu)$ for the group velocity, generalized to complex $\nu$. Passing to a frame moving with speed $s_\mathrm{env}$, we find the dispersion relation 
\begin{equation}\label{e:dco}
D_s^\mathrm{co}(\lambda,\nu):=D(\lambda -s\nu,\nu),
\end{equation}
with $s=s_\mathrm{env}$. Roots $(\lambda,\nu)$ in the steady frame translate to roots $(\lambda+s\nu,\nu)$ in the comoving frame. In particular, group velocity follows Galilean transformation laws and $\Re s_\mathrm{g}^\mathrm{co}=s_\mathrm{g}-s_\mathrm{env}$ in the frame moving with the linear spreading speed. 

Beyond order preserving systems, we may allow modulations of the envelope and would then require that this minimum is taken over the maximal (with respect to modulations, that is, variations in $\Im\nu$) envelope speed 
\begin{equation}\label{e:minmax}
s_\mathrm{lin}=\min_{\Re\nu}\  \max_{\Im\nu}\ (s_\mathrm{env}(\nu)).
\end{equation}
As a consequence, $s_\mathrm{g}\in\R$ and $\lambda'(\nu)=0$ in a comoving frame, which is the classical double root criterion. We note that the min-max criterion can be obtained more systematically from a contour analysis of the Green's function in the complex plane.

The real part of the group velocity  is often interpreted as the speed at which a ``localized patch'' of the mode $e^{\nu x}$ spreads. In dispersive media, the group velocity provides the speed at which wave packets propagate, see for example \cite{whitham}.  It plays a similar role in the Fisher-KPP equation where the group velocity gives the slope of the ray in space-time for which solutions have a particular exponential decay rate, see \cite{booty,anomalousII}. From this viewpoint, if the group velocity of a mode exceeds the envelope velocity then perturbations overtake the solution and marginal stability is not attained.  On the other hand, if the group velocity is slower than the envelope velocity then the solution spreads faster than the perturbation and marginal stability is again not achieved.  With this interpretation, the linear spreading speed is the speed at which the group velocity equals the envelope velocity. These conditions: that the group velocity is real and equal to the envelope velocity, in turn imply that the mode leading to these conditions is a critical point of the envelope velocity; see \cite{vansaarloos88}. The group velocity of the mode $\nu$ can also be interpreted as giving the speed of the region in space-time for which the solution resembles the mode $\nu$\footnote{This interpretation appears to be valid when the group velocity is real, but no longer so for complex group velocities.  The interpretation of complex group velocities is less well understood, although headway has been made in several articles \cite{muschietti93,sonnenschein98,gerasik10}.   When the group velocity is complex, the ray in space time for which the mode is conserved is no longer a straight line.}. 

In order to justify this min-max characterization somewhat more explicitly, we start with the Fourier representation of solutions to the linear constant-coefficient equation $u_t = \mathcal{L} u$,
\[
u(t,x)=\frac{1}{2\pi} \int_\R \rme^{\rmi \ell x+\lambda(\rmi\ell) t}\hat{u}(0,i\ell)\rmd\ell,
\]
where the Fourier transform of the initial condition $\hat{u}(0,i\ell)$ is analytic in $\ell$. Under suitable assumptions on $\lambda(\nu)$, we can deform the contour in the complex plane,
\[
u(t,x)=\frac{1}{2\pi} \int_\R \rme^{(\rmi \ell +\eta)x+\lambda(\rmi\ell+\eta) t}\hat{u}(0,i\ell+\eta)\rmd\ell,
\]
which can in turn be estimated as 
\[
|u(t,x)|\leq C \sup_\ell \rme^{\Re\lambda(\rmi\ell+\eta)t},
\]
again using mild assumptions on $\lambda(\nu)$. We can now optimize over $\eta$ and obtain the optimal exponential growth  estimates 
\[
|u(t,x)|\leq C \inf_\eta\sup_\ell \rme^{\Re\lambda(\rmi\ell+\eta)t}.
\]
The spreading speed is obtained by replacing $\lambda(\nu)\mapsto \lambda(\nu)+s \nu$ and finding the largest speed for which growth vanishes, 
\[
s_*=\sup\left\{s|\inf_\eta\sup_\ell(\Re(\lambda(\rmi\ell+\eta))+s\eta)=0\right\}.
\]
Introducing $\lambda_\mathrm{max}(\eta):=\sup_\ell \Re(\lambda(\rmi\ell+\eta))$, this simplifies to 
\[
s_*=\sup\left\{s|\inf_\eta(\lambda_\mathrm{max}(\eta)+s\eta)=0\right\}.
\]
Geometrically, $-s$ is the slope of the least steep line through the origin that touches the graph of $\lambda_\mathrm{max}(\eta)$. On the other hand, this slope can also be obtained as the minimum of $\lambda_\mathrm{max}(\eta)/(-\eta)$, which is of course the min-max criterion that we introduced in \eqref{e:minmax}. 

We remark that such min-max characterizations of spreading speeds go back to at least \cite{hadrot}, for particular scalar examples, providing however also nonlinear characterizations of spreading speeds in these cases. 

\subsection{Linear speeds based on quadratic mode interaction --- definition of $s_\mathrm{quad}$}\label{s:def}

Going back to the possibility of quadratic interaction of modes, consider two modes $\nu_{2,3}\in\mathbb{C}$.  Quadratic terms in the partial differential equation will couple these two modes and this interaction will potentially lead to amplification of the mode $\nu_1=\nu_2+\nu_3$ and faster spreading speeds.  The temporal behavior of $\nu_1$ will depend on the temporal behavior of the modes $\nu_{2,3}$ and the temporal behavior of $\nu_1$ by itself. 

We identify the following criterion to predict the spreading speeds induced by this quadratic interaction.  For simplicity, we state the criterion in the scalar case and will discuss generalization to systems in Section~\ref{s:sys}. 

\begin{defi}[$2:1$-resonant spreading speeds]\label{defi:quad} 
The spreading speed $s_\mathrm{quad}$ induced by quadratic interaction of modes is a critical point of the envelope velocity $s_\mathrm{env}$ associated with pinched, space-time resonant modes $\nu_2,\nu_3$, 
\[ s_\mathrm{quad}=\min_{\Re(\nu_2+\nu_3)}\ \max_{\Im(\nu_2+\nu_3)}\ \left\{ s_\mathrm{env}(\nu_2+\nu_3)\,\right|\left.\; \nu_2,\nu_3 \mbox{  space-time resonant and pinched}\right\},
\]
where
\[
s_\mathrm{env}(\nu_2+\nu_3)=-\frac{\Re(\lambda(\nu_2+\nu_3))}{\Re(\nu_2+\nu_3)},
\]
and space-time resonance and pinching constraints on $\nu_2,\nu_3$ are 
\begin{enumerate}
\item (space-time- resonance) $\nu_1=\nu_2+\nu_3$ and  $\lambda(\nu_1)=\lambda(\nu_2)+\lambda(\nu_3)$;
\item (pinching) solving $D_s^\mathrm{co}(\lambda,\nu)$ for $\nu_j=\nu(\lambda_j)$, $s=s_\mathrm{quad}$,  we require $\Re(\nu_1(\lambda))\to +\infty$ as $\Re(\lambda)\to+\infty$ and $\Re(\nu_{2,3}(\lambda))\to -\infty$ as $\Re(\lambda)\to+\infty$.
\end{enumerate} 
The corresponding quadratic coupling condition is
\[
e^{-\nu_1 x}N_2[e^{\nu_2 x},e^{\nu_3 x}]\not\equiv 0.
\]
\end{defi}
\begin{rmk}
For pointwise functions $N(u)(x)=f(u(x))$, quadratic coupling simply requires that quadratic terms do not vanish, $f''(0)\neq 0$. On the other hand, the quadratic coupling condition is presumably not strictly necessary, coupling of almost-resonant modes $\tilde{\nu_j}=\nu_j+\e_j$, with $\e_j$ arbitrarily small, appears to be sufficient; see the discussion for more details. 
\end{rmk}
In other words, we mimic the procedure for scalar equations, but rather than combining two modes $\nu_1$ and $\nu_2$ ``linearly'' via a double root, we combine $\nu_1$ and $\nu_2+\nu_3$, where the latter is obtained from the quadratic interaction of modes $\nu_2$ and $\nu_3$.

\begin{rmk}
Nonlinear resonant interaction is of course a well known phenomenon, in nonlinear dynamics as well as in the context of nonlinear waves \cite{whitham}. There, one usually considers dispersive, Hamiltonian systems with dispersion relation $\omega(k)\in\R$, $\omega=\Im\lambda$, $k=\Im\nu$. Resonant triads correspond precisely to our space-time resonance condition, $\omega_1=\omega_2+\omega_3$, $k_1=k_2+k_3$. In this sense, our criterion could be viewed as an extension of the theory to complex wavenumbers.
\end{rmk}

\subsection{Criticality versus interaction in space-time}\label{s:int}

Critically of the envelope speed subject to the space-time resonance constraint can be expressed as a Lagrange multiplier problem. In fact, deriving the Euler-Lagrange equations associated with the minimization problem, one obtains an interesting formulation of criticality as an interaction condition. 

\begin{prop}[Criticality $\Rightarrow$ Complex Interaction]\label{l:el}
Suppose that $s_\mathrm{env}(\nu_2+\nu_3)\neq s_\mathrm{g}(\nu_2+\nu_3)$ and $s_\mathrm{g}(\nu_2+\nu_3)\neq s_\mathrm{g}(\nu_2)$ at a finite constrained critical point of the envelope velocity, defined in Definition \ref{defi:quad}. Then the Euler-Lagrange equations associated with the constrained minimization problem from Definition \ref{defi:quad} reduce to 
\[
s_\mathrm{g}(\nu_2)=s_\mathrm{g}(\nu_3),
\]
in addition to ``space-time resonance''. 
\end{prop}
The term ``complex interaction' refers to the fact that (complex) equality of group velocity implies in particular that patches of modes $\nu_2$ and $\nu_3$ travel with the same speed such that they can interact over long time intervals. Equality of the imaginary part of the group velocities encodes criticality as in the linear $1:1$-interaction. 

\begin{Proof}[of Proposition \ref{l:el}]
Define the envelope velocity $S$ and space-time resonance $C$ as functions of modes $\nu_2,\nu_3$,
\begin{align*}
S(\nu_2,\nu_3)&=-\frac{\Re(\lambda(\nu_2)+\lambda(\nu_3))}{\Re(\nu_2+\nu_3)},\\
C(\nu_2,\nu_3)&=\lambda(\nu_2+\nu_3)-\lambda(\nu_2)-\lambda(\nu_3).
\end{align*}
Extremality is encoded in the Euler-Lagrange equations with Lagrange multipliers $\mu_1,\mu_2$ associated with the constraints $\Re(C)=0$, $\Im(C)=0$,
\begin{equation}\label{e:E}
\nabla_{\nu_2,\nu_3} S=\mu_1 \nabla_{\nu_2,\nu_3} (\Re(C))+\mu_2 \nabla_{\nu_2,\nu_3} (\Im(C)).
\end{equation}
Here, $\nabla_{\mu_2,\mu_3}$ is interpreted as a a real 4-vector. Writing $\iota:\C\to\R^2, z\mapsto (\Re(z),\Im(z))^T$, and using the Cauchy-Riemann equations for analytic functions $f$, $\nabla_\nu (\Re (f(\nu)))=\iota(f'(\nu))$,  $\nabla_\nu (\Im (f(\nu)))=\iota(i f'(\nu))$,  we find after a short calculation that \eqref{e:E} is equivalent to the two complex equations
\begin{subequations}
\begin{align}
\frac{1}{\Re(\nu_2+\nu_3)^2}\left(\Re(\nu_2+\nu_3)\lambda'(\nu_2)-\Re(\lambda(\nu_2)+\lambda(\nu_3))\right)&=\mu\left(\lambda'(\nu_2+\nu_3)-\lambda'(\nu_2)\right),\label{e:E'1}\\
\frac{1}{\Re(\nu_2+\nu_3)^2}\left(\Re(\nu_2+\nu_3)\lambda'(\nu_3)-\Re(\lambda(\nu_2)+\lambda(\nu_3))\right)&=\mu\left(\lambda'(\nu_2+\nu_3)-\lambda'(\nu_3)\right),\label{e:E'2}
\end{align}
\end{subequations}
where, $\mu=\mu_1+i\mu_2\in\C$.
Subtracting $\eqref{e:E'2}$ from $\eqref{e:E'1}$ gives
\[
\frac{\lambda'(\nu_2)-\lambda'(\nu_3)}{\Re(\nu_2+\nu_3)} = -\mu(\lambda'(\nu_2)-\lambda'(\nu_3)).
\]
Hence, we have either $\lambda'(\nu_2)=\lambda'(\nu_3)$ or $\mu=-(\Re(\nu_2+\nu_3))^{-1}$. The latter implies that 
\[
\lambda'(\nu_2+\nu_3)=\frac{\Re(\lambda(\nu_2+\nu_3))}{\Re(\nu_2+\nu_3)},
\]
using space-time resonance, which implies that $\nu_1=\nu_2+\nu_3$ simply corresponds to a double root at $\nu_1$, hence does not actually involve quadratic interaction. We are therefore left with the second case, $\lambda'(\nu_2)=\lambda'(\nu_3)$. Having solved $\eqref{e:E'1}-\eqref{e:E'2}=0$, we can now solve $\eqref{e:E'1}=0$ since $\lambda'(\nu_2+\nu_3)\neq \lambda'(\nu_2)$. 
\end{Proof}

\subsection{Generalization to systems of equations}\label{s:sys}
Consider a system of equations
\[ u_t=\mathcal{L} u+N(u),\]
with $u\in\mathbb{R}^n$, $N(u)=N_2(u)+\cO(|u|^3)$, and linear part $\mathcal{L}$ defined through its Fourier symbol $A( i \ell)$.  Applying the Fourier-Laplace transform to the linear equation $u_t=\mathcal{L}u$, solutions are obtained for any triple $\nu,\lambda,v_\nu$ for which
\[ \left(A(\nu)-\lambda I\right)v_\nu=0.\]
Note that such a solution exists precisely when $(\nu,\lambda)$ is a root of the dispersion relation,
\begin{equation}\label{e:dr}
D(\lambda,\nu)=\mathrm{det}\left(A(\nu)-\lambda I\right).
\end{equation}
Assuming that a mode $(\lambda,\nu)$ is simple, that is,  $\partial_\lambda D\neq 0$ at  $(\lambda,\nu)$, we can solve 
\[ \left(A^*(\nu)-\bar{\lambda} I\right)w_\nu=0, \quad (w_\nu,v_\nu)=1,\]
where $(\cdot,\cdot)$ denotes the hermitian scalar product.
Again using $\partial_\lambda D\neq 0$,  we find a smooth family $\lambda(\nu)$ and expressions for envelope and group velocities of the mode $\nu$, 
\[
s_\mathrm{env}(\nu)=-\frac{\Rea (A(\nu)v_\nu,w_{\nu})}{\Rea \nu},\qquad  s_\mathrm{g}(\nu)= -\frac{d}{d\nu}  (A(\nu)v_\nu,w_{\nu}).
\]
\begin{defi}[$2:1$-resonant spreading speeds --- quadratic coupling in systems]\label{defi:quadsystem} 
The quadratic speed in systems is defined as for scalar systems, via the dispersion relation \eqref{e:dr}. The quadratic coupling condition for extremal, space-time resonant, pinched modes $(\lambda_j,\nu_j),\ j=1,2,3,$ is\footnote{Again, for pointwise evaluation nonlinearities $f(u)$, the exponentials $e^{\nu_jx}$ can be omitted. }
\[
\left(N_2[e^{\nu_2 x}v_{\nu_2},e^{\nu_3 x}v_{\nu_3}],e^{-\bar{\nu}_1 x}w_{\nu_1},\right)\not\equiv 0.
\]
\end{defi}

\section{Unidirectionally coupled amplitude equations --- $s_\mathrm{quad}$ gives the invasion speed}\label{sec:uni}

Our goal here is to validate the criterion from Section \ref{sec:criterion} in a simple case of unidirectionally coupled amplitude equations. We shall first motivate the particular set of equations that we will work with, and state our main theorem in Section \ref{s:res}. We will then compute the predicted speed $s_\mathrm{quad}=s_{AU}$ informally, and from Definition \ref{defi:quadsystem} in Section \ref{s:sau}. Section \ref{s:proof} contains the proof of our main theorem. 

\subsection{Quadratic spreading speeds  --- main theorem}\label{s:res}

A natural starting point for the analysis of resonant interaction is near simultaneous  onset of a Turing and a pitchfork bifurcation. In this regime, amplitude equations can be derived which reduce the problem to the study of coupled reaction-diffusion equations.  
Our primary analytical result pertains to spreading speeds in coupled amplitude equations of the form,
\begin{subequations}
\label{eq:CGLone}
\begin{align}
U_T&= dU_{XX}+(\alpha -6A^2)U-U^3+2\gamma A^2  \\
A_T&= 4A_{XX}+A-3A^3.
\end{align}
\end{subequations}
Here, $U$ stands for the amplitude of a homogeneous instability, and $A$ represents the amplitude of a Turing mode, which we restricted here to real values. It turns out that the spreading speeds observed in (\ref{eq:CGLone}) are determined by the linearization about the unstable zero state.  Depending on the parameter values $(d,\alpha)$, the spreading speed observed in (\ref{eq:CGLone}) will be one of three speeds: the speed of the zero mode $s_U=2\sqrt{d\alpha}$, the speed of the Turing mode $s_A=4$, or the speed of the zero mode induced by the Turing mode through the quadratic interaction $\gamma A^2$ as defined in Definition \ref{defi:quadsystem}.  We call this speed $s_{AU}$. 
\begin{lem}\label{l:sau}
The speed $s_\mathrm{quad}$ induced by coupling modes $\nu_{2,3}$ from the equation for the Turing mode to modes $\nu_1$ from the equation for the homogeneous mode $U$ through the quadratic term $\gamma A^2$, $\gamma\neq 0$, is faster than the single-mode speeds $s_A$ and $s_U$ in the region 
\[ \mathrm{P}=\left\{ (d,\alpha)\ | \ 4-d<\alpha \ (0< d\leq 1),\ 4-d<\alpha<\frac{d}{d-1}\  (1<d<2)\right\}.\]
\end{lem}
We shall prove Lemma \ref{l:sau} in Section \ref{s:sau}. 
\begin{thm}\label{thm:CGLone} Choose $(\alpha,d)\in \mathrm{P}$, such that $s_{AU}>\max\{s_A,s_U\}$. 
Define the invasion point,
\[ \kappa(t)=\sup_{x\in\mathbb{R}} \{ x \ | \ u(t,x)>\sqrt{\alpha-2}\},\]
and the selected speed 
\[ s_\mathrm{sel}=\lim_{t\to\infty} \frac{\kappa(t)}{t}.\]
For $(d,\alpha)\in\mathrm{P}$, and $\gamma\neq 0$, the solution of (\ref{eq:CGLone}) with initial data consisting of compactly supported perturbations of Heaviside step functions will spread with speed
\[ s_{AU}=d\sqrt{\frac{\alpha-2}{2-d}}+\alpha\sqrt{\frac{2-d}{\alpha-2}},\]
i.e. $s_\mathrm{sel}=s_{AU}.$
\end{thm}

We note that in the complement of the region $\mathrm{P}$, one observes 
\[
s_\mathrm{sel}=\max\{s_A,s_U\},
\]
see Figure \ref{fig:speedAMP}. We suspect that this could be established using similar methods as employed here but we do not pursue this direction. 

The proof itself relies on the the construction of sub and super solutions and is based on ideas in  \cite{anomalousI,anomalousII}.  We note that for the particular case of amplitude equations in (\ref{eq:CGLone}), the dynamics of the Turing mode is independent of the dynamics of the zero mode. This fact, that is, the absence of a back-coupling as seen in (\ref{eq:amp}), is essential to the proof of Theorem~\ref{eq:CGLone}, it is, however not generic for amplitude equations near Turing/pitchfork instabilities.  On the other hand, as we shall demonstrate later, this skew-product nature of the equation does not appear to be  relevant to the underlying phenomenon and should be thought of as a technical assumption that allows the use of comparison principles.

We briefly comment on the  mechanism leading to faster speeds in (\ref{eq:CGLone}), which is similar to the one identified in \cite{anomalousI,anomalousII}.  Starting from compactly supported initial data, the $A$ component forms a traveling front propagating with speed $4$.  Ahead of the front interface, the solution decays to zero faster than any exponential.  Through the quadratic coupling term, the $A$ component acts as a source in the $U$ equation with decay rates twice those of the original solution.  Under the evolution of the equation governing $U$, these "steep" decay rates may actually be "weak"  and lead to faster invasion speeds and the front profile will converge to a super-critical traveling front with weak exponential decay.

In this section, we focus on the system of equations (\ref{eq:CGLone}).  First, motivate the  speed $s_{AU}$ directly using linear envelope speeds. We then show that this speed can be obtained directly from Definition \ref{defi:quad}. The bulk of this section is devoted to the proof of Theorem~\ref{thm:CGLone}.  

\subsection{Derivation of the speed $s_{AU}$}\label{s:sau}

We first motivate the speed $s_{AU}$ and the region $\mathrm{P}$. 
Consider (\ref{eq:CGLone}) and linearize about the unstable zero solution.  Using the ansatz $e^{\lambda t+\nu x}$ we obtain the dispersion relation
\[ D(\lambda,\nu)=(d\nu^2+\alpha-\lambda)(4\nu^2+1-\lambda).\]
We generalize the approach in \cite{anomalousI} where linear coupling was considered.  In a moving coordinate frame, the roots of the dispersion relation $D^\mathrm{co}_s(\lambda,\nu)$ can be calculated explicitly,
\begin{subequations}
\begin{align}
\nu_U^\pm (\lambda,s)&= -\frac{s}{2d}\pm\frac{1}{2d}\sqrt{s^2-4d\alpha+4d\lambda},  \\
\nu_A^\pm (\lambda,s)&= -\frac{s}{8}\pm\frac{1}{8}\sqrt{s^2-16+16\lambda}.
\end{align}
\label{eq:roots}
\end{subequations}
Taking a different approach, one can solve for the speed $s$ for which the dispersion relation has a root at $\lambda=0$.  This speed is referred to as the envelope velocity and captures the speed of the moving reference frame for which a pure exponential is a stationary solution.  The envelope velocities associated to modes $\nu\in\mathbb{R}^-$ are,
\begin{eqnarray*}
s_U(\nu)&=& -d\nu-\frac{\alpha}{\nu}, \\
s_A(\nu)&=& -4\nu-\frac{1}{\nu}.
\end{eqnarray*}
For linear coupling, one seeks pinched double roots of the dispersion relation.  These pinched double roots can be found by finding intersections of the envelope velocity curves, i.e. $\nu$ values for which $s_U(\nu)=s_A(\nu)$.  When the coupling is quadratic, as in (\ref{eq:CGLone}), then the decay rates of the $A$ component are doubled and a prediction is obtained by computing intersections of the curves $(\nu,s_U(\nu))$ and $(2\nu,s_A(\nu))$ in $s-\nu$ space.  The factor of two is required due to the quadratic coupling.  These curves intersect for a subset of $(d,\alpha)$ parameter space and give a predicted spreading speed,
\begin{equation}\label{e:senv} s_{A \to U}=d\sqrt{\frac{\alpha-2}{2-d}}+\alpha\sqrt{\frac{2-d}{\alpha-2}}.\end{equation}
We now proceed to a more formal derivation of $s_{AU}$ from  Definition~\ref{defi:quad}.
\begin{Proof}[of Lemma \ref{l:sau}]
We use the Euler-Lagrange formulation, Lemma \ref{l:el}. Group velocities in the second equation are simply  $s_\mathrm{g}=-\lambda'(\nu)=8\nu$, so that ``complex interaction'' implies $\nu_2=\nu_3$. Space-time resonance then implies that 
\[
2(4\nu_2^2+1)=d(2\nu_2)^2+\alpha,
\]
and therefore that
\[ \nu_2=-\frac{1}{2}\sqrt{\frac{\alpha-2}{2-d}}.\]
This implies that $\nu_1=2\nu_2$ and we calculate the envelope velocity $s_U(\nu_1)$ which yields the speed in \eqref{e:senv}. 

In order to find the restrictions on parameters $(\alpha,d)\in \mathrm{P}$ as stated in the theorem, we check the pinching condition which imposes restrictions on the parameter values. We first note that $\Re\nu=0$ would give infinite envelope speed, certainly not a minimum in the definition of $s_\mathrm{quad}$. Now $\Re\nu_2< 0$  implies that, either (i), $\alpha>2$ and $d<2$, or (ii), $\alpha<2$ and $d>2$.  Since $\nu_1=2\nu_2$, we see that $\nu_1$ is a root of the dispersion relation in the comoving frame $D_{s_{AU}}(\lambda,\nu)$ with $\lambda=0$. In order to verify the pinching condition, we need to track this root as $\mathrm{Re}(\lambda)\to +\infty$ and verify that this root tends to $+\infty$ as well.

Since we have explicit representations of roots, we only need to show that $\nu_1=\nu_U^+(0,s_{AU})$. This is true if $\nu_1+\frac{s_{AU}}{2d}>0$, or, equivalently, if
\[
\nu_1+\frac{s_U(\nu_1)}{2d}=\frac{\nu_1}{2}-\frac{\alpha}{2d\nu_1}>0.
\] 
Since $\nu_1<0$, this is equivalent to $\nu_1^2<\frac{\alpha}{d}$.  Expand this condition, 
\[ \frac{\alpha-2}{2-d}<\frac{\alpha}{d},\]
and solve for $\alpha$ to find,
\[
\alpha(d-1)<d \mbox{ and } d<2, \quad \mbox{ or,} \quad  \alpha(d-1)>d \mbox{ and } d>2. 
\]
This condition holds automatically if $d\leq 1$ and gives a condition on $\alpha$ if $d>1$. 

In a similar fashion, we require $\nu_2=\nu_A^-(0,s_{A}(\nu_2))$.  Once again referencing (\ref{eq:roots}), this is equivalent to the requirement that $\nu_2+\frac{s_{A}}{8}<0$.  Expanding we find that 
\[
\alpha>4-d \mbox{ and }  d<2,\quad  \mbox{   or, }\quad  \alpha<4-d \mbox{ and } d>2.
\]
Finally, note that requirements that $\frac{d}{d-1}<\alpha$ and $\alpha<4-d$ are not compatible for $d>2$. As a consequence, we are left with the restrictions $d<2$, $\alpha<d/(d-1)$, $\alpha>4-d$, which delimits precisely the region $\mathrm{P}$.

One can check that the min-max criterion actually gives a finite value for $s,\nu_{2/3}$, which then necessarily coincides with the value of the unique critical point that we computed here.

\end{Proof}

\subsection{Proof of Theorem~{\ref{thm:CGLone}}}\label{s:proof}

Consider (\ref{eq:CGLone}) with $U$ and $A$ real valued and take $\gamma>0$.  We consider the set of parameters,
\[ \mathrm{P}=\left\{ (d,\alpha)\ | \ 4-d<\alpha \ (0< d\leq 1),\ 4-d<\alpha<\frac{d}{d-1}\  (1<d<2)\right\}.\]
We will prove that for $(d,\alpha)\in\mathrm{P}$, Heaviside step function initial data will spread with speed $s_{AU}$.  

The idea of the proof  is to construct sub and super solutions that constrain the evolution of the initial data.  The construction of these sub and super-solutions is motivated by a similar construction in \cite{anomalousII} for system of Fisher-KPP equations with coupling through a linear, rather than quadratic, coupling term. The primary differences here are that the coupling is quadratic and also affects the strength of the linear instability of the zero state in the $U$ equation.  Since the proof is a modification of the one in \cite{anomalousII}, we do not aim to give a full treatment, but instead outline the approach with an emphasis on where the arguments are different.  Consider 
\[ N(U)=U_T-dU_{XX}-\left(\alpha-6A^2\right)U+U^3-2\gamma A^2.\]
The proof strategy is to construct sub-solutions, i.e. functions for which $N(\underbar{U}(T,X))\leq 0$, and super-solutions where the functional is non-negative.  Any initial data lying between a sub-solution and a super-solution will remain so and therefore if Heaviside step functions can be constrained in this way then bounds on the spreading speeds are obtained.  

\paragraph{Super-solutions.}

Let $s>s_{AU}$ and note that $\bar{A}(T,X)=\max\left\{\frac{1}{\sqrt{3}}, C_A e^{\nu_A^-(0,s)(X-sT)}\right\}$ is a super-solution for the $A$ component.  We focus on the $U$ component and consider $C_A$ to be fixed.  Consider,
\begin{equation} \bar{U}(T,X)=\left\{\begin{array}{cc} U_m & X<sT+\theta  \\ C_U e^{\nu_U^+(0,s)(X-sT)}+\frac{\gamma C_A^2}{4d(\nu_A^-(0,s))^2+2s\nu_A^-(0,s)+\alpha}e^{2\nu_A^-(0,s)(X-sT)}  & X>sT+\theta, \end{array}\right.\label{eq:Usuper} \end{equation}
where $\theta$, $C_A$ and $U_m$ remain to be determined.  Note that $4d(\nu_A^-(0,s))^2+2s\nu_A^-(0,s)+\alpha$ is negative for parameters in $\mathrm{P}$.  A calculation then  reveals that 
\[ N(\bar{U})=6\bar{A}^2\bar{U}+\bar{U}^3+2\gamma\left( \bar{A}^2-A^2\right),\]
and since $\bar{A}$ is a super-solution this term is always positive for $X>sT+\frac{1}{\nu_A^-(0,s)} \log \frac{1}{C_A\sqrt{3}  }$. Next, chose $U_m$ large enough so that $-(\alpha-2)U_m+U_m^3-\frac{2}{3}\gamma>0$.  We require two conditions on $C_U$.  First, we select $C_U$ sufficiently large so that the maximum of the sum of exponentials in (\ref{eq:Usuper}) exceeds $U_m$.  Secondly, we require again that $C_U$ is chosen sufficiently large so that the intersection point between these exponentials and the constant $U_m$ occurs for $X>sT+\frac{1}{\nu_A^-(0,s)} \log \frac{1}{C_A\sqrt{3}  }$.  This defines $\theta$ and we have a one-parameter family of super-solutions. 
\paragraph{Sub-solutions.}

We proceed as follows. The sub-solution is constructed by breaking space-time into pieces and utilizing sub-solutions that approximate the $A$ component in those regions.  We first find an exponential sub-solution for the decoupled $A$ equation.  Then, we use this sub-solution to find an exponential sub-solution for the $U$ component.  

Fix $\max\{4,2\sqrt{d\alpha}\}<\sigma<s_{AU}$.  In a frame of reference moving with speed $\sigma$, the dispersion relation has four roots: $\nu_{U}^\pm$ and  $\nu_A^\pm$, see (\ref{eq:roots}).  We have that $\nu_U^+<2\nu_A^-$ since $\sigma<s_{AU}$.

\paragraph{An exponential sub-solution for the $A$ component.} Fix initial data $A_0(X)$ satisfying $0\leq A_0(x)\leq \frac{1}{\sqrt{3}}$, a compactly supported perturbation of the Heaviside step function $\frac{1}{\sqrt{3}}H(-x)$.  We require a result from \cite{hamel13}, where sub-solutions for $A(T,X)$ are constructed from solutions of the linear equation
\[ A_T=4A_{YY}+\sigma A_Y+A,\]
for $(T,Y)\in\mathbb{R}^+\times\mathbb{R}^+$ where $Y=X-\sigma T$ and with Dirichlet boundary condition, $A(0,T)=0$ imposed at the left of the boundary.  This sub-solution can be expressed as
\[ \underbar{A}(T,Y)=e^{\left(1-\frac{\sigma^2}{16}\right)T}e^{-\frac{\sigma}{8}Y}e^{-\frac{Y^2}{16T}} G(T,Y),\]
with $G(T,0)=0$ and $|G(T,Y)|<C$ for some $C>0$. With the nonlinearity included, $\underbar{A}(T,Y)$ is no longer a sub-solution.  Multiplying by an unknown function $\zeta(T)$ we find that  $\zeta(T)\underbar{A}(T,Y)$ is a sub-solution if
\[ \zeta_T=-3\underbar{A}^2\zeta^3.\]
Since $\sigma>4$, we have that $\underbar{A}<Ce^{-\omega T}$ with $\omega=\frac{\sigma}{16}-1$ and we find the explicit solution
\[ \zeta(t)=\zeta(0)\sqrt{\frac{\omega}{\omega +3C^2\left(1-e^{-2\omega T}\right)}}.\]  
The explicit form of the sub-solution allows us to find a region in space-time for which the exponential $e^{\nu_A^-Y-\delta T}$ is also a sub-solution.  The implicit function theorem then implies that for any $\delta>0$, there is a value of $T_\delta>0$ such that $e^{\nu_A^-Y-\delta T}$  is a sub-solution for all $t>T_\delta$ and for $Y\in[\tau_-(T),\tau_+(T)]$, where $\tau_\pm(T)=-8\nu_A^-T-\sigma T \pm \sqrt{16\delta}T +o(T)$.  Note that $-8\nu_A^--\sigma$ is precisely the group velocity of the mode $\nu_A^-$.  

\paragraph{An exponential sub-solution for the $U$ component.}
The next step is to show that, for sufficiently large times, there exists an exponential sub-solution for the $U$ component on the same interval where there exists an exponential sub-solution for the $A$ component. To do this, we consider the linear inhomogeneous equation
\[ U_T=dU_{YY}+\sigma U_Y+\alpha U+2\gamma e^{2\nu_A^-Y-2\delta T}.\]
This has solution
\[ c_1 e^{\nu_U^+Y}-\frac{2\gamma}{4d(\nu_A^-)^2+2\sigma\nu_A^-+\alpha+2\delta} e^{2\nu_A^-Y-2\delta T}.\]
Now define
\[ \phi(T,Y)=c_1(T) e^{\nu_U^+Y}-\frac{2\gamma}{4d(\nu_A^-)^2+2\sigma\nu_A^-+\alpha+2\delta} e^{2\nu_A^-Y-2\delta T},\]
where we have modified the constant term $c_1(T)$ so that it evolves in time such that $\phi(T,0)=1$ for all $T\geq 0$.  To be precise
\[ c_1(T)=1+\frac{2\gamma}{4d(\nu_A^-)^2+2\sigma\nu_A^-+\alpha+2\delta} e^{-2\delta T}.\]
We also note that $\phi(T,X-\sigma T)$ as a single zero occurring at $X=\Theta(T)$, where
\[ \Theta(T)=\sigma T+\frac{1}{2\nu_A^--\nu_U^+}\left(2\delta T+\log{\frac{c_1(T)(4d(\nu_A^-)^2+2\sigma\nu_A^-+\alpha+2\delta)}{2\gamma}}\right).\]

\paragraph{A sub-solution for the $U$ component.}
We now consider the function
\begin{equation} \underbar{U}(T,X)=\left\{\begin{array}{cc} \kappa U_r(X-sT) & X<\sigma T \\ \kappa U_r((\sigma-s)T)\phi(T,X-\sigma T) &  \sigma T \leq X\leq  \Theta(T) \\
0 & X>\Theta(T)\end{array}\right.\label{eq:Ubar} \end{equation}
Here $\kappa>0$ is a constant, $U_r(X-sT)$ is a traveling front solution with $A=0$,  moving with speed $s<\sigma$ and satisfying $U_r(r)=\frac{\sqrt{\alpha}}{2}$. Note the three parameters: $r$, $s$ and $\delta$.  The parameter $\delta$ will be chosen so that the wedge of existence of the pure exponential sub-solution for the $A$ component coincides with the wedge $\sigma T \leq X\leq  \Theta(T)$.  The parameter $s$ will be chosen sufficiently close to $\sigma$ so as to guarantee that $\underbar{U}$ constitutes a sub-solution.  Finally, $r$ will remain as a free parameter that can be adjusted depending on the initial data of the $U$ component. 

Select 
\[ \delta=\frac{(2\nu_A^--\nu_U^+)\sqrt{\sigma^2-16}}{2}.\]
This ensures that there exists a $T^*\geq T_\delta$ such that $\sigma T+\tau_-(T)<\Theta(T)<\sigma T+\tau_+(T)$.

Now consider (\ref{eq:Ubar}).  First consider the region $X<\sigma T$.  We compute
\[ N(\kappa U_r(X-sT))=\kappa U_r^3(\kappa^2-1)+2A^2(3\kappa U_r-\gamma).\]
Note that if $\kappa<\min\{1, \frac{\gamma}{3\sqrt{\alpha}}\} $ then  $N(\kappa U_r)<0$.  
Next, we compute 
\begin{eqnarray} N\left(\kappa U_r((\sigma-s)T)\phi(T,X-\sigma T)\right)&=&\kappa (\sigma-s)U_r'\phi+\kappa U_r c'_1(T) e^{\nu_U^+Y}+6\kappa A^2 U_r \phi +\kappa U_r^3 \phi^3 \nonumber \\
& +& 2\kappa U_r e^{2\nu_A^-Y-2\delta T}-2\gamma A^2 \nonumber \\
&=& 
\kappa U_r \phi \left( \nu_U^+(s,0)(\sigma-s)+h(U_r)U_r +6A^2+\kappa^2U_r^2\phi^2\right)  \nonumber \\
&+& \kappa U_r c'_1(T) e^{\nu_U^+Y}+ 2\kappa U_r e^{2\nu_A^-Y-2\delta T}-2\gamma A^2 \label{eq:subcalc}
\end{eqnarray}
Here we have replaced $U_r'((\sigma-s)T)$ by $\nu_U^+(0,s)U_r((\sigma-s)T)\left(1+h(U_r)\right)$, since $U_r$ is a traveling front solution and when $U_r$ is small there exists an almost linear relationship between $U_r$ and its derivative $U_r'$.  Continuing, we must argue in two pieces.  Note that $c'_1(T)<0$ when the exponential $e^{\nu_A^-Y-\delta T}$ is a sub-solution, then the final two terms are negative.  That leaves the first term in the expression, for which by decreasing $r$ the magnitude of the positive terms $h(U_r)U_r+\kappa^2 U_r^2\phi^2$ can be made arbitrarily small.  Note also that since $\sigma$ exceeds the linear spreading speed of the $A$ component we have that $A$ will converge pointwise exponentially fast to zero as well.  Finally, since $\sigma>s$ and $\nu_U^+<0$, we find that the first term in (\ref{eq:subcalc}) is negative and hence $N(\kappa U_r((\sigma-s)T)\phi(T,X-\sigma T))<0$ for all $\sigma T+\tau_-(T)<X<\Theta(T)$.

It remains to consider the region $\sigma T< X< \sigma T+\tau_-(T)$.  We need to control the term $2\kappa U_r e^{2\nu_A^-Y-2\delta T}$.  To do this, we include it in the factor
\[  \left( \nu_U^+(s,0)(\sigma-s)+h(U_r)U_r +6A^2+\kappa^2U_r^2\phi^2+\frac{2\kappa U_r e^{2\nu_A^-Y-2\delta T}}{\phi(T,X-\sigma T)}\right), \]
and note that for $\sigma T< X< \sigma T+\tau_-(T)$ the term $\phi(T,X-\sigma T)$ is bounded strictly away from zero.  Since the exponential term in the numerator can be made arbitrarily small for large values of $T$, we can argue as before and offset this small positive value with the negative constant term $\nu_U^+(s,0)(\sigma-s)$.

Finally, for $\underbar{U}(T,X)$ to be a sub-solution we must consider the matching point at $X=\sigma T$ and require that the derivative from the left is more negative than the derivative from the right.  This is equivalent to the condition,
\begin{equation} U_r'((\sigma-s)T)< U_r((\sigma-s)T)\left( c_1(T)\nu_U^+(0,\sigma)-2\frac{\nu_A^-(0,s)e^{-2\delta T}}{4d(\nu_A^-)^2+2\sigma\nu_A^-+\alpha+2\delta}\right).\label{eq:slopecond} \end{equation}
The term on the left can again be replaced by $\nu_U^+(0,s)U_r((\sigma-s)T)\left(1+h(U_r)\right)$, from which we note that as $T\to\infty$ condition (\ref{eq:slopecond}) holds since $\nu_U^+(0,s)<\nu_U^+(0,\sigma)$. 

\paragraph{Completing the proof.}   
We now turn to the proof of Theorem~\ref{thm:CGLone}.  Consider positive initial data for the $U$ and $A$ components consisting of compactly supported perturbations of Heaviside step functions.  Let $s>s_{AU}$.  It is straightforward to find $C_A$ and $C_U$ so that $\bar{U}(T,X)>U(T,X)$ and $\bar{A}(T,X)>A(T,X)$ for all $T>0$ and all $X\in \mathbb{R}$.  By construction, the super-solutions propagate at speed $s$, providing an upper bound for the spreading speed of the solution.   Now let $s<s_{AU}$.  Fixing the sub-solution $\underbar{A}(T,X)$, we find a one-parameter family of sub-solutions for the $U$ component.  These sub-solutions only hold for $T$ sufficiently large, so in order to apply them we must bound the solution for finite amounts of time and show that the sub-solutions can still be selected so that $\underbar{U}(T,X)\leq U(T,X)$.  This requires using alternate sub-solutions for finite time and is similar to the argument used in \cite{anomalousII}, so we omit the specifics.  Since the spreading speed associated to the sub-solution $\underbar{U}(T,X)$ is $s$, we have established Theorem~\ref{thm:CGLone}.

\section{ Linear instability of invasion fronts slower than $s_\mathrm{quad}$}\label{sec:linunstable}

In order to motivate our Definition \ref{defi:quad} in a more general context, we propose to investigate the stability of ``coherent invasion fronts'', depending on the speed of propagation. This criterion is well known to give the correct invasion speed in scalar cases. One can argue that it also gives correct criteria in more complicated situations, but we will not pursue this argument, here. More interestingly, it is important to make the notion of stability precise. Since invasion fronts invade an unstable state, perturbations grow, by definition, and stability can only be obtained by restricting to a special class of initial condition, and measuring the growth of perturbations in a suitable topology. For linear equations $u_t=\mathcal{L} u$, where $\mathcal{L}$ is the linearization at a front solution, say, the evolution can be determined using the inverse Laplace transform, which expresses the ``heat kernel''
$H$ as an integral over resolvents. More precisely,
\[
H(t,x,y)=\int_\Gamma \rme^{\lambda t} G_\lambda(x,y)\rmd \lambda,
\]
where $G_\lambda(x,y)$ is the Green's function to the resolvent equation, $(\lambda - \mathcal{L})G_\lambda(\cdot,y) = \delta(\cdot-y)$.  This leads us to define linear stability of an invasion front through the absence of singularities of $G_\lambda(x,y)$ for any fixed $x,y$ as a function of $\lambda$ in $\Re\lambda>0$. Since pointwise decay of the heat kernel $H(t,x,y)$ would imply analyticity of the Laplace transform $G_\lambda(x,y)$ in $\Re\lambda>0$, singularities do indeed imply linear pointwise instability. 

The remainder of this section is organized as follows. We first review linear instability of invasion fronts with speed $s<2$ in the KPP equation, Section \ref{s:inst1}.  In Section \ref{s:inst2} and  Section \ref{s:inst3}, we illustrate our ideas in simple example of coupled transport equations where one can explicitly calculate the Green's function and locate singularities.  In Section \ref{s:inst1} we review the instability mechanism outlined in \cite[Sec 8.3]{holzsch} for relevant and irrelevant coupling in double double roots, in particular interpreting it as a $1:1$-resonance.  In  Section \ref{s:inst3} we calculate a singularity induced by a $2:1$ resonance.   Section \ref{s:inst4} then explains how this same singularity arises more generally near fronts, relating it to a singularity of the Evans function at the boundary of validity of the Gap Lemma. In summary, the results in this section exhibit a pointwise instability mechanism near invasion fronts that propagate slower than the resonant spreading speed in systems that exhibit a skew-product structure, in particular the amplitude equations that we studied in Section \ref{sec:uni}.

\subsection{Instability of slow KPP fronts}
\label{s:inst1}
In order to motivate our instability criterion, we briefly review fronts in the KPP equation
\[
u_t=u_{xx}+u-u^3,
\]
with $u(x-st;s)$, $u(\xi;s)\to 0$, $\xi\to\infty$, and $u(\xi;s)\to 1$, $\xi\to-\infty$, for any $s>0$. We claim that the linearization 
\[
u_t=\mathcal{L} u = u_{xx}+s u_x+ u-3u_*^2(x;s)u,
\]
is pointwise stable for any $s>2$ and pointwise unstable for any $s<2$. In fact, the resolvent $G_\lambda(x,y)$ can be obtained after conjugation with the exponential weight $\rme^{-sx/2}$, from the self-adjoint operator 
\[
\tilde{\mathcal{L}}u=u_{xx}+(1-s^2/4)u - 3u_*^2(x;s)u. 
\]
Weyl's Lemma immediately gives that the essential spectrum of this self-adjoint operator is the line $\lambda<1-s^2/4$ and one could continue along these lines in order to establish pointwise instability for $s<2$. Here, in order to compute the pointwise Green's function, we recast the equation $\mathcal{L}u=\lambda u + \delta(\cdot-y)$ as a first-order differential equation
\begin{equation}\label{e:kppl}
u_x=v,\qquad v_x=-sv-u+3u_*^2(x;s)u+\lambda u +\delta_y,
\end{equation}
where $\delta_y=\delta(\cdot-y)$ simply implies that left- and right-sided limits of $v$ differ by 1. For $\lambda>1$, there exist unique (up to scalar multiples) solutions $(u,v)_\pm(x;\lambda)$ that decay exponentially for $x\to \pm\infty$. In fact, 
\eqref{e:kppl} induces a flow on the complex Grassmanian, which can be written in coordinates $z=u/v$ and $\zeta=v/u$ as a Riccati equation
\[
z_x=1+sz+(1-3u_*^2-\lambda)z^2,
\qquad \zeta_x=-(1-3u_*^2-\lambda)-s\zeta -\zeta^2.
\]
For $\lambda\to\infty$, the equilibria $\zeta\sim\pm\sqrt{\lambda}$ are stable ($+\sqrt{\lambda}$) and unstable ($-\sqrt{\lambda}$), respectively. The unstable equilibrium at $x=+\infty$ therefore possesses a unique stable manifold $\zeta_+(x;\lambda)$ in the non-autonomous dynamics. Since $\Re\zeta<0$ for $x\to\infty$ and $u_x=\zeta_+ u$, $u$ decays exponentially and corresponds to the unique bounded solution. Similarly, there exists a unique unstable manifold $\zeta_-(x;\lambda)$ for the equilibrium  at $-\infty$, corresponding to solutions with exponential decay as $\xi\to -\infty$. These stable and unstable manifolds are by standard ODE theory analytic in the parameter $\lambda$, as long as the equilibria remain hyperbolic. Inspecting the quadratic Riccati nonlinearity with $u_*=0$, that is, at $x=+\infty$, we find bifurcations of equilibria when $\zeta=-s/2$, $\lambda=1-s^2/4$. Introducing $\gamma^2=\lambda-1+s^2/4$ as new spectral parameter, we find that the equilibrium at $x=+\infty$ is analytic in $\gamma$ and the stable manifold can be continued analytically in a vicinity of $\gamma=0$ as a function of $\gamma$, exploiting exponential convergence of $u_*$, with a nontrivial leading-order term in $\gamma$. As a consequence, the solution $u$ will exhibit a leading-order decay rate $\rme^{\gamma x}$ for $x\to\infty$ and therefore will not be analytic in $\lambda$ --- regardless of the prefactor induced by the matching solutions $u_+$ and $u_-$ with jump in $u_{\pm,x}$ at $x=y$.

The construction outlined is, phrased slightly differently, known as the Gap Lemma \cite{gap1,gap2}. Our point here is slightly different from the analysis there, where eigenvalues are tracked into the essential spectrum.

Of course, the singularity of the bounded solution $u_+(x)$ as a function of $\lambda$ induced by the bifurcation of equilibria at $\pm\infty$ is not the only possible singularity of the Green's function. The most common other type of singularity arises when matching of $u_+$ and $u_-$ at $x=y$ up to the jump fails. Since $u_\pm$ are determined up to scalar multiples only, such singularities arise precisely when $(u_\pm,u_{\pm,x})$ are collinear, that is, when $\zeta_+(0)=\zeta_-(0)$\footnote{Collinearity is independent of $x$ by linearity of the equation.}. Such values of $\lambda$ yield poles of the pointwise Green's function and correspond to eigenvalues for $\lambda$ to the right of the essential spectrum. Particular examples for such poles arise in the context of invasion fronts when nonlinearities amplify growth, that is, for pushed fronts, such as in $u_t=u_{xx}+u+\gamma u^3-u^5$, $\gamma>2/\sqrt{3}$.

\subsection{Instability induced by $1:1$-resonances}
\label{s:inst2}

We present a simple mock-example that illustrates the effect of a 1:1 resonance on the pointwise Green's function. This example was discussed in \cite{holzsch}. Consider therefore the system of transport equations
\begin{subequations}
\begin{align}
u_t&=u_x+\alpha_u u + v,\\
v_t&=-v_x + \alpha_v v,
\end{align}
\end{subequations}
on $x>0$, with boundary conditions $v=\beta u$ at $x=0$. 

The dispersion relation gives $\lambda_u=\nu_u + \alpha_u$, $\lambda_v=-\nu_v+\alpha_v$, such that $\nu_u\to\infty$, $\nu_v\to -\infty$, when $\lambda\to\infty$. Resonances arise when $\lambda_u=\lambda_v$, $\nu_u=\nu_v$, that is, 
\[
\lambda=\lambda_\mathrm{res}=(\alpha_u+\alpha_v)/2,\qquad  \nu=(\alpha_v-\alpha_u)/2.
\]
In particular, one expects instabilities for $\alpha_v+\alpha_u>0$. The Green's function can readily be computed, finding the bounded solutions for $x\to\infty$ in 
\[
u_x=(\lambda-\alpha_u)u - v ,\qquad v_x=(-\lambda + \alpha_v)v,
\]
which are 
\begin{equation}\label{l:ln0}
v(x;\lambda)=v_0(\lambda)\rme^{(\alpha_v-\lambda)x},\qquad u(x;\lambda)=\frac{1}{2(\lambda-\lambda_\mathrm{res})} v_0(\lambda)\rme^{(\alpha_v-\lambda)x},\ \lambda\neq \lambda_\mathrm{res}.
\end{equation}
In order for the limit  $\lim_{\lambda\to 0} u(x;\lambda)$ to exist, we need $v_0(\lambda_\mathrm{res})=0$, such that 
\begin{equation}\label{l:ln00}
v(x;\lambda)=0,\qquad u(x;\lambda)=\frac{1}{2}v_0'(\lambda)\rme^{(\alpha_v-\lambda)x},\ \mbox{ at }\lambda= \lambda_\mathrm{res}.
\end{equation}
As a consequence, one finds an eigenvalue $\lambda=\frac{\beta}{2}+\lambda_\mathrm{res}$ as described in section 3 of \cite[Sec 3]{holzsch}. On the other hand, in the case of a skew-coupled system, $\beta=0$, we see that the solution \eqref{l:ln00} satisfies the boundary condition, thus generating a singularity of the pointwise Green's function at $\lambda=\lambda_\mathrm{res}$. 

On the Grassmanian, we find for $\zeta=v/u$,
\[
\zeta_x=-2(\lambda-\lambda_\mathrm{res})\zeta + \zeta^2,
\]
with a transcritical bifurcation at $\lambda=\lambda_\mathrm{res}$. In other words, as noticed explicitly above, the solution at a fixed value $x$ can be extended in an analytic fashion across the resonance, but flips into the direction of the $u$-component at $\lambda=\lambda_\mathrm{res}$, thus generating a singularity as explained before. 

\subsection{Instability induced by $2:1$-resonances --- an example}
\label{s:inst3}

The toy example from the previous section can easily be adapted to $2:1$-resonances. Consider 
\begin{subequations}
\begin{align}
u_t&=u_x+\alpha_u u  + v^2,\\
v_t&=-v_x + \alpha_v v,
\end{align}
\end{subequations}
on $x>0$, with boundary conditions $v=g(u)$ at $x=0$ for some $g(u)$. To fix ideas, we restrict to 
\[
\alpha_u>0>\alpha_v,
\]
such that stationary profiles will be of the form $v_*(x)=v_0\rme^{\alpha_v x}$. The dispersion relation is as in the previous example, and $2:1$-resonances correspond to $2\lambda_v=\lambda_u$, $2\nu_v=\nu_u$, which gives
\[
\lambda_v^\mathrm{res}=\frac{2\alpha_v+\alpha_u}{4},\quad \lambda_u^\mathrm{res}=\frac{2\alpha_v+\alpha_u}{2},
\quad \nu_v^\mathrm{res}=\frac{2\alpha_v-\alpha_u}{4},\quad \nu_u^\mathrm{res}=\frac{2\alpha_v-\alpha_u}{2}. 
\]
In particular, $\alpha_u>-2\alpha_v$ corresponds to an unstable resonance. With the absence of linear coupling in the equation, we conclude immediately from the equation that the linearization at the trivial state exhibits pointwise decay. One can also verify that pointwise Green's functions possess analytic extensions in $\lambda\in\C$. 

However, considering the linearization at a nontrivial profile $v_*$ with expansions $v_*\sim v_*^0\rme^{\alpha_v x}$, we find
\begin{subequations}
\begin{align}
u_x&=\lambda u-\alpha_u u  - 2v_*^0\rme^{\alpha_v x}v,\\
v_x& = -\lambda v + \alpha_v v,
\end{align}
\label{e:21expl}
\end{subequations}
with a linearized boundary condition $v=\beta u$, $\beta=g'(u_0)$, at $x=0$. Analytic families of bounded solutions satisfy
\[
v(x;\lambda)=v_0(\lambda)\rme^{(\alpha_v-\lambda)x},\qquad u(x;\lambda)=\frac{v_0(\lambda)}{\lambda-\lambda_u^\mathrm{res}}v_*^0 \rme^{(2\alpha_v-\lambda)x}.
\]
Analyticity at $\lambda=\lambda_u^\mathrm{res}$ implies $v_0(\lambda)=0$, and enforces a singularity of the pointwise Green's function in the skew-coupled case. 

\begin{figure}
\includegraphics[width=0.47\textwidth]{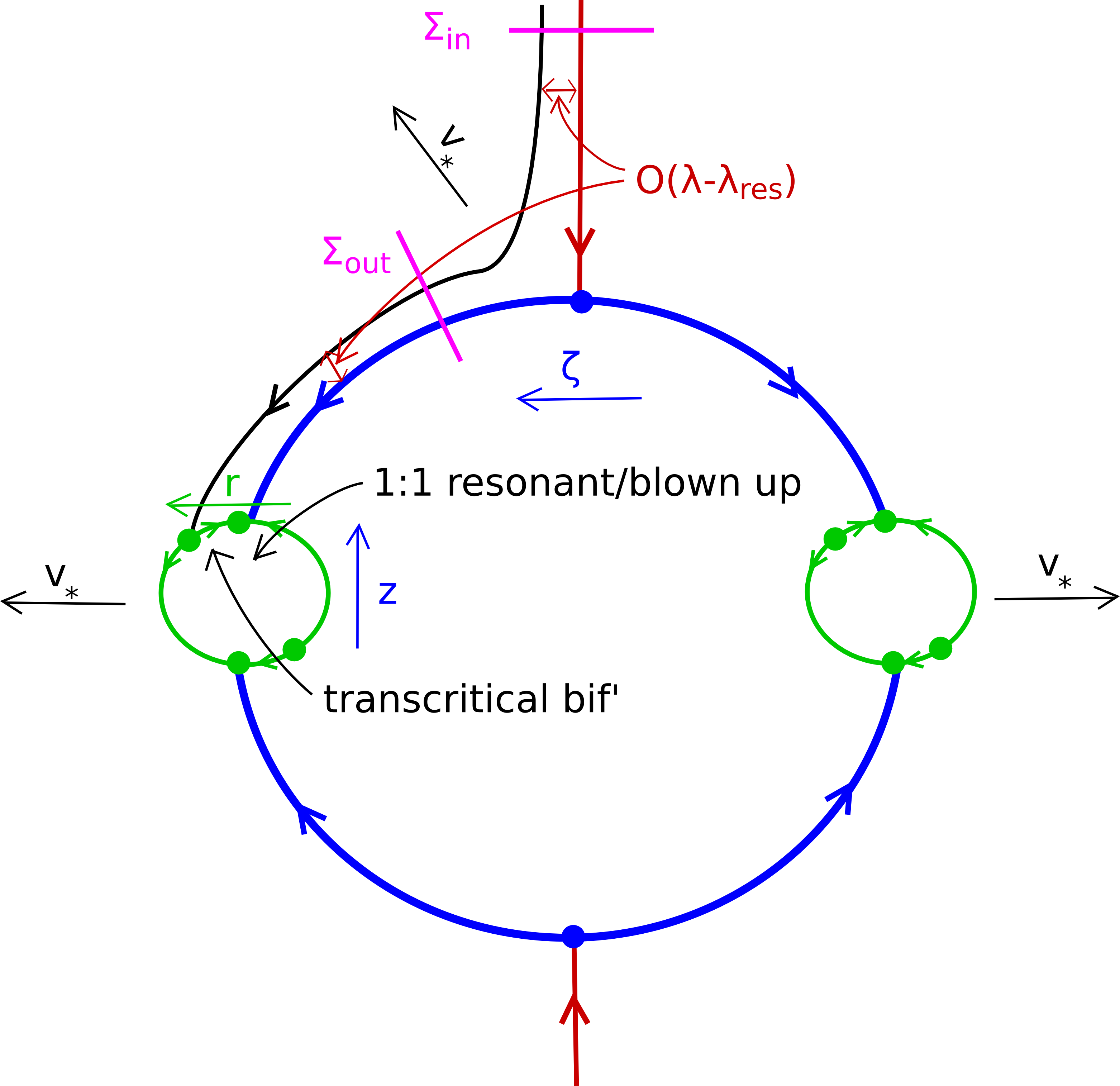}\hfill
\includegraphics[width=0.47\textwidth]{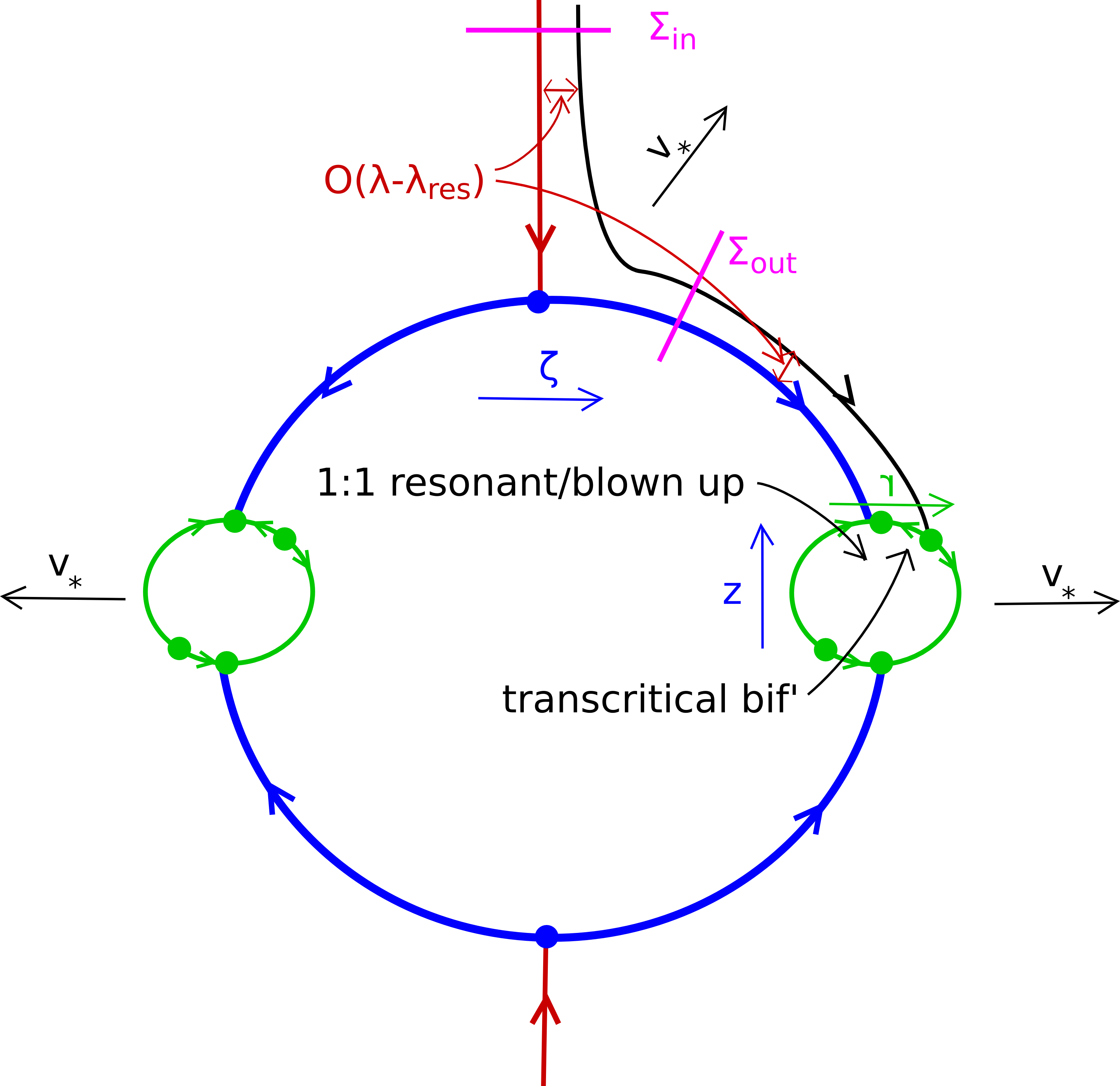}
\caption{Geometric blowup of the passage through the $1:1$-resonance; phase portraits before (left) and after (right) passage through resonance. Note the rotation symmetry by $\pi$ induced by $\R P^1\sim S^1/\Z^2$. }\label{f:grass}
\end{figure}

We next illustrate how this calculation translates into dynamics on the Grassmanian; see Figure \ref{f:grass} for a schematic picture. The equation becomes, again writing $\zeta=v/u$, $z=u/v$,
\[
\zeta_x=-(2(\lambda - \lambda_u^\mathrm{res})+\alpha_v)\zeta +2v_*\zeta^2,\qquad v_{*,x}=\alpha_v v_*,
\]
and, in the complementary chart,
\[
z_x=(2(\lambda - \lambda_u^\mathrm{res})+\alpha_v)z - 2v_*,\qquad v_{*,x}=\alpha_v v_*,
\]
For $\lambda\gg 1$, the stable subspace at $x=\infty$, $v_*=0$, is $z=0$. The stable manifold of this subspace can be parameterized over the $v_*$ component and can be continued analytically while decreasing $\lambda$ until the $1:1$-resonance. As explained above, one can continue this stable manifold past the $1:1$-resonance as a strong stable manifold, a fact that is central to the Gap Lemma \cite{gap1,gap2}. Note that in the present case, due to the absence of linear coupling, the eigenspace is unchanged through the resonance and no eigenvalue is enforced in skew-product systems. In fact, at the $1:1$-resonance, the Grassmannian flow at $v_*=0$ consists entirely of equilibria. The strong stable manifold ceases to be the strong stable manifold when $\lambda$ is further decreased to $\lambda=\lambda_u^\mathrm{res}$, where $z_x=\alpha z -2v_*$, $v_{*,x}=\alpha_v v_*$. 

At this point, a geometric blowup construction as described in \cite{blowup} can elucidate further the dynamics. Introducing $r=v_*/z$ as projective coordinate near $z=v_*=0$, we find 
\[
r_x=-2(\lambda-\lambda_u^\mathrm{res})r + 2 r^2,
\]
with equilibria $r=\lambda-\lambda_u^\mathrm{res}$ and $r=0$. The stable manifold of the nontrivial equilibrium $r=\lambda-\lambda_u^\mathrm{res}$ can be continued through the resonance as follows. At the resonance, the stable manifold is contained in $v_*=0$, which connects on the singular sphere to the equilibrium $\zeta=0,v_*=0$. It can therefore be tracked in backward time following this singular heteroclinic and then the stable manifold of $\zeta=0,v_*=0$, which is simply $\zeta=0$. Past the resonance, $\zeta$ is negative. The connection via the singular heteroclinic on the Grassmannian encodes the ``flip'' of the stable subspace that one can also see in the explicit calculation. 
More precisely, one can track the stable subspace past the singular equilibrium $\zeta=0$ for nonzero $\lambda-\lambda_u^\mathrm{res}$ in a resonant normal form near this equilibrium, which is of the form, after rescaling time,
\begin{align*}
\zeta_x&=-(1+ a(\lambda))\zeta + \zeta g(\zeta v_*,\lambda),\\
v_{*,x}&=v_*,
\end{align*}
where $a(0)=g(0)=0$, $a'(0)=g'(0)\neq0$. In order to obtain the location of the subspace $\zeta(\lambda)$ at a finite distance, $v_*>0$, one analyzes the passage map near $\zeta=0,v_*=0$. Shooting backwards in $x$, we consider a section $\Sigma^\mathrm{out}=\{\zeta=\delta,\ v_*\geq 0\}$, for some fixed $\delta>0$, small. The strong stable manifold intersects $\Sigma^\mathrm{out}$  in a point $v_*=Z(\lambda)>0$, where $Z(\lambda)$ is an analytic function with nonzero derivative at the origin. We would now like to track this strong stable manifold backwards in time $x$ through a vicinity of the equilibrium to a section $\Sigma^\mathrm{in}=\{v_*=\delta\}$ with $\zeta=\zeta_*(\lambda)$. In order to compute $\zeta_*(\lambda)$, we notice that the normal form equation can be reduced to a scalar equation for $w=v_*\zeta$,
\[
w_x=-a(\lambda) w + wg(w;\lambda).
\]
Note that $ w(T)/\delta=Z(\lambda)$, and $\zeta_*(\lambda)=w(0)/\delta$. Analyticity  of the stable subspace is equivalent to analyticity of $w(0)$. The time of flight $T$ is simply $\log(Z(\lambda))\sim \log\lambda$ since $Z'(0) \neq 0$. These log-terms propagate as singularities into the value of $\zeta_*(0)$ unless $w_x=0$. That is, analyticity of the stable subspace relies on the fact that the stable manifold gives rise to an initial shooting condition for the Dulac map which corresponds to a precise balance of the linear detuning $a(\lambda)$ in $\lambda$ and the effect of resonant nonlinearity $g$.

\subsection{Instability induced by $2:1$-resonances --- coupled amplitude equations and the boundary of the Gap Lemma}
\label{s:inst4}
We will show how the geometric picture identified above arises in more general contexts. Consider therefore the system
\begin{align*}
u_t&=d u_{xx}+\alpha_u u - u^3 + v^2 + s u_x,\\
v_t&=v_{xx}+\alpha_v v + h(v)+sv_x,
\end{align*}
where we assume $\alpha_u,\alpha_v>0$. Assume for simplicity that the $v$-equation possesses a pushed front, that is, a stationary solution with steep decay. To make this more precise, we introduce the dispersion relations 
\begin{equation}\label{e:duv}
D_u(\lambda_u,\nu_u)=d\nu_u^2 + s \nu_u + \alpha_u-\lambda_u, \qquad D_v(\lambda_v,\nu_v)=\nu_v^2 + s\nu_v + \alpha_v-\lambda_v,
\end{equation}
with roots $\nu_u^\mathrm{ss}<\nu_u^\mathrm{s}<0$, $\nu_v^\mathrm{ss}<\nu_v^\mathrm{s}<0$ for $s>\max\{2\sqrt{d\alpha_u},2\sqrt{\alpha_v}\}$. We shall assume the existence of a stationary profile $v_*(x)\sim \rme^{\nu_v^\mathrm{ss} x}$ for $x\to\infty$, and an associated $u$-profile $u\sim \rme^{\nu_u^\mathrm{s}x}$. Note that the decay rates are obtained from the spatial eigenvalues $\nu_{u/v}^{\mathrm{s/ss}}$  calculated at $\lambda_{u/v}=0$.  Such profiles can be constructed under suitable assumptions on the nonlinearity $h$. In particular, the $v$-profile can be stable within the $v$-equation. Our goal is to explain how an instability in the full system is generated by the quadratic coupling in the case that the $2:1$-resonance is unstable. We therefore define $2:1$-resonances through the system
\[
2\nu_v^\mathrm{ss}=\nu_u^\mathrm{s},\qquad 2\lambda_v=\lambda_u,
\]
A neutrally stable $2:1$-resonances therefore occurs when $\nu_v=-\sqrt{\frac{2\alpha_v-\alpha_u}{4d-2}}$ for a resonant speed $s_{\mathrm{res}}=-\nu_v - \alpha_v/\nu_v$. We can continue the resonance in $s$, denoting $\lambda^{\mathrm{res}}_u(s)$  and find, differentiating the equations implicitly, that $\partial_s\lambda_u^\mathrm{res}<0$. In other words, profiles with $0<s<s_\mathrm{res}$ possess an unstable resonance.

Returning to the pushed front, the linearization at such a profile is 
\begin{align*}
\lambda u&=d u_{xx}+su_x +\alpha_u u  +2v_* v -3u_*^2 u,\\
\lambda v&=v_{xx}+sv_x +\alpha_v v + h'(v_*)v.
\end{align*}
For simplicity, we drop the terms $-3u_*^2$ and $h'(v_*)v$. The following analysis can readily be extended to incorporate those. Keeping only the leading-order decay term in $v_*$, we arrive at
\begin{subequations}
\begin{align}
\lambda u&=d u_{xx}+su_x +\alpha_u u  +2v_* v,\\
\lambda v&=v_{xx}+sv_x +\alpha_v v,\\
v_{*,x}&=\nu_0^\mathrm{ss} v_*,
\end{align}
\label{e:linres}
\end{subequations}
where $\nu_0^\mathrm{ss}:=\nu_v^\mathrm{ss}(0)$ solves the second equation in \eqref{e:duv} with $\lambda_v=0$. Linearization of (\ref{e:linres}) about the origin we observe resonances between the eigenvalues now occur when the decay rate of $2v_* v$ equals $\nu_u^\mathrm{s}$, that is, when 
\[
\nu_v^\mathrm{ss}(\lambda_*)+\nu_0^\mathrm{ss}=\nu_u^\mathrm{s}(\lambda_*).
\]
Comparing with the equation for resonances, 
\[
2\nu_v^\mathrm{ss}(\lambda_u^\mathrm{res}/2)=\nu_u^\mathrm{s}(\lambda_u^\mathrm{res}),
\]
one finds that 
\[
\lambda_\mathrm{res}^u(s)=\lambda_*(s)+\cO(s-s_\mathrm{res})^2,
\]
such that for $s\lesssim s_\mathrm{res}$, $\lambda_*>0$. The system \eqref{e:linres} can now be solved explicitly near,by thus obtaining asymptotics of the bounded solutions, in a fashion completely analogous to the previous section. In particular, $v=v_0(\lambda)\rme^{\nu_v^\mathrm{ss}(\lambda)x}$, and 
\[
u(x)=u_0(\lambda)\rme^{\nu_u^\mathrm{ss}(\lambda)x}-\frac{v_0(\lambda)}{D_u(\lambda,\nu_0^\mathrm{ss}+\nu_v^\mathrm{ss}(\lambda))}v_*^0\rme^{(\nu_v^\mathrm{ss}(\lambda)+\nu_0^\mathrm{ss})x},
\]
such that analyticity implies $v_0(\lambda_*)=0$ and the stable subspace is entirely contained in the $u$-component. On the other hand, the two-dimensional subspace of solutions bounded at $x=-\infty$ is not entirely contained in the $v$-component, which implies the existence of an intersection and therefore a singularity of the pointwise Green's function. 

One can generalize this reasoning to cases that are not explicitly integrable or those that do include coupling terms $v_* u$ in the $v$-equation using variations of the techniques described previously. Dynamics of the stable subspace can be described within the Grassmannian $\textbf{Gr}(4,2)$ of 2-dimensional subspaces in 4-dimensional space. Writing \eqref{e:linres} as a first-order equation and diagonalizing the linear part at $v_*=0$, we find an equation of the form
\begin{align*}
U_x^\mathrm{s}&=(A^\mathrm{s}+\kappa C^\mathrm{s})U^\mathrm{s}+\kappa B^\mathrm{s} U^\mathrm{ss},\\
U_x^\mathrm{ss}&=(A^\mathrm{ss}+\kappa C^\mathrm{ss})U^\mathrm{s}+\kappa B^\mathrm{ss} U^\mathrm{s}.
\end{align*}
Here, $U^\mathrm{s/ss}=(u^\mathrm{s/ss},v^\mathrm{s/ss})$ are coordinates in the respective stable and strong stable eigenspaces, $A^\mathrm{s/ss}=\mathrm{diag}\,(\nu_u^\mathrm{s/ss},\nu_v^\mathrm{s/ss})$. Writing $U^\mathrm{s}=ZU^\mathrm{ss}$ with $2\times 2$-matrix $Z$, we obtain the matrix Riccati equation
\[
Z_x=(A^\mathrm{s}+v_* C^\mathrm{s})Z-Z(A^\mathrm{ss}+v_*C^\mathrm{ss})+v_*(B^\mathrm{s}-ZB^\mathrm{ss}Z),
\]
with equilibrium $Z=0$, $v_*=0$. The flow at $v_*=0$ is explicit, with equilibrium $Z=0$ and eigenvalues $\nu_j^\mathrm{s}-\nu_k^\mathrm{ss}$, $j\neq k$, $j,k\in \{u,v\}$. We see that there is precisely one negative eigenvalue $\nu_u^\mathrm{s}-\nu_v^\mathrm{ss}$ near the $2:1$-resonance. At the resonance, this stable eigenvalue becomes the strong stable eigenvalue compared to the eigenvalue in the direction of $v_*$, in complete analogy to the previous example. As a consequence, the strong stable manifold needs to be tracked following a singular heteroclinic, which corresponds to the stable manifold of $Z=0$. This stable manifold connects to the unique completely unstable equilibrium on the Grassmannian, which is the strong stable subspace of the $U$-system, given by $v=0$. The linearization at this subspace possesses the equivalent $1:-1$ resonance as seen in the previous simple example and can be analyzed in a similar fashion.

\section{Quadratic resonance speeds in numerical simulations}\label{sec:num}

We consider four classes of problems and compare numerically observed spreading speeds to the prediction obtained from the criterion in Definition~\ref{defi:quad}.  All examples center around the common theme of coupling a Turing mode (or its amplitude) to a homogeneous mode. 

\begin{itemize}
\item{\bf Unidirectional coupled amplitude equation.} This class of equations includes  (\ref{eq:CGLone}) where the dynamics of the Turing mode decouple from those of the zero mode.  We compare results with Theorem \ref{thm:CGLone}.
\item{\bf Bi-directionally coupled amplitude equation.} More general systems near the onset of instability will lead to amplitude equations where the zero mode couples into the Turing mode and vice-versa.  An example of this situation is provided by the amplitude equations
\begin{subequations}
\begin{align}
U_T&= k_0 U_{XX}+U\left(\delta_0+ \frac{U}{2}-\frac{U^2}{3}-2|A|^2\right)+|A|^2,  \\
A_T&= k_cA_{XX}+A\left(\delta_c-|A|^2+U-U^2\right). 
\end{align}
\label{eq:CGLtwopreview}
\end{subequations}
We note that the linearization about the unstable zero state is diagonal and the quadratic coupling term appears as in the unidirectional case in (\ref{eq:CGLone}).   Since linear spreading speeds are determined in the leading edge these two features are sufficient to allow for predictions to be made, although we are unable to give a proof that spreading speeds are as predicted and observed.

\item{\bf Swift-Hohenberg  coupled to Nagumo's equation.}
As a simple  example of a (non-normal form) pitchfork-Turing instability, we will consider the Swift-Hohenberg equation coupled to Nagumo's equation, 
\begin{subequations}
\begin{align}
u_t &= du_{xx}+\e^2\alpha u-u^3+\e \gamma v^2  \\
v_t &= -(\partial_x^2+1)^2v +\e^2 v-v^3. 
\end{align}
\label{eq:SHKPPpreview}
\end{subequations}
For  $\alpha$ positive, $\e^2$ detunes past the onset of a simultaneous Turing/pitchfork bifurcation. The quadratic term in the first equation induces the desired quadratic coupling between unstable modes.  For $0<\e\ll 1$, this system can be well approximated by the amplitude equations (\ref{eq:CGLone}) for which Theorem~\ref{thm:CGLone} provides exact results on spreading speeds. Continuing to larger values of $\e$, the amplitude equation formalism introduces significant errors.  We predict invasion speeds using our criterion from Definition~\ref{defi:quad} both for the amplitude equation and for the full system and compare to direct numerical simulations.  We find that speeds predicted from amplitude equations and from the full system agree well with numerical simulations for $\e$ small. For moderate $\e$, the results for amplitude equations deviate significantly from the numerical simulations of the full system, but quadratic speeds computed from the dispersion relation of the full system agree well with the numerics, thus validating the linear approximation even for finite amplitudes. 

\item{\bf A neural field model.}
As a more ``generic'' example, without particular simplifying structure,  we consider the following single-layer neural field model \cite{amari77,wilsoncowan73},
\begin{equation} 
u_t = - \mu_\e u + \K_\e*S_\e(u), 
\label{eq:NFpreview}
\end{equation}
where 
\bqs
\K_\e*S_\e(u)(x):=\int_\R \K_\e(x-y)S_\e(u(y))\mathrm{d}y, \quad \text{ for all } x\in\R.
\eqs
We assume that the firing rate function $S_\e$ is of sigmoidal form such that $u=0$ is always a homogeneous stationary state for all $\mu_\e>0$. Here, $\K_\e$ is the connectivity function and will be chosen such that it features local excitations and lateral inhibitions which are spatially modulated. From a modeling perspective, such connectivity functions encode the functional architecture of cortical areas and have been used to study short term working memory  in the prefrontal cortex \cite{laingetal02} and cortical spreading properties in visual areas \cite{buzasetal01,rankinetal14}. We refer the interested reader to the comprehensive review \cite{bressloff12} for further explanations on neural field models. Our main hypothesis on $\K_\e$ and $S_\e$ implies the presence of a a simultaneous Turing/pitchfork bifurcation at $u=0$, $\e=0$. This hypothesis can easily be satisfied by imposing that the Fourier transform of $\K_\e$ is maximal at the modes $\ell=0$ and $\ell=\pm \ell_c$, for some $\ell_c>0$, when evaluated at $\e=0$. When $\e\ll 1$, amplitude equations can be derived that take the form of (\ref{eq:CGLtwopreview}).  Continuing to larger values of $\e$ these equations do not provide good approximations and we use the quadratic speed criterion to make predictions.  We emphasize that equation (\ref{eq:NFpreview}) is scalar and  amplitude equations do not exhibit a skew-product structure as present in the toy model (\ref{eq:SHKPPpreview}). Again, we predict quadratic interaction speeds using Definition~\ref{defi:quad} and find good agreement with numerical simulations, also for $\e$ not necessarily small. 
\end{itemize}

\subsection{Unidirectionally coupled amplitude equations}

\begin{figure}[ht]
\centering
\includegraphics[width=0.4\textwidth]{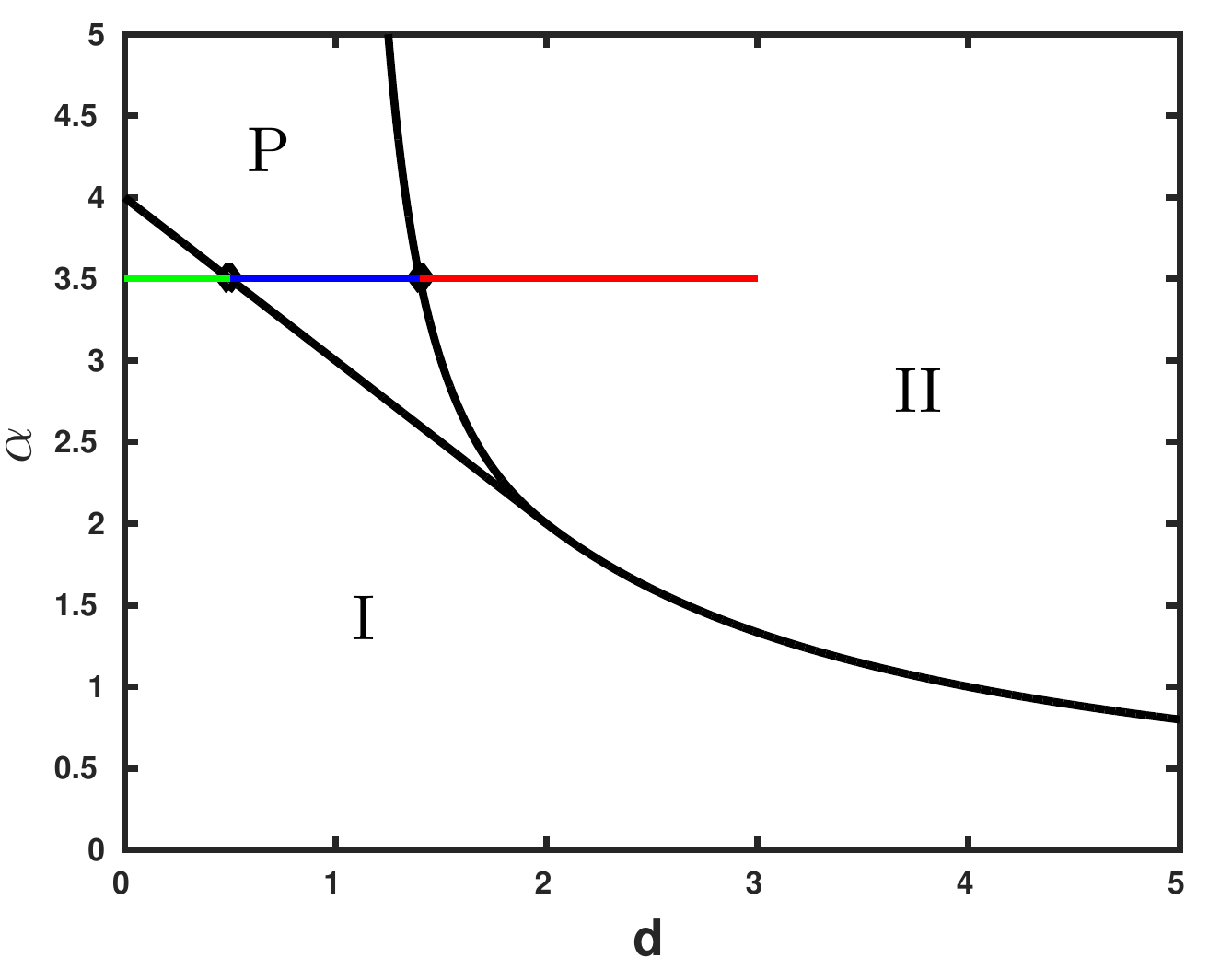}
\includegraphics[width=0.44\textwidth]{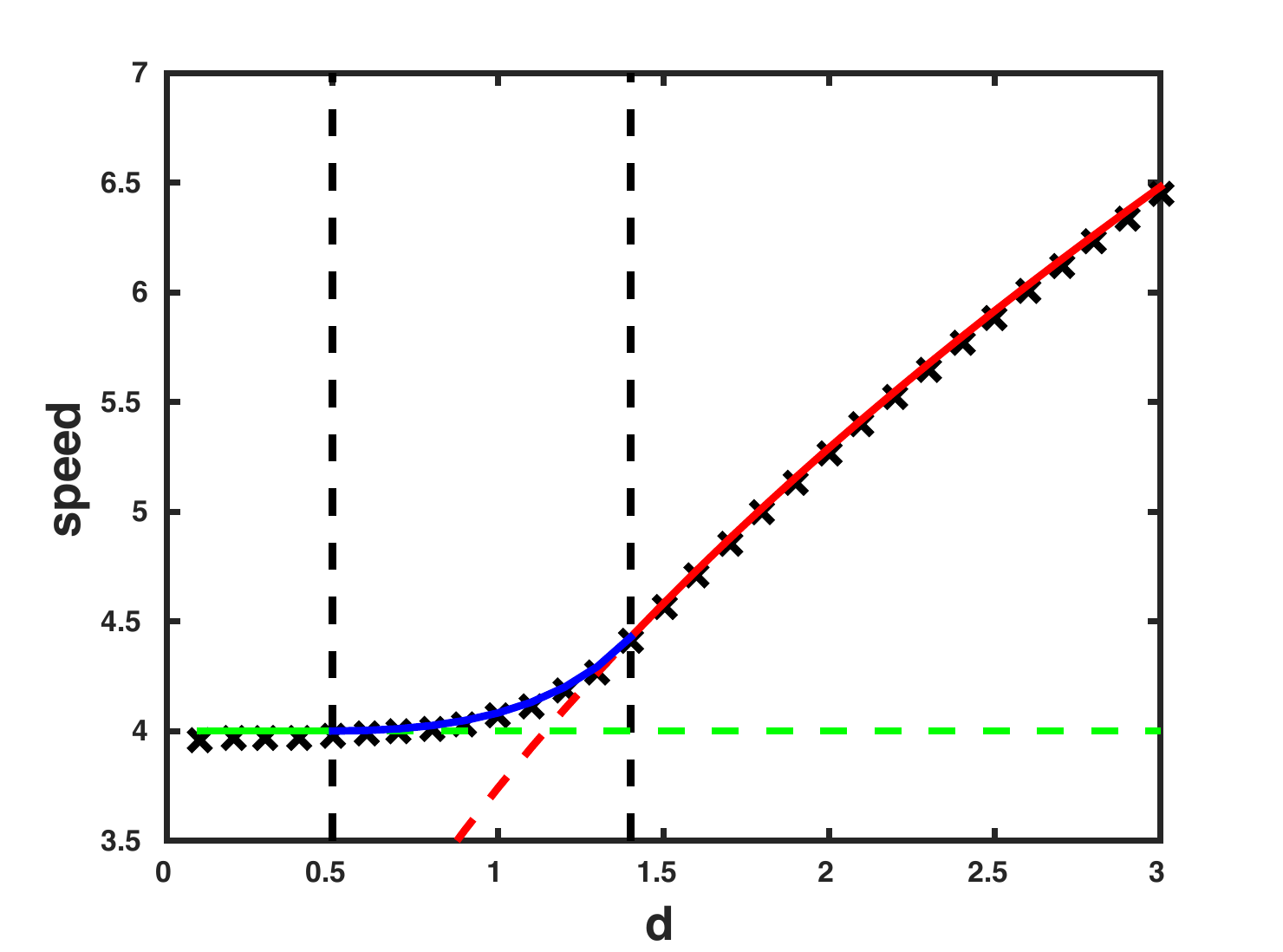}
\caption{Comparison of numerically observed spreading speeds (crosses) with theoretical predictions (colored lines). Theory predicts transitions from speeds $s_A$ (green) to $s_{AU}$ (blue)  to $s_U$  (red) as $d$ is increased (left). }
\label{fig:speedAMP}
\end{figure}

\begin{figure}[ht]
\centering
 \subfigure[Space-time plot of $U$.]{\includegraphics[width=0.4\textwidth]{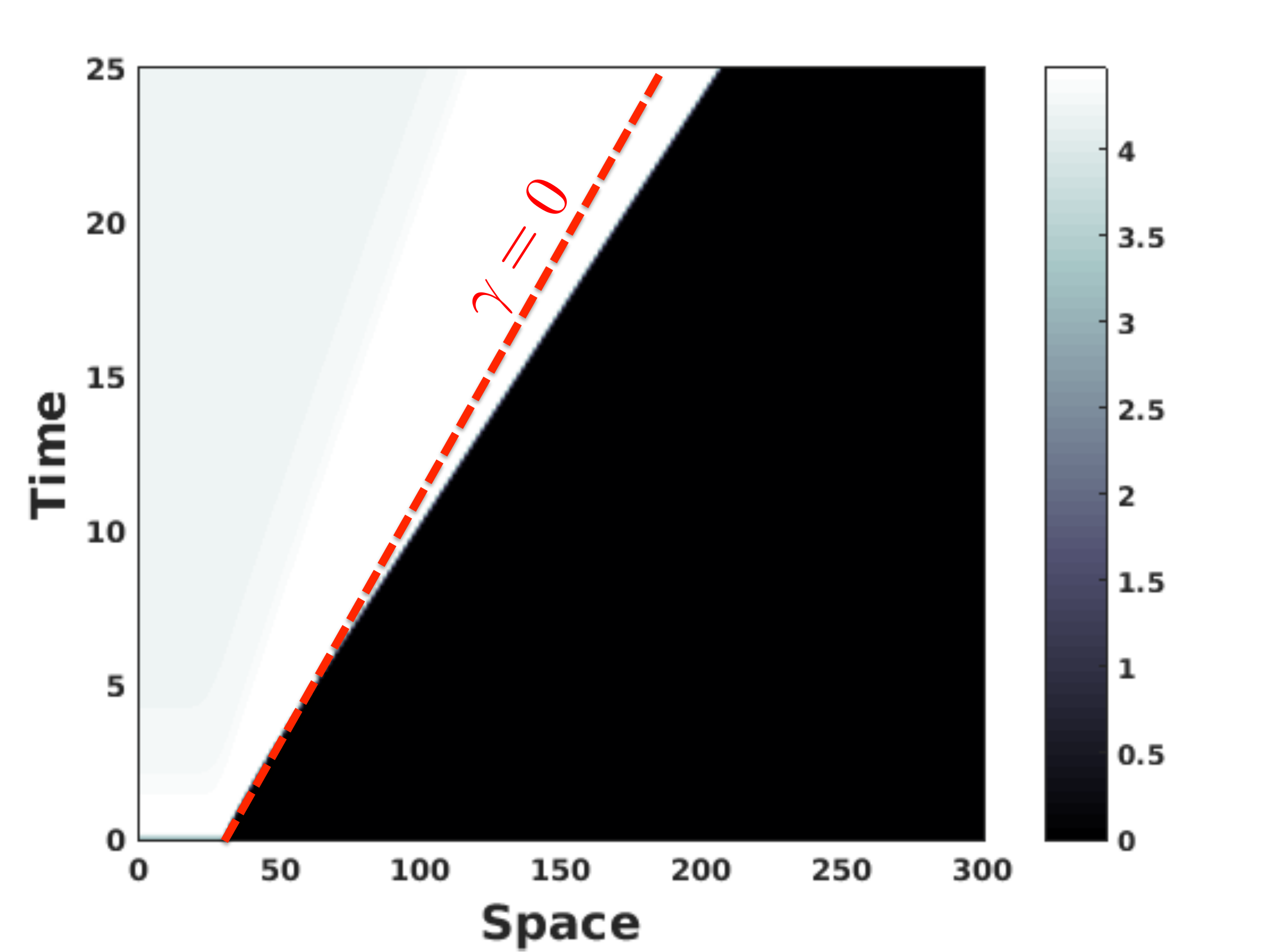}}
 \subfigure[Space-time plot of  $A$.]{\includegraphics[width=0.4\textwidth]{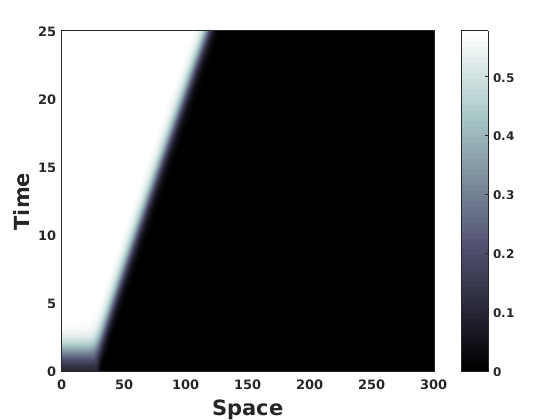}}
 \subfigure[Space-time plot of $U$ when $\gamma=0$.]{\includegraphics[width=0.4\textwidth]{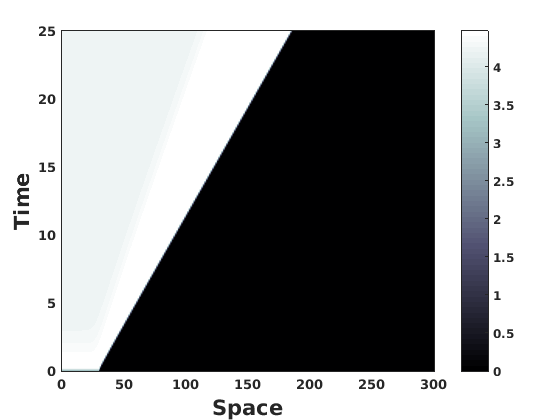}}
 \subfigure[Profiles at $t=25$.]{\includegraphics[width=0.4\textwidth]{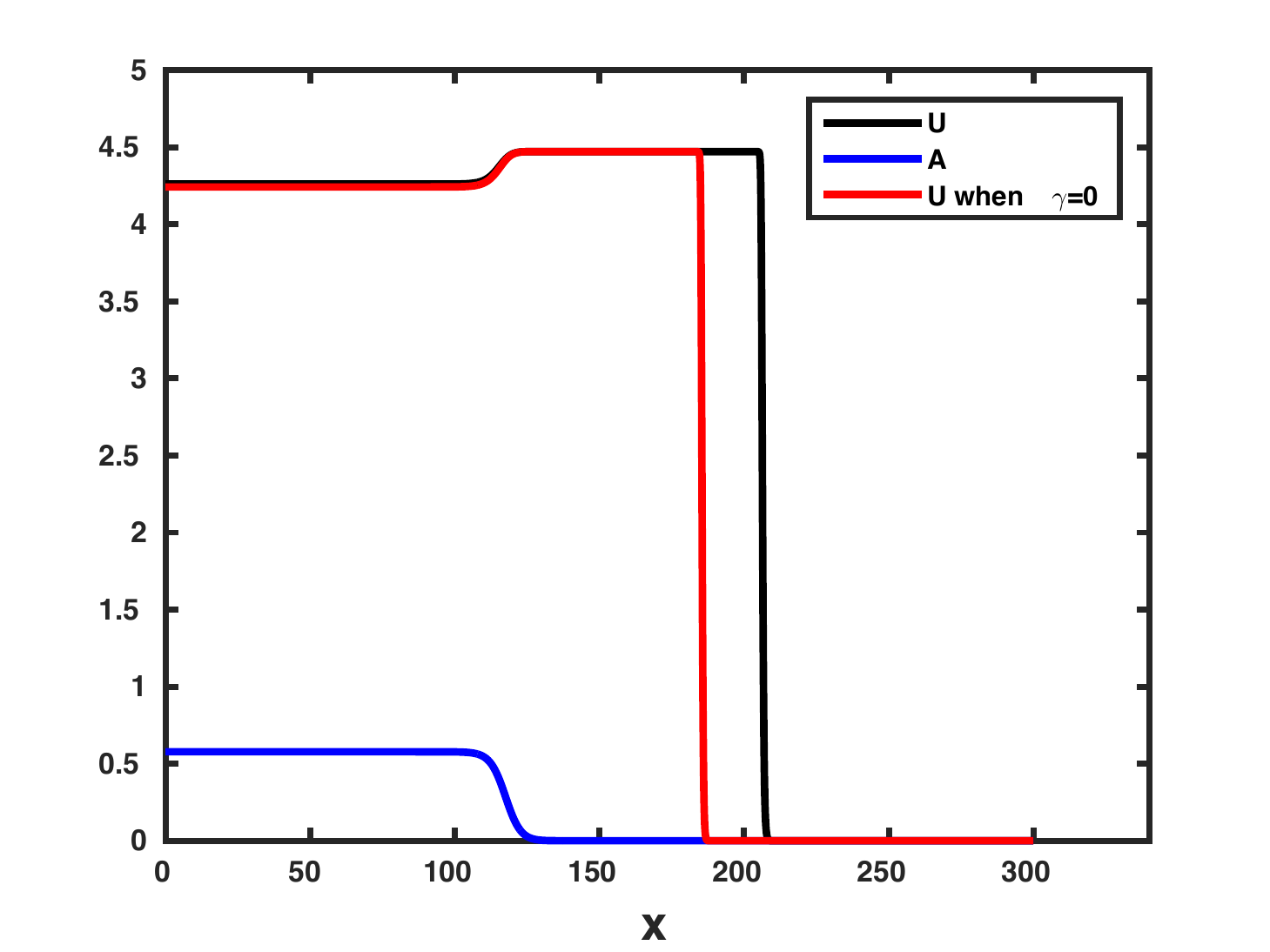}}
\caption{(a)-(b) Space-time plots of the solutions of equations \eqref{eq:CGLone} in the case $\gamma =1$ where the selected wave speed for the $U$ component is $s_{AU}$.(c) Space-time plot of the $U$-component of equations \eqref{eq:CGLone} in the uncoupled case $\gamma=0$ where the selected wave speed is $s_U$.(d) Plot of the profiles at $t=25$ of the solutions taken from each space-time plot. We recover the fact that in the coupled case the U component (black) spreads at a faster wave speed $s_{AU}$ than in the uncoupled case (red) and that of the $A$ component (blue). Values of the parameters are fixed to $d=1/2$ and $\alpha=20$ so that $(d,\alpha)\in \mathrm{P}$ from Figure~\ref{fig:speedAMP}. Note that in this regime we have $s_A<s_U<s_{AU}$.}
\label{fig:STplot}
\end{figure}

\begin{figure}[ht]
\centering
 \subfigure[Space-time plot of $\log(U)$.]{\includegraphics[width=0.4\textwidth]{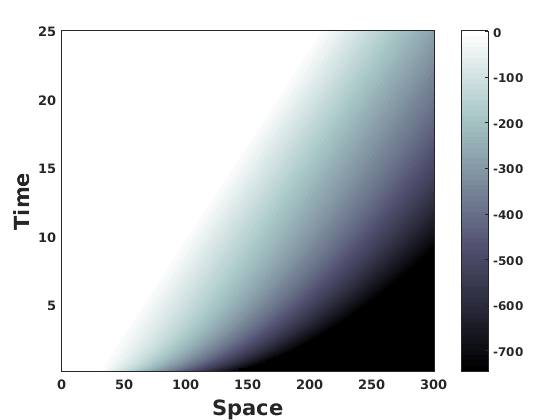}}
 \subfigure[Space-time plot of  $\log(A)$.]{\includegraphics[width=0.4\textwidth]{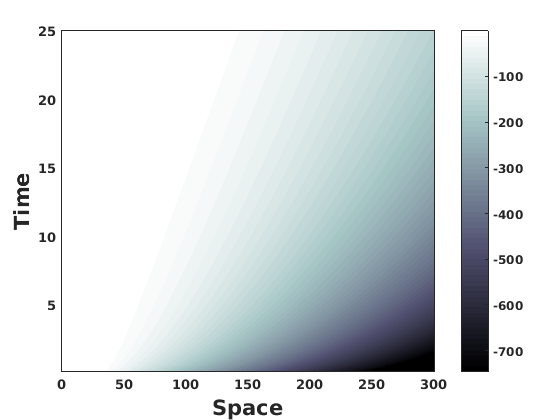}}
 \subfigure[Space-time plot of $\log(U)$ when $\gamma=0$.]{\includegraphics[width=0.4\textwidth]{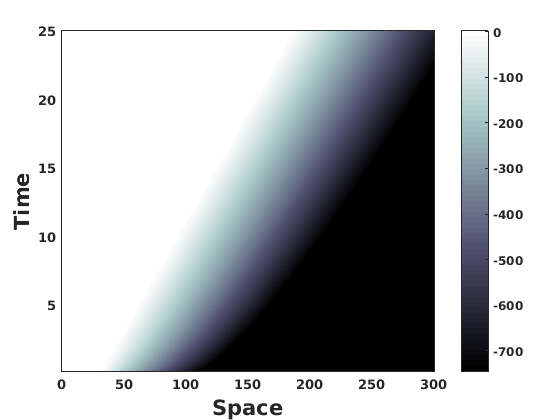}}
 \subfigure[Log-profiles at $t=25$.]{\includegraphics[width=0.4\textwidth]{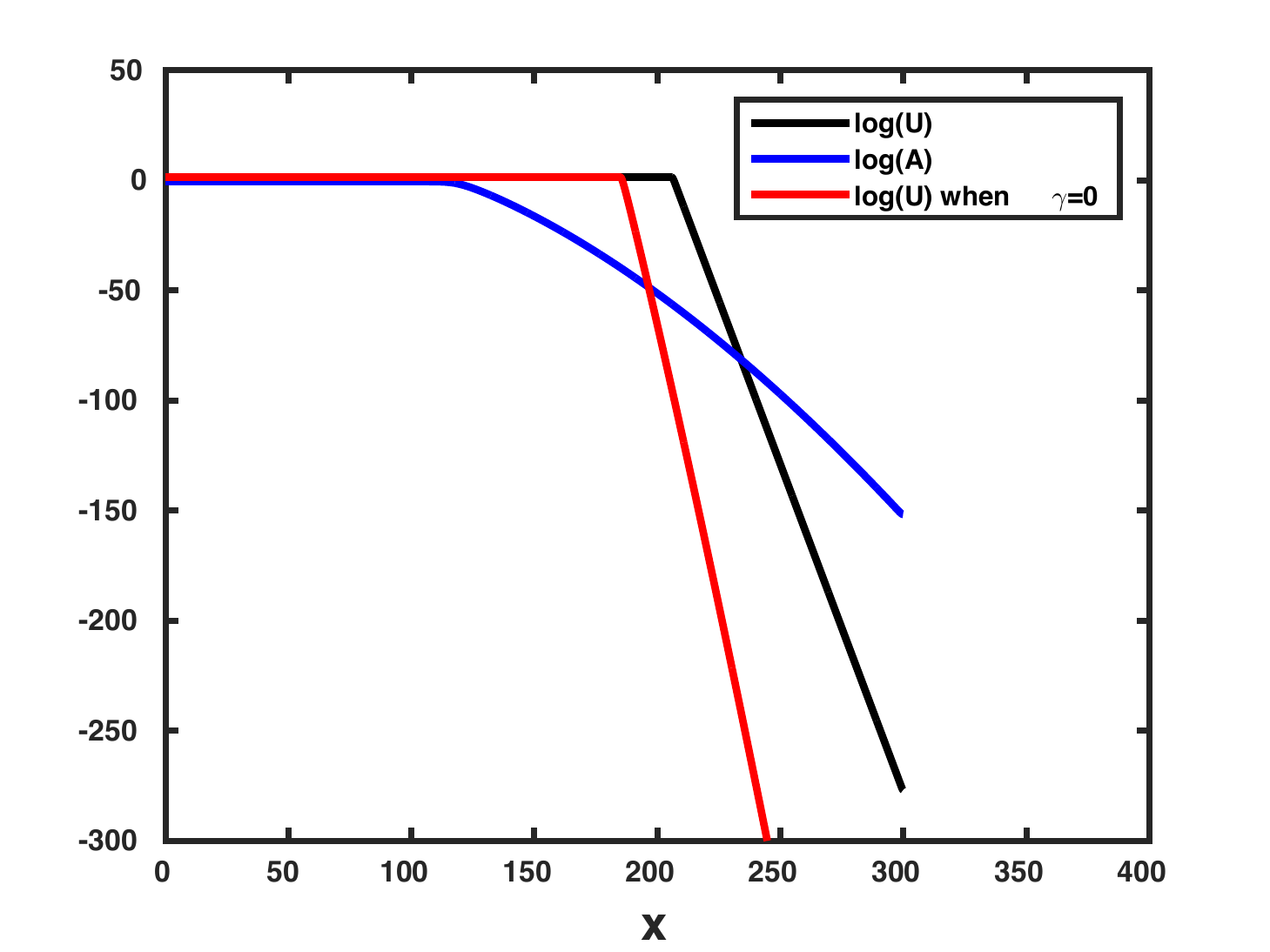}}
\caption{(a)-(b)-(c) Space-time plots of the $\log$ of the amplitude of the profiles from Figures~\ref{fig:STplot}(a)-(b)-(c). (d) Log-plot of the profiles at $t=25$ of the solutions taken from each space-time plot. We observe the faster decay of the $U$ component in the uncoupled case. Values of the parameters are fixed to $d=1/2$ and $\alpha=20$ so that $(d,\alpha)\in \mathrm{P}$ from Figure~\ref{fig:speedAMP}.}
\label{fig:STlogplot}
\end{figure}

We observed the spreading speeds established in Theorem~\ref{thm:CGLone} in numerical simulations as demonstrated  in Figure~\ref{fig:speedAMP}; note the transition of speeds from $s_A$ to $s_{AU}$ to $s_U$ as $d$ is increased. We have also reported in Figure~\ref{fig:STplot}  space-time plots of the solutions of equations \eqref{eq:CGLone} in the regime where the quadratic spreading speed established in Theorem~\ref{thm:CGLone} is selected.  Note the faster invasion speed in the presence of coupling.  This is made more apparent in Figure~\ref{fig:STlogplot} where space-time plots for the logarithm of the solution are compared.  

\subsection{Example: Swift-Hohenberg coupled to Nagumo's equation}\label{sec:SH}

In this section, we consider 
\begin{subequations}
\begin{align}
u_t &= du_{xx}+\e^2\alpha u-u^3+\e \gamma v^2 \\
v_t &= -(\partial_x^2+1)^2v +\e^2 v-v^3, 
\end{align}
\label{eq:SHKPP}
\end{subequations}
consisting of a Swift-Hohenberg equation (for $v$) coupled to a Nagumo equation (for $u$) with inhomogeneous quadratic coupling.  When $\e=0$, this system undergoes a simultaneous Turing/pitchfork bifurcation and amplitude equations can be derived via a multiscale analysis.  In this section, we compare spreading speeds derived from the amplitude equations to those given by the linear criterion and to direct numerical simulations.  

Comparisons of spreading speeds were performed as follows.  System (\ref{eq:SHKPP}) was solved using finite differences with Heaviside step function initial data and speeds were calculated by computing how much time elapsed between the solution passing a threshold at two separate points in the spatial domain.  These spreading speeds were compared with predictions using the linear criterion in Definition~\ref{defi:quad}.  These predictions were found using numerical continuation.  

\begin{figure}[t]
\centering
   \includegraphics[width=0.4\textwidth]{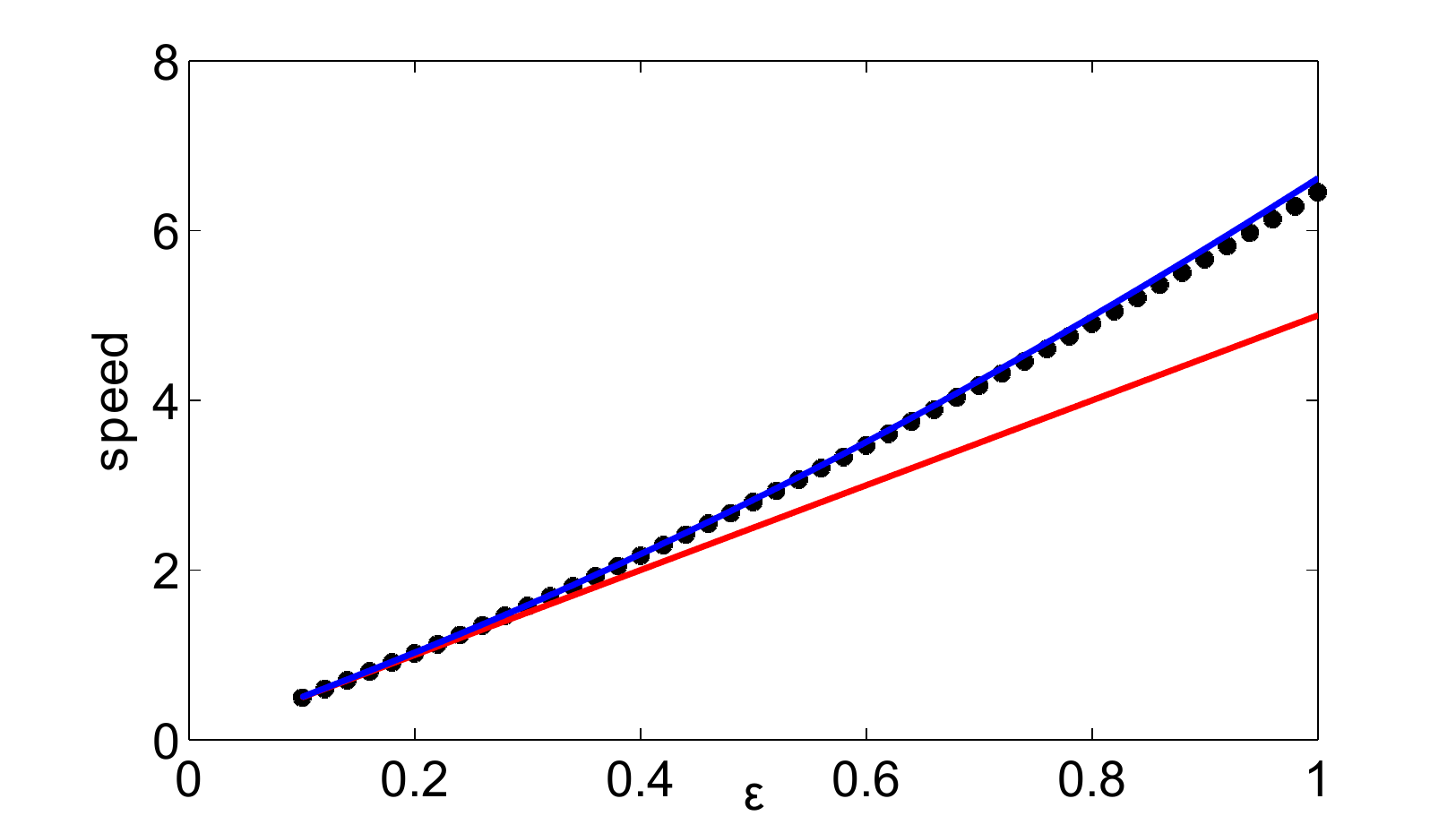}
   \includegraphics[width=0.4\textwidth]{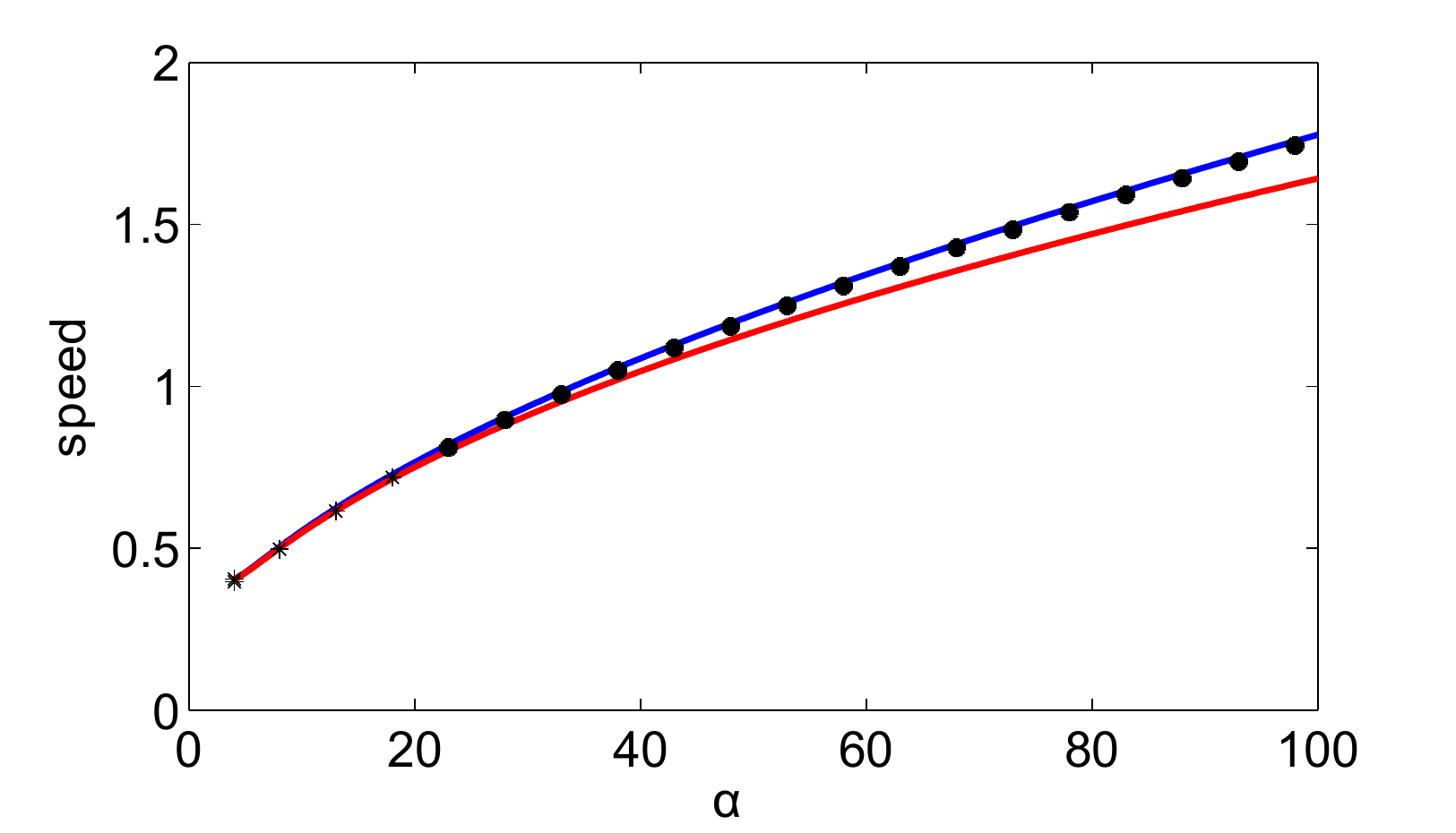}  
\caption{Comparison of numerically observed spreading speeds (black) for equation (\ref{eq:SHKPP}), compared to linear predictions generated by the amplitude equations (red) and the criterion in Definition~\ref{defi:quad}.  On the left, we vary $\e$ and fix all other parameters to $d=0.5$, $\alpha=8.0$ and $\gamma=1.0$.   Note that all three speeds agree for small values of $\e$ while the linear criterion remains valid for larger values of $\e$.  On the right, we vary $\alpha$ while fixing all other parameters to $d=0.5$, $\e=0.1$ and $\gamma=1.0$.}
\label{fig:SHKPPspeeds}
\end{figure}

Typical results are plotted in Figure~\ref{fig:SHKPPspeeds}.  The prediction from the linear criterion ($s_\mathrm{quad}$) matches well with the speed from the amplitude equation when $\e$ is small.  For larger values of $\e$, the two speeds deviate and the spreading speeds for (\ref{eq:SHKPP}) observed in direct numerical simulations match closely with the speed $s_\mathrm{quad}$ in this regime.  

\begin{figure}[ht]
\centering
 \subfigure[$\epsilon=0.1$]{
\label{fig:svsq1}
\includegraphics[width=0.4\textwidth]{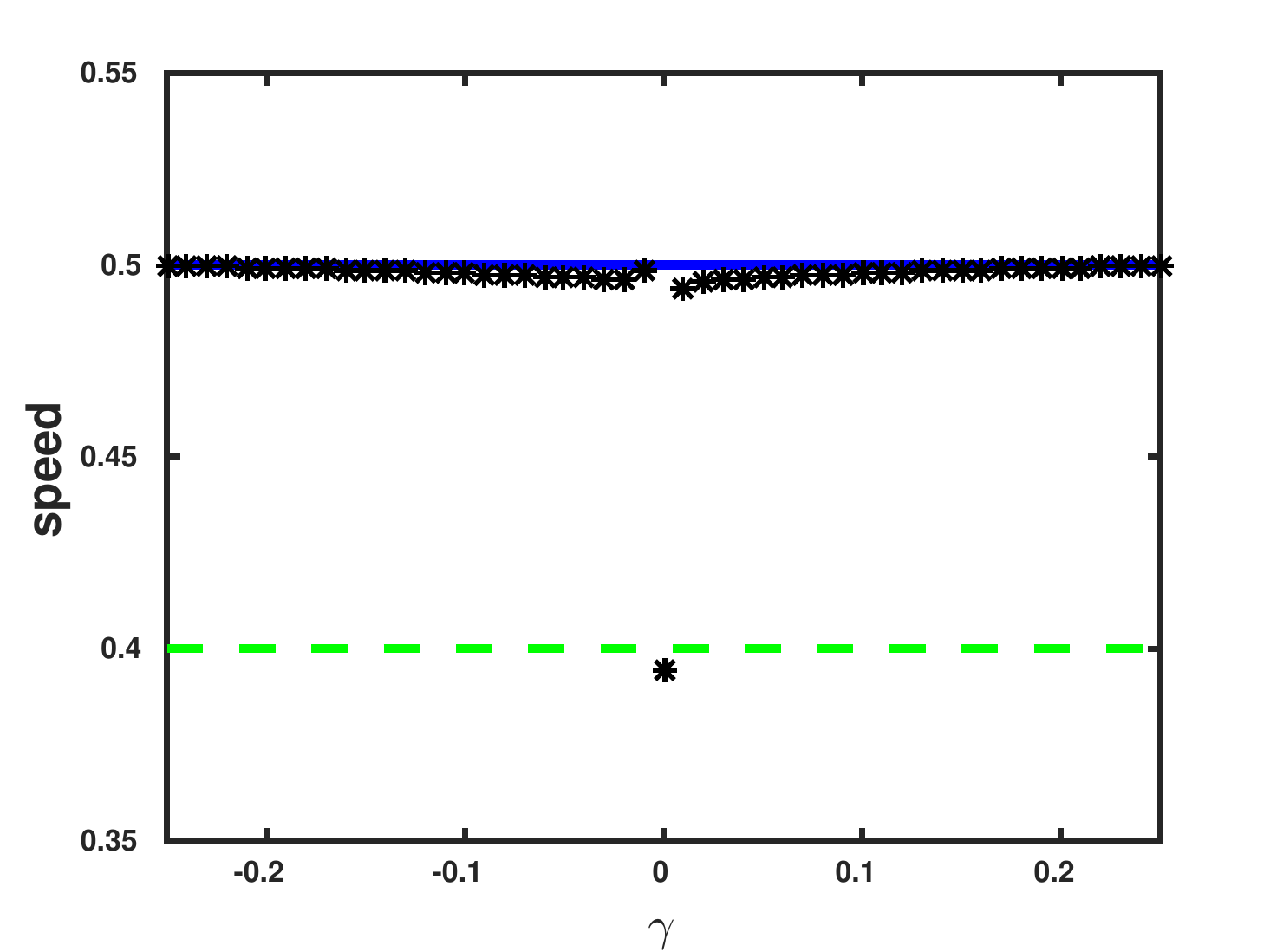}}
\hspace{.3in}
\subfigure[$\epsilon=1$]{
\label{fig:svsq2}
\includegraphics[width=0.4\textwidth]{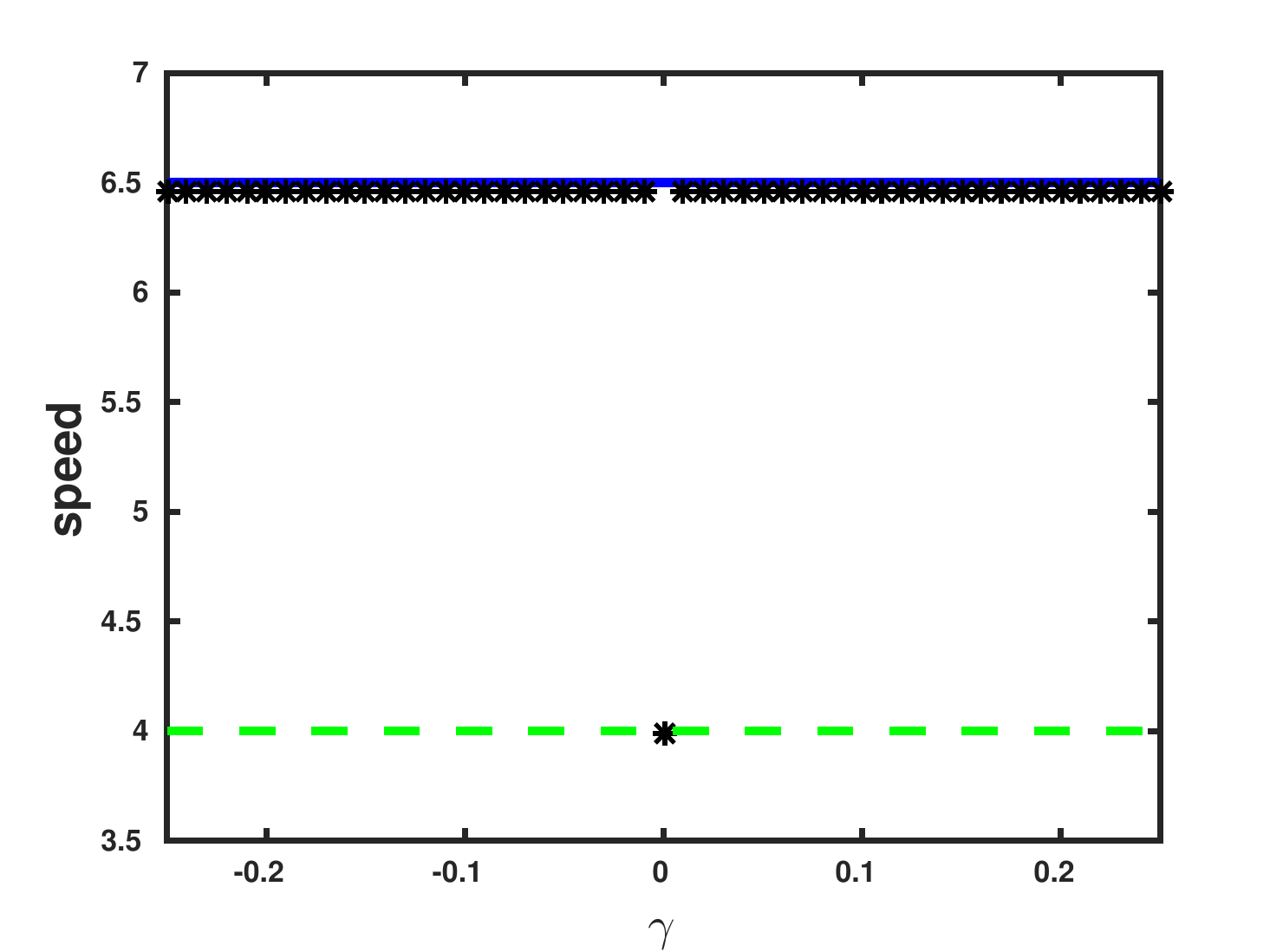}}
\caption{Speed vs quadratic terms. We vary $\gamma$ in \eqref{eq:SHKPP} from $-0.25$ to $0.25$ and compare the numerically observed spreading speeds (black) to the criterion in Definition~\ref{defi:quad} (blue) for $\e=0.1$ (a) and $\e=1$ (b). The dotted green line represents the linear spreading speed of the $u$ component of \eqref{eq:SHKPP} in isolation. We remark that our predicted linear spreading speed  from the criterion in  Definition~\ref{defi:quad} is independent of the strength of the quadratic coupling. The other values of the parameters are set to $d=0.5$, $\alpha=8$ and $\mu=1$.}
\label{fig:speedVSquad}
\end{figure}

We also investigated the role of $\gamma$ in the speed selection.  These results are depicted in Figure~\ref{fig:speedVSquad}.  From our analysis, we expect the value of $\gamma$ to not be relevant to the spreading speed selection.  The exception occurs at $\gamma=0$, where the quadratic term that enforces the faster spreading speeds ceases to exist.  Thus, we expect that the spreading speeds of the $u$ component in (\ref{eq:SHKPP}) should be discontinuous at $\gamma=0$.  Indeed, this is what is observed.

\subsection{Example: A scalar neural field model}\label{sec:NFE}

We now turn our attention to our second main example in this article which takes the form of a scalar neural field equation.  Similar to the previous section, our main goal here is to compare spreading speeds predicted by the amplitude equation, the linear criterion and those observed in direct numerical simulations.  We first describe the model and explain how the amplitude equation (\ref{eq:CGLtwopreview}) is derived.  Next we compute and compare spreading speeds in the full model, with those in the amplitude equation and from the linear criterion.

\begin{figure}[t]
\centering
\subfigure[$S_\e$]{
\label{fig:sigmoid}
\includegraphics[width=0.4\textwidth]{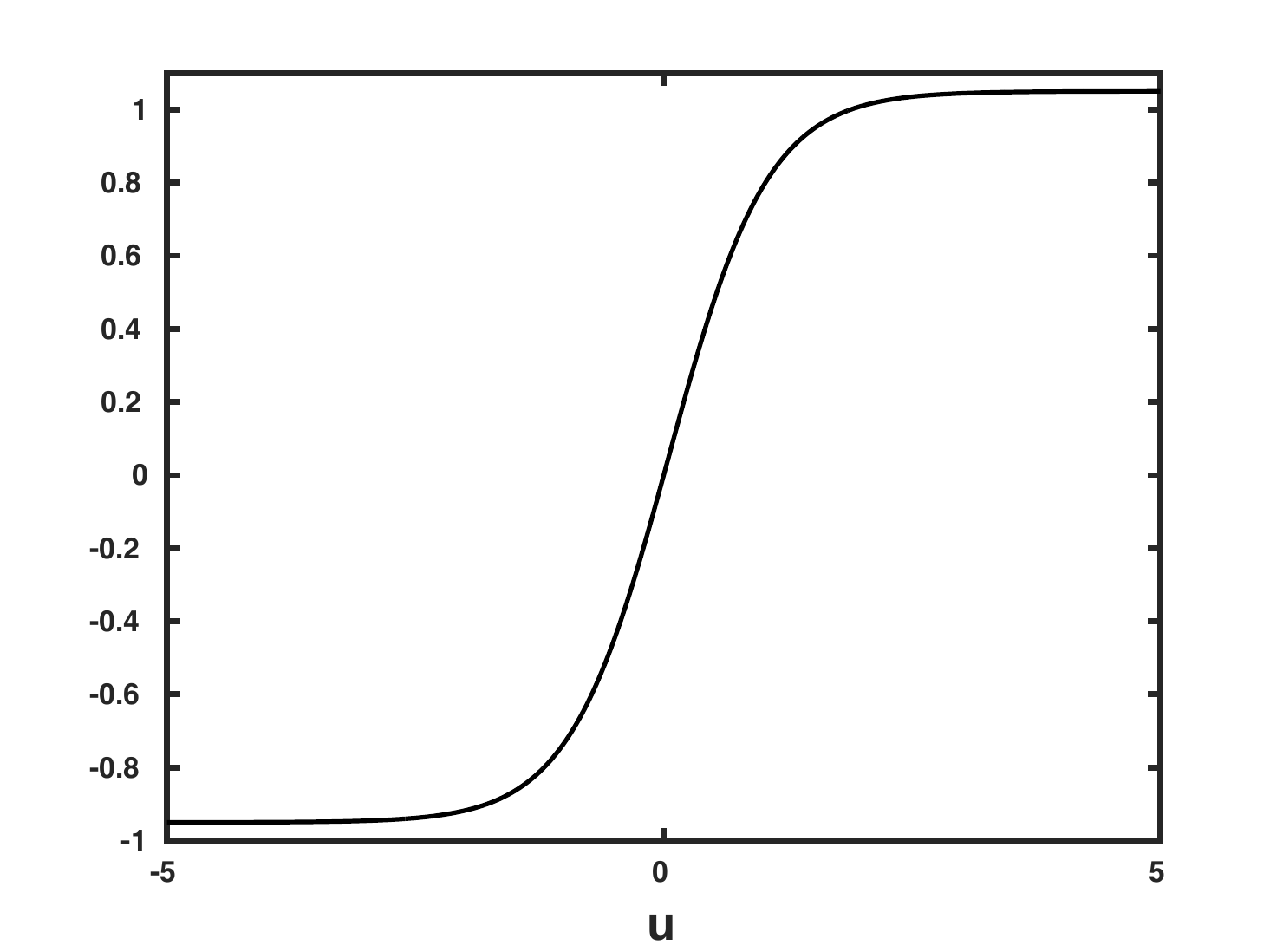}}
\hspace{.3in}
\subfigure[$\K_\e(x)$]{
\label{fig:kernel}
\includegraphics[width=0.4\textwidth]{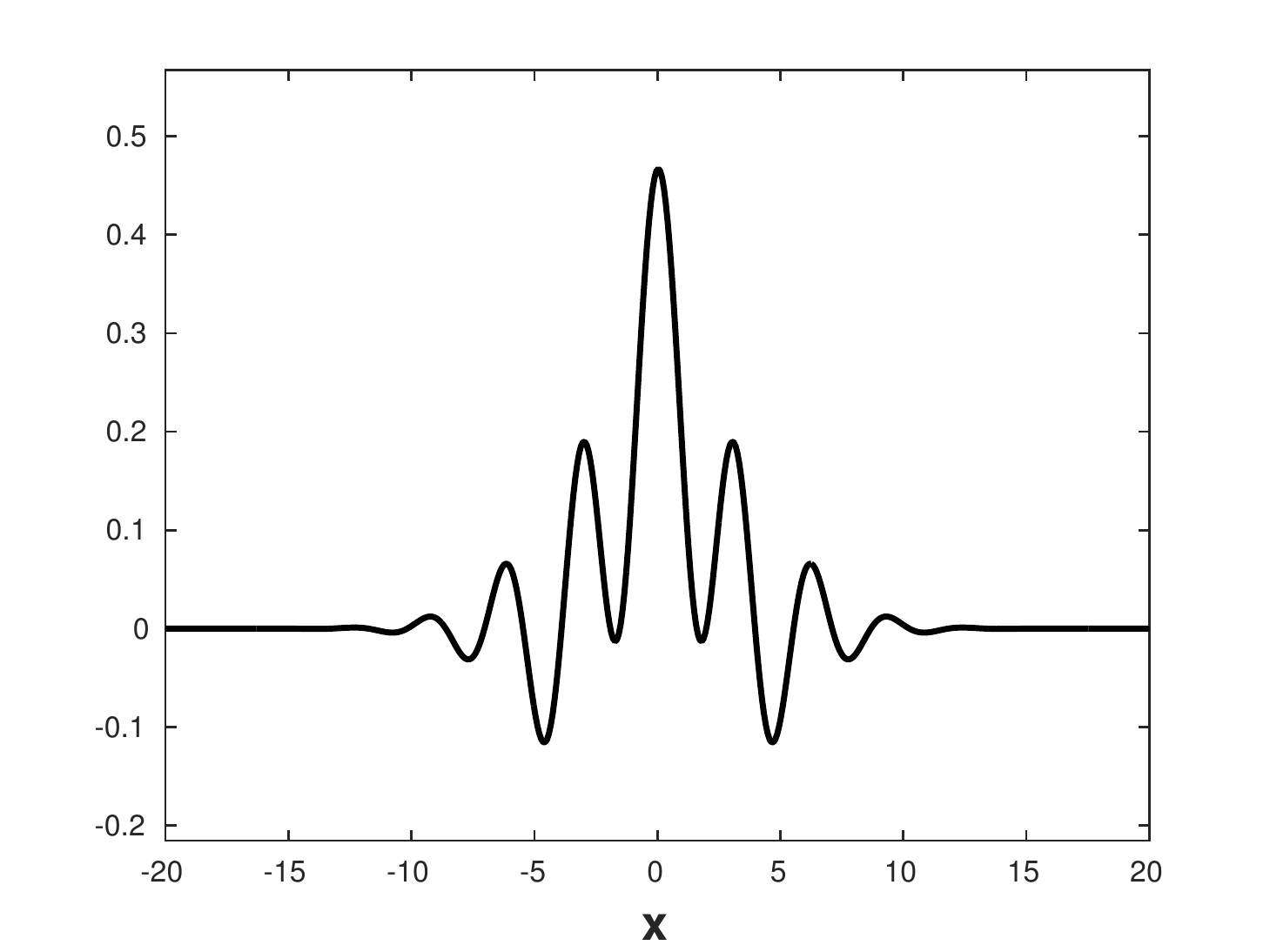}}
\subfigure[$\widehat{\K_\e}(\ell)$ -- Model 1.]{
\label{fig:kerF1}
\includegraphics[width=0.4\textwidth]{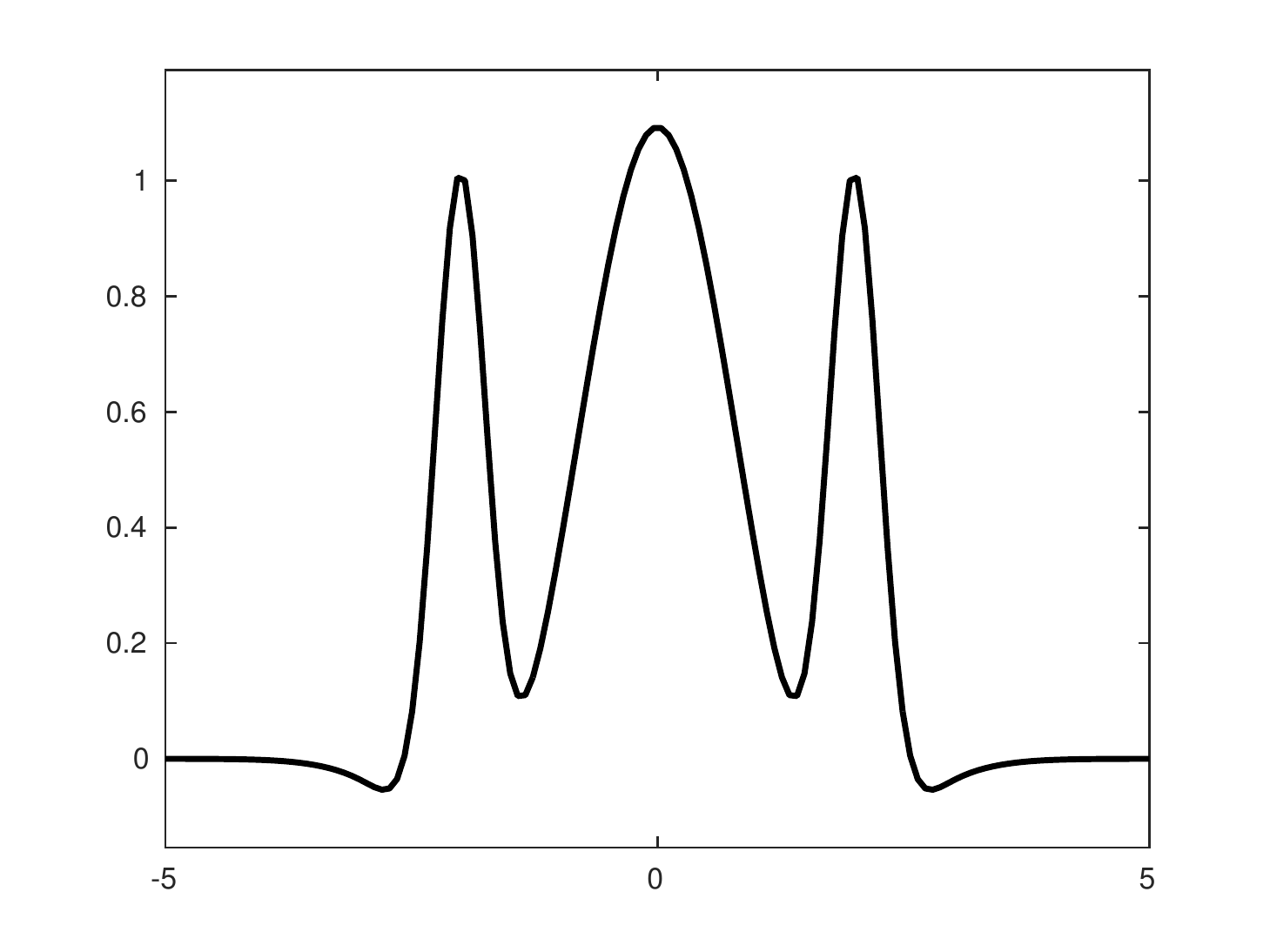}}
\hspace{.3in}
\subfigure[$\widehat{\K_\e}(\ell)$ -- Model 2.]{
\label{fig:kerF2}
\includegraphics[width=0.4\textwidth]{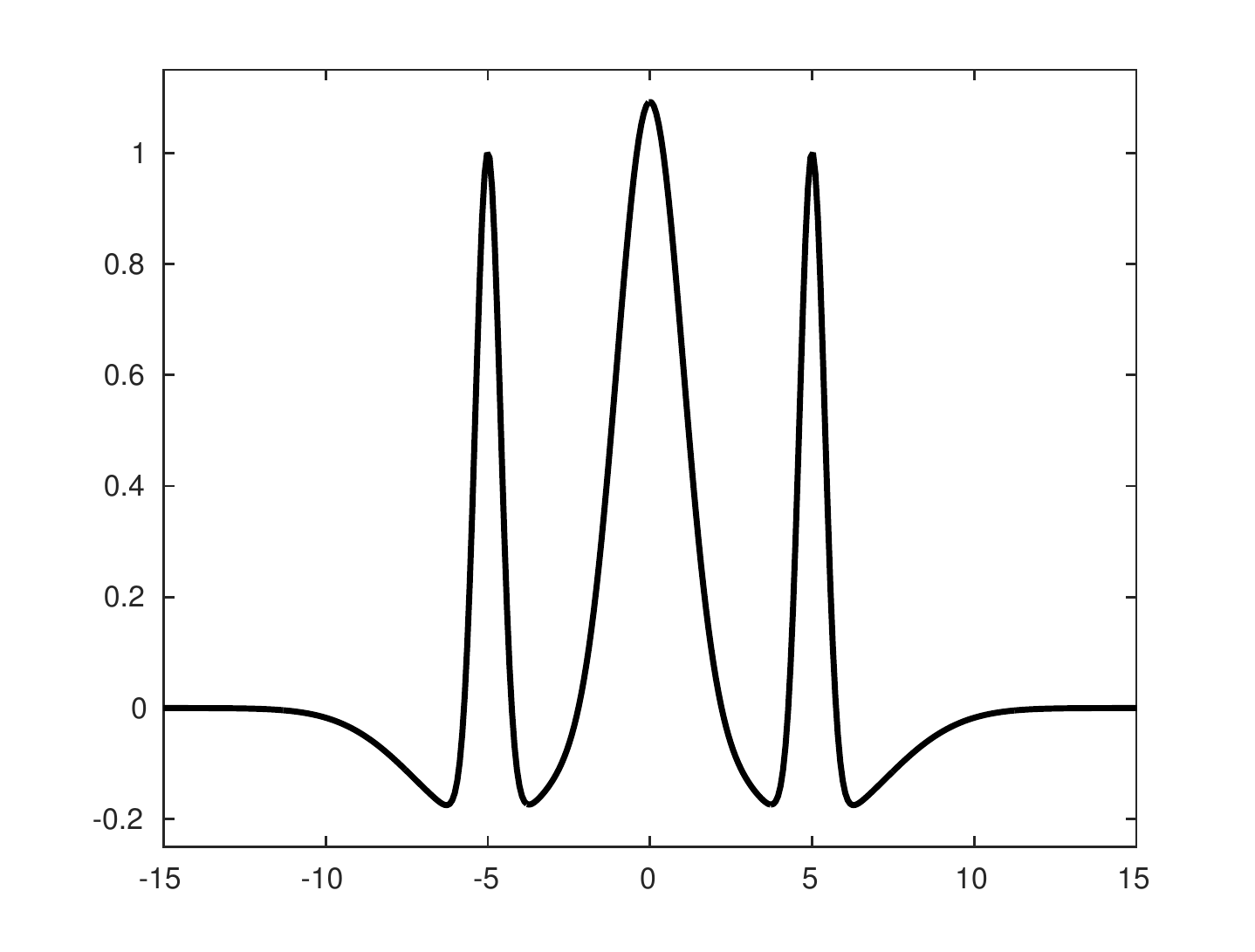}}
\caption{ (a) Plot of the nonlinear function $S_\e$ for $\e=0.1$.
(b) Illustration of a kernel $\K_\e$ satisfying Hypothesis (H). Values of the parameters are to: $(\e,\beta_0,\beta_c)=(0.1,9.25,0.25)$ and $(a_1,a_3,b_1,b_2,b_3)=(0.5,0.51,7,0.2,4)$. Note that condition \eqref{condK0} automatically imposes the value of the other parameters. (c)-(d) Plot of the Fourier transforms of the two different kernels used in our numerical simulations for equation \eqref{nfe}. For each figure, we set $(\e,\beta_0,\beta_c)=(0.1,9.25,0.25)$. (a) For Model 1 we have set $(a_1,a_3,b_1,b_2,b_3)=(0.5,0.51,7,0.2,4)$ and $\ell_c=2$. (b) For Model 2 we have set $(a_1,a_2,b_1,b_3)=(0.5,0,3.317,0.1)$ and $\ell_c=5$. Note that condition \eqref{condK0} automatically imposes the value of the other parameters.}
\label{fig:KS}
\end{figure}

\paragraph{Description of model.}
Consider the following scalar one-dimensional neural field equation
\bqq
\label{nfe}
u_t = -\mu_\e u + \K_\e*S_\e(u), \quad (x,t)\in \R\times\R^+,
\eqq
where the nonlinearity $S_\e$ is defined as
\bqq
\label{sigmoid}
S_\e(u) := \frac{2}{1+e^{-2(u-\e/2)}}-\frac{2}{1+e^{\e}}.
\eqq 
Note that this specific form of the nonlinearity, see Figure~\ref{fig:sigmoid} for an illustration, implies that $u=0$ is always a homogeneous stationary state of the neural field equation \eqref{nfe} for all $\mu_\e>0$. From now on, we set $\mu_e:=1-\epsilon^2 \sigma$ for some $\sigma>0$. For the connectivity function $\K_\e$ we assume the following hypotheses.
\begin{Hypothesis}{(H)}
We suppose that the kernel $\K_\e$ satisfies the following conditions:
\begin{itemize}
\item[(i)] $\K_\e \in W_\eta^{1,1}(\R)$\footnote{Here $W_\eta^{1,1}(\R)$ denotes the Banach space of exponentially localized functions which are in $W^{1,1}(\R)$. More precisely, we define $W_\eta^{1,1}(\R):=\left\{u:\R\mapsto \R~|~ \int_\R u(x)e^{\eta |x|}\mathrm{d}x<+\infty \text{ and } \int_\R \frac{\mathrm{d}}{\mathrm{d}x} u(x)e^{\eta |x|}\mathrm{d}x<+\infty  \right\}$.} for some $\eta>0$ and $\K_\e$ is even;
\item[(ii)] if $\widehat \K_\e$ denotes the Fourier transform of $\K_\e$, then there exists a unique $\ell_c>0$ such that at $\e=0$
\bqq
\label{condK0}
\widehat \K_0(\pm \ell_c) = \widehat \K_0(0) =1, \text{ and } \widehat \K_0(\ell) < 1 \text{ for all } \ell \notin\{0,\pm\ell_c \},
\eqq
with
\begin{subequations}
\label{expK}
\begin{align}
\widehat \K_\e(\ell) &= 1+\e^2\beta_0 - k_0 \ell ^2 +\scriptsize{o}(\ell^2+\e^2), \quad \text{ as } \ell \rightarrow 0, \e\rightarrow0 \\
\widehat \K\e(\ell) &= 1+\e^2\beta_c - k_c (\ell \pm \ell_c) ^2 +\scriptsize{o}(|\ell\pm\ell_c|^2+\e^2), \quad \text{ as } \ell \rightarrow \pm \ell_c,\e\rightarrow0
\end{align}
\end{subequations}
for some $\beta_0,\beta_c\in\R$ and 
\bqq
\label{defk0c}
k_0:=\frac{1}{2}\int_\R x^2 \K(x)\mathrm{d}x, \quad k_c:=\frac{1}{2}\int_\R x^2 \K(x)e^{-i \ell_c x}\mathrm{d}x.
\eqq
\end{itemize}
\end{Hypothesis}
The first condition ensures that the Fourier transform is analytic within a stripe $\mathcal{S}_{\eta_0}$\footnote{$\mathcal{S}_{\eta_0}:=\left\{ z \in \mathbb{C} ~|~ |\Re(z)|<\eta_0 \right\}$.} centered on the imaginary axis of the complex plane of width $\eta_0$ for some $0<\eta_0<\eta$. Thus, the expansions \eqref{expK} are well defined. The second condition implies that when $\epsilon=0$, the homogeneous state $u=0$ is marginally stable with respect to constant perturbations and spatially periodic perturbations of the form $e^{\pm i \ell_c x}$. As a consequence, we have at the same time a Turing instability for \eqref{nfe} and a pitchfork bifurcation for the associated  kinetic equation:
\bqq
\label{kinetic}
\dot u = -\mu_\e u + (1+\e^2\beta_0)S_\e(u).
\eqq

Note that the above conditions of the connectivity kernels $\K_e$ can be easily satisfied for kernels whose Fourier transforms are of the form
\bqq
\label{kernel}
\widehat \K_\e(\ell) = \widehat{\mathcal{A}}_\e(\ell)+\widehat{\mathcal{B}}_\e(\ell),
\eqq
where
\begin{subequations}
\label{ABeps}
\begin{align}
\widehat{\A}_\e(\ell)&:=(a_0+\e^2\beta_0)e^{-a_1 \ell^2}-a_2e^{-a_3\ell^2},\\
\widehat{\B}_\e(\ell)&:=(b_0+\e^2\beta_c)\left(e^{-b_1 (\ell-\ell_c)^2}+e^{-b_1 (\ell+\ell_c)^2}\right)-b_2\left(e^{-b_3(\ell-\ell_c)^2}+e^{-b_3 (\ell+\ell_c)^2}\right),
\end{align}
\end{subequations}
for some parameters $(a_0,a_1,a_2,a_3,b_0,b_1,b_2,b_3)\in \R^6$. In Figure~\ref{fig:kernel}-\ref{fig:kerF1}-\ref{fig:kerF2}, we present kernels of the form \eqref{kernel} and \eqref{ABeps} which satisfy Hypothesis (H). Note that generically, our conditions on the Fourier transform of connectivity kernel $\K_\e$ imply that, in real space, the kernel is locally excitatory while it presents lateral modulations of inhibition and excitation. This specific form of kernels have already been used in the literature \cite{laingetal02,rankinetal14}  to analyze stationary multi-bump solutions of equation \eqref{nfe} and is also in agreement with experimentally recorded cortical connections in cat visual areas \cite{buzasetal01}.

\paragraph{Amplitude Equations}

As previously stated, at $\e=0$, there is a bifurcation of the stationary state $u=0$ where we have at the same time a Turing instability and a pitchfork bifurcation for the kinetics \eqref{kinetic}. Linearizing equation \eqref{nfe} around its homogeneous state $u=0$ and looking for perturbations of the $e^{\lambda t} e^{\nu x}$ for complex values of the parameters $\lambda$ and $\nu$, we obtain the following dispersion relation 
\bqq
\label{dispersionNFE}
D_\e(\lambda, \nu) := \e s \nu +\e^2\sigma + \widetilde{\K}_\e(\nu)S'_\e(0)-1 -\lambda,
\eqq
where we have set
\bqs
\widetilde{\K}_\e(\nu) := \int_\R \K_\e(x) e^{-\nu x}\mathrm{d}x, \quad \text{ for } \nu \in \mathbb{C}.
\eqs
Note that $\widetilde{\K}_\e(\nu)$ is well defined for any $\nu \in \mathcal{S}_{\eta_0}$
and that with this notation we have $\widetilde{\K}_\e(i\ell)=\widehat{\K}_\e(\ell)$. When $\e=0$, we know that for $\lambda=0$ there exists roots of the dispersion relation \eqref{dispersionNFE} at $\nu=0$ and $\nu=\pm i \ell_c$ as condition \eqref{condK0} implies $D_0(0,0)=0$ and $D_0(0,\pm i\ell_c)=0$. We now want to track these roots for $0<\e\ll1$. We suppose the following expansion 
\begin{eqnarray*}
\nu_0&=& \e p +\cO(\e^2), \\
\nu_c&=& \pm i\ell_c+\e q +\cO(\e^2), \\
\lambda_0&=& \e \lambda_1+\e^2\lambda_2 +\cO(\e^3), \\
\lambda_c&=& \e  \gamma_1+\e^2\gamma_2+\cO(\e^3).
\end{eqnarray*}
Using our Hypothesis (H)-(ii) for the kernel, we have 
\bqs
\widetilde{\K}_\e(\nu) \sim 1 +\e^2\beta_0+ k_0 \nu^2
\eqs
close to $\nu = 0$, and also
\bqs
\widetilde{\K}(\nu) \sim 1 +\e^2\beta_c+ k_c (\nu\pm i\ell_c)^2 
\eqs
close to $\nu = \pm i \ell_c$. Furthermore, we have the asymptotic expansion for the nonlinearity of the form
\bqs
S_\e (u)\sim u-\frac{\e^2u}{4}+\frac{\e u^2}{2}-\frac{u^3}{3}+\text{h.o.t}
\eqs
for $(u,\e)\sim(0,0)$. Then, solving successively the equations at order $\cO(\e)$ and $\cO(\e^2)$, we find
\begin{eqnarray*}
\lambda_1 &=&0,\\
\gamma_1&=&\pm i\ell_c s,\\
\lambda_2&=&k_0p^2+sp+\sigma-\frac{1}{4}+\beta_0,\\
\gamma_2&=&k_cq^2+sq+\sigma-\frac{1}{4}+\beta_c.
\end{eqnarray*}

As a consequence, at the linear level, the amplitude equations associated to the neural field equation \eqref{nfe} around the stationary homogeneous state $u=0$ should be of the form
\begin{eqnarray*}
U_T&=& k_0 U_{XX}+\delta_0U \\
A_T&=& k_cA_{XX}+\delta_cA,
\end{eqnarray*}
where we have set
\[\delta_0:= \sigma-\frac{1}{4}+\beta_0, \quad \delta_c:=\sigma-\frac{1}{4}+\beta_c, \]
and $U$ is associated to the $0$th mode and $A$ is associated to the Fourier modes $e^{\pm i \ell_c x}$. One still needs to compute the nonlinear terms. Postulating an Ansatz for the solution of the form 
\bqs
u(t,x)=\e A(T,X)e^{i\ell_c x} +\e U(T,X)+c.c.+\cO(\e^2), \quad X=\e x \text{ and } T=\e^2 t,
\eqs
we find, after straightforward expansions, the following system of equations
\begin{subequations}
\label{ampNFE}
\begin{align}
U_T&= k_0 U_{XX}+U\left(\delta_0+ \frac{U}{2}-\frac{U^2}{3}-2|A|^2\right)+|A|^2, \\
A_T&= k_cA_{XX}+A\left(\delta_c-|A|^2+U-U^2\right).
\end{align}
\end{subequations}
Note that in order to get for example terms like $k_0 U_{XX}$ in the expansion, one needs to do the following formal type of computations:
\begin{align*}
-U(\e x)+\int_\R\K(x-y)U(\e y)\mathrm{d}y&= \int_\R\K(y)\left(U(\e x-\e y)-U(\e x)\right)\mathrm{d}y\\
&=\int_\R\K(y)\left(-\e yU_X(\e x)+\frac{\e^2y^2}{2}U_{XX}(\e x)+\text{h.o.t}\right)\mathrm{d}y\\
&=\e^2k_0 U_{XX}(\e x) +\cO(\e^3).
\end{align*}

\paragraph{Spreading speed induced by the quadratic mode interactions}

As a consequence of the instability of the stationary state $u=0$ in equation \eqref{nfe}, we have that $(U,A)=(0,0)$ is an unstable homogeneous solution of the amplitude equations \eqref{ampNFE}. Linearizing \eqref{ampNFE} at that unstable and using the Ansatz $e^{\lambda t+\nu x}(U_0,A_0)$ for some nonzero vector $(U_0,A_0)\in \mathbb{C}^2$, we find a dispersion relation
\bqq
\label{dispersionAmpNFE}
D(\lambda,\nu)=(k_0\nu^2+\delta_0-\lambda)(k_c\nu^2+\delta_c-\lambda).
\eqq
Using Definition~\ref{defi:quad}, we find the spreading speed induced by the quadratic mode interactions for the amplitude equations \eqref{ampNFE} is given by
\bqq
\label{speedampbi}
s_{c\rightarrow 0}=2k_0\sqrt{\frac{\delta_0-2\delta_c}{2(k_c-2k_0)}}+\frac{\delta_0}{2}\sqrt{\frac{2(k_c-2k_0)}{\delta_0-2\delta_c}},
\eqq
for $k_c>2k_0$ and $\delta_0>2\delta_c$. This speed is maximal for those parameter values
\bqq
 \mathbf{P}=\left\{ (k_{0,c},\delta_{0,c} )\ | \ \frac{4\delta_c(k_c-k_0)}{k_c}<\delta_0 \  (k_c>4k_0),  \  \frac{4\delta_c(k_c-k_0)}{k_c}<\delta_0 < \frac{4\delta_ck_0}{4k_0-k_c}, \  (4k_0>k_c) \right\}.
 \label{eq:PampBi}
 \eqq

\begin{figure}[t]
\centering
\subfigure[]{
\label{fig:CurvesAmpBi}
\includegraphics[width=0.4\textwidth]{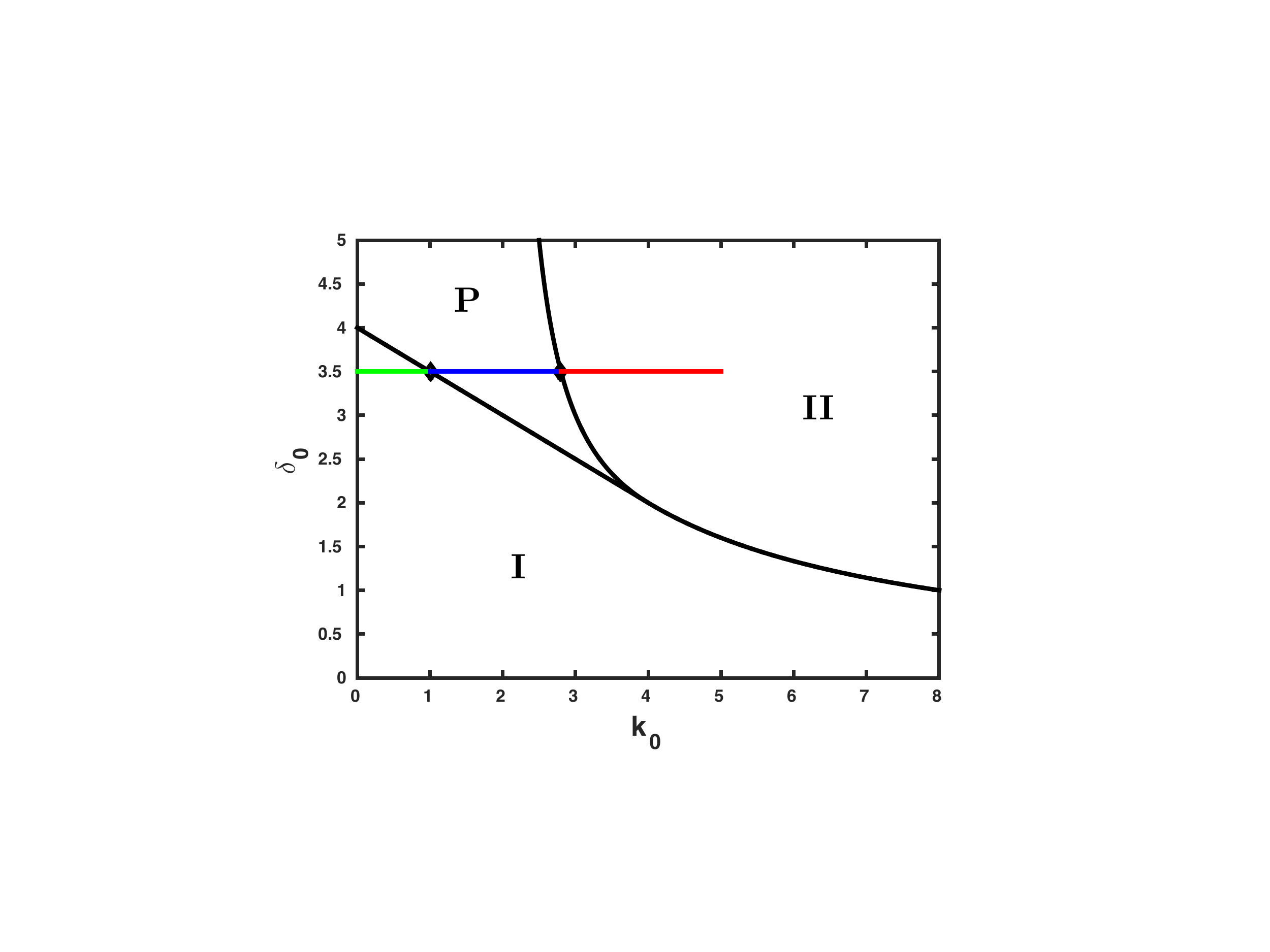}}
\hspace{.3in}
\subfigure[]{
\label{fig:SpeedAmpBI}
\includegraphics[width=0.425\textwidth]{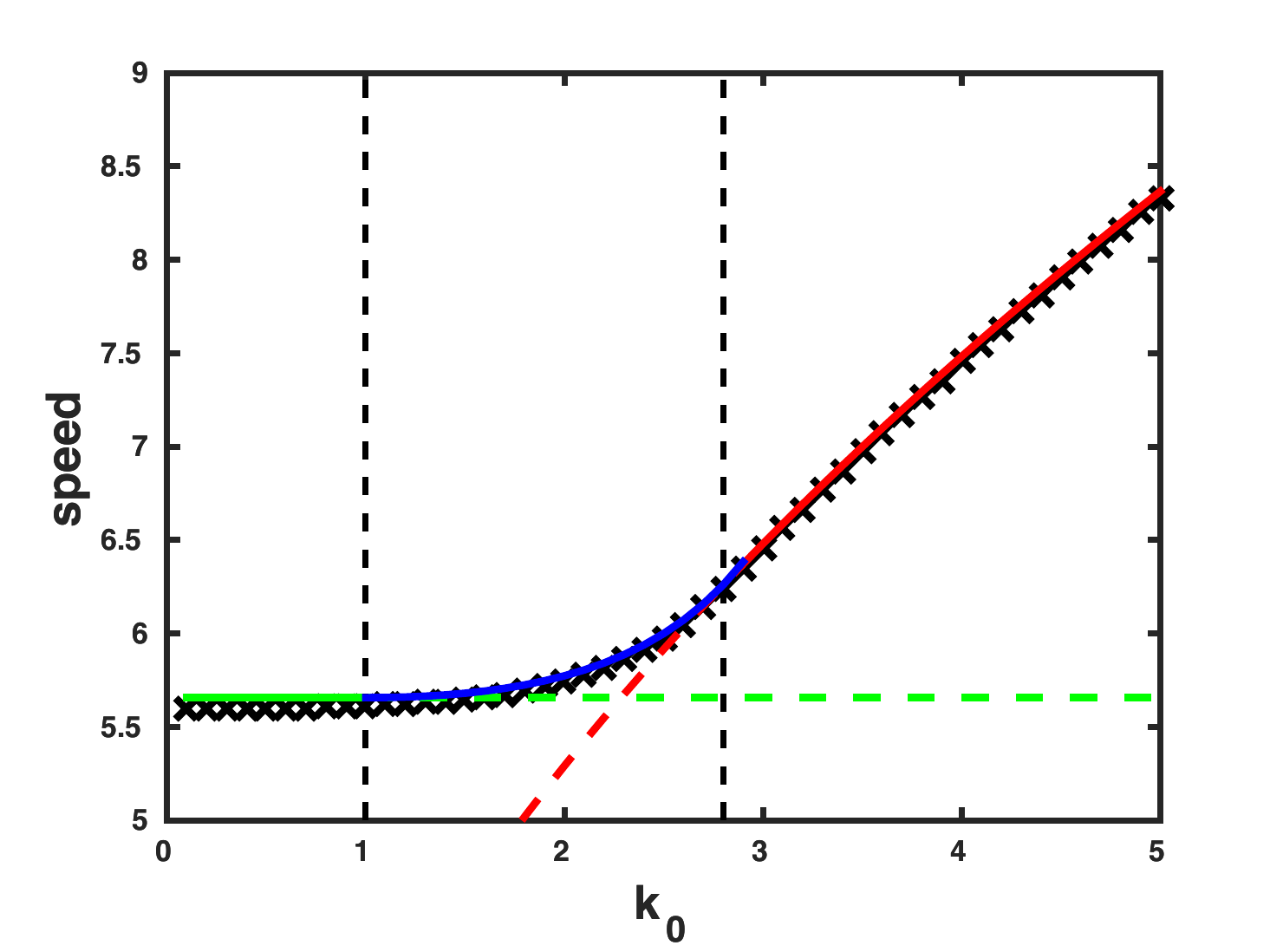}}
\caption{Comparison of numerically observed spreading speeds (black crosses) for the amplitude equations \eqref{ampNFE} to the linear predictions when parameters are in the region $\textbf{I}$ (green), region $\textbf{P}$ (blue) and region $\textbf{II}$ (red) in parameter space $(k_0,\delta_0)$ when $k_c=8$ and $\delta_c=1$. In regions $\textbf{I}$ and  $\textbf{II}$ the linear spreading speeds are $2\sqrt{k_c\delta_c}$ and $2\sqrt{k_0\delta_0}$ respectively, while in region $\textbf{P}$ the linear spreading speed is $s_{c\rightarrow0}$ given by formula \eqref{speedampbi}. Note that the boundary between regions $\textbf{I}$ and $\textbf{P}$ is given by $\delta_0=4-k_0/2$, the boundary between regions $\textbf{P}$ and $\textbf{II}$ is given by $\delta_0=k_0/(k_0-2)$ and the boundary between regions $\textbf{I}$ and $\textbf{II}$ is given by $\delta_0=8/k_0$.}
\label{fig:speedAMPbi}
\end{figure}

The boundaries of these two regions occur for those speeds where the anomalous speed occurs at the linear spreading speed of one of the two components.  In other words, the lower bound occurs when
\[ s_{\mathrm{env},0}\left(-2\sqrt{\frac{\delta_c}{k_c}}\right)=2\sqrt{k_c\delta_c},\]
and the upper bound occurs when
\[ s_{\mathrm{env},c}\left(-2\sqrt{\frac{\delta_0}{k_0}}\right)=2\sqrt{k_0\delta_0},\]
where
\begin{align*}
s_{\mathrm{env},0}(\nu)&:=-k_0\nu-\frac{\delta_0}{\nu},\\
s_{\mathrm{env},c}(\nu)&:=-k_c\nu-\frac{\delta_c}{\nu}.
\end{align*}

We report in Figure~\ref{fig:speedAMPbi} numerically observed spreading speeds for the amplitude equations \eqref{ampNFE} for various values of the parameters and compare with our linear predictions. System \eqref{ampNFE} was solved using finite differences with Heaviside step function initial data and speeds were calculated by computing how much time elapsed between the solution passing a threshold at two separate points in the spatial domain. We see that our prediction for the spreading speed induced by the quadratic mode interactions is still able to accurately describe the numerically observed spreading speeds in the case of bi-directionally coupled amplitude equations. This confirms that the linear criterion in Definition ~\ref{defi:quad} is general enough to correctly predict spreading speeds for systems of the form of the amplitude equations \eqref{ampNFE}.

\paragraph{Numerical simulations: full model}

Comparisons of spreading speeds were performed as follows.  The neural field equation (\ref{nfe}) was solved using Fast Fourier Transform with symmetric compactly supported initial data of the form
\bqs
u_0(x) \propto \mathds{1}_{[-L,L]}\cos(\ell_c x), \quad \text{ for an } L>0 \text{ and } x\in\R,
\eqs
with nonlinearity given in \eqref{sigmoid} and connectivity kernels $\K$ satisfying Hypothesis (H). The periodic modulation in the initial condition is necessary in order to feeds in the critical modes $e^{\pm i \ell_c x}$ as suggested by the form of the amplitude equations \eqref{ampNFE} and the $A=0$ invariance. Speeds were calculated by computing how much time elapsed between the solution passing a threshold at two separate points in the spatial domain.  These spreading speeds were compared with predictions using the linear criterion in Definition~\ref{defi:quad}. These predictions were found using numerical continuation.  

\begin{figure}[t]
\centering
\subfigure[]{
\label{fig:speed1eps}
\includegraphics[width=0.4\textwidth]{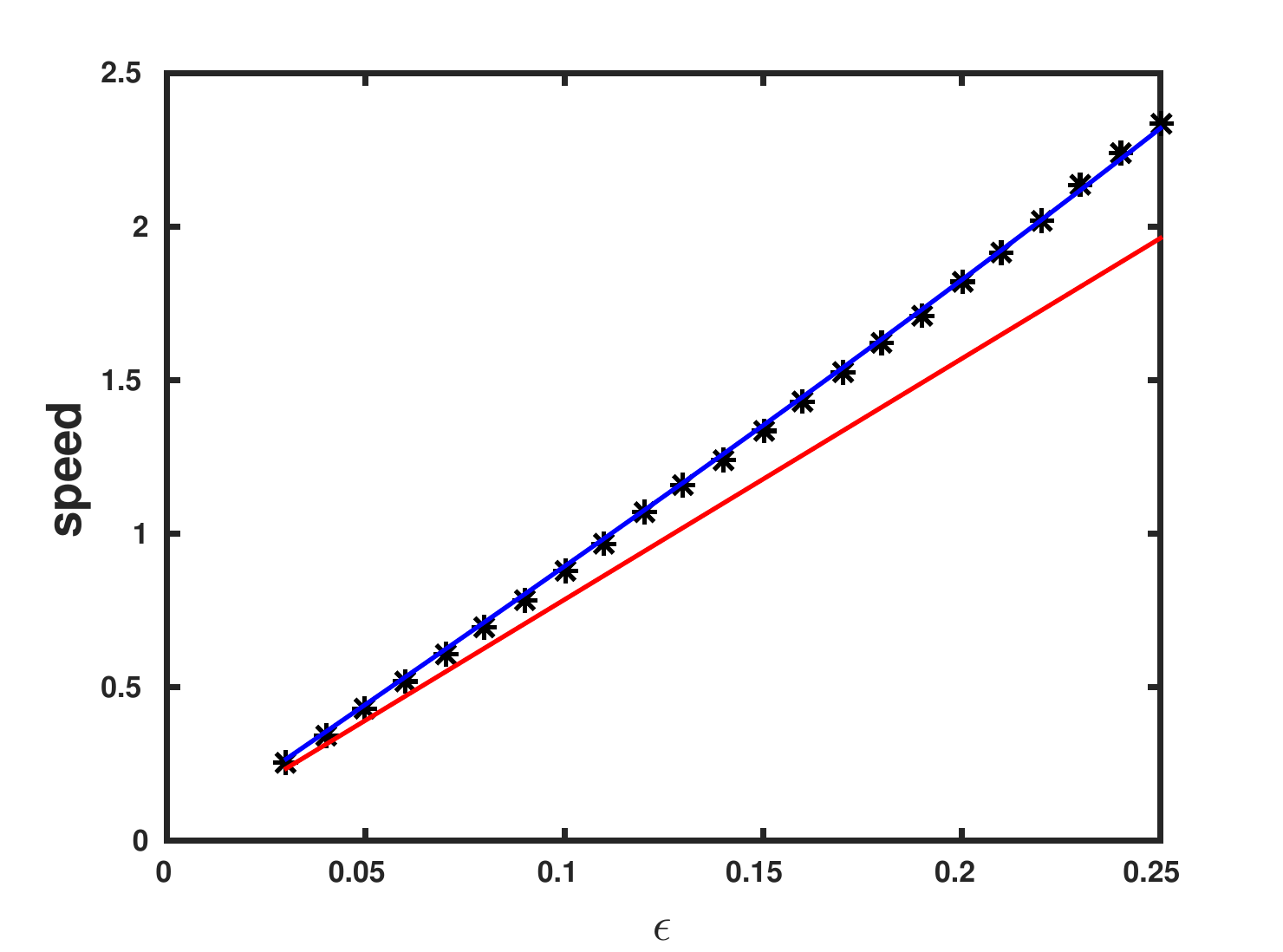}}
\hspace{.3in}
\subfigure[]{
\label{fig:speed1beta}
\includegraphics[width=0.4\textwidth]{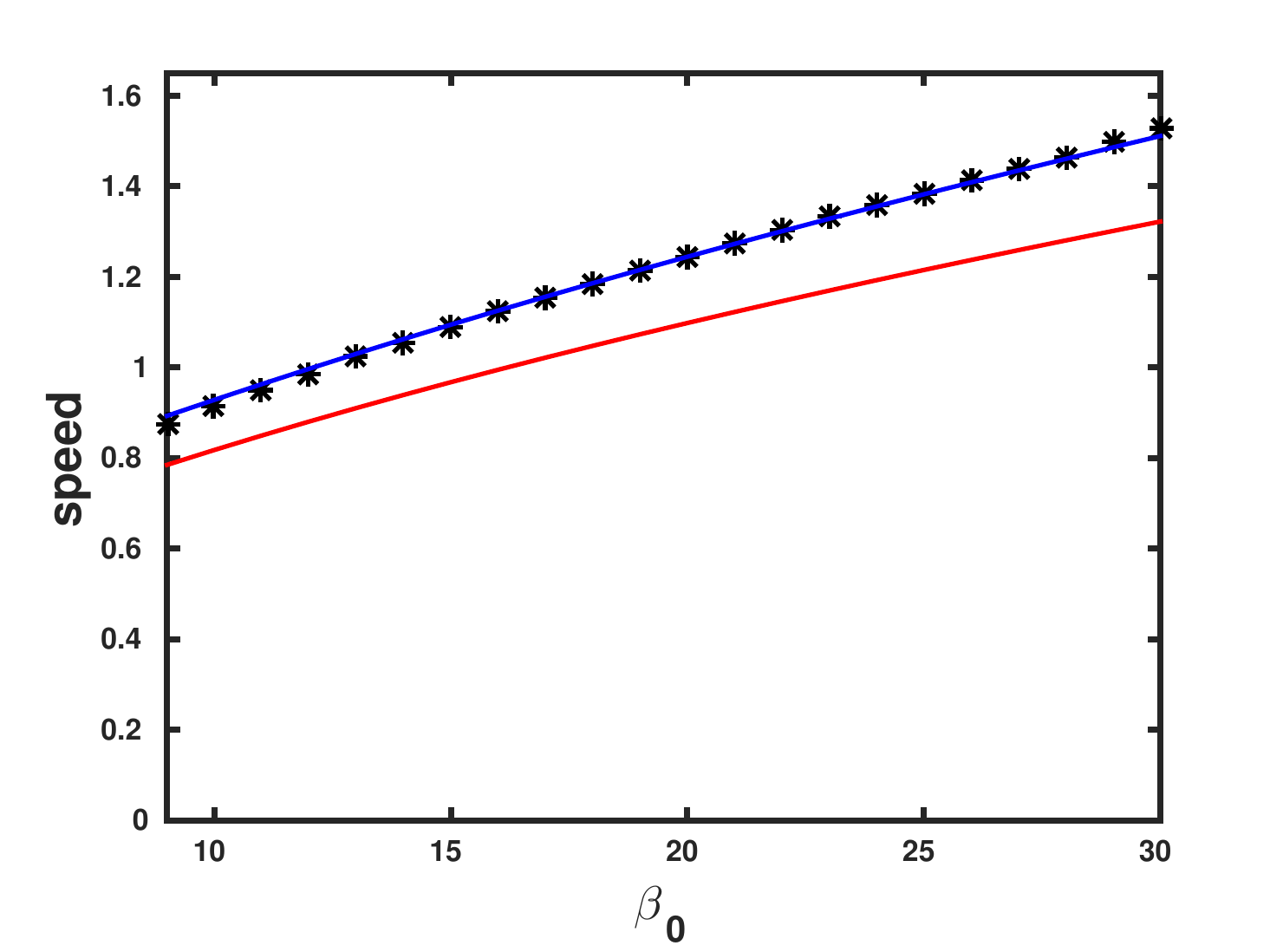}}
\subfigure[]{
\label{fig:speed2eps}
\includegraphics[width=0.4\textwidth]{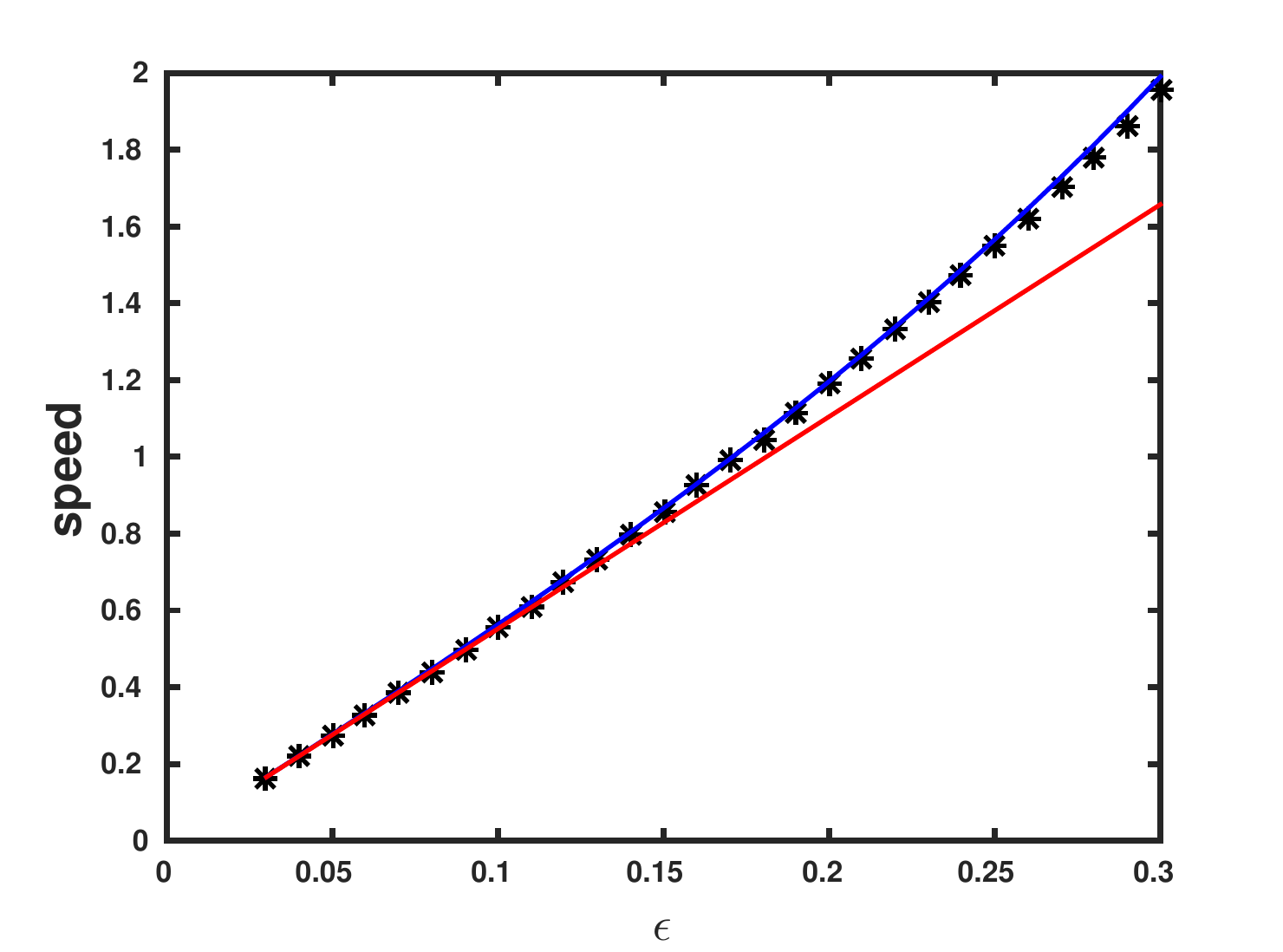}}
\hspace{.3in}
\subfigure[]{
\label{fig:speed2beta}
\includegraphics[width=0.4\textwidth]{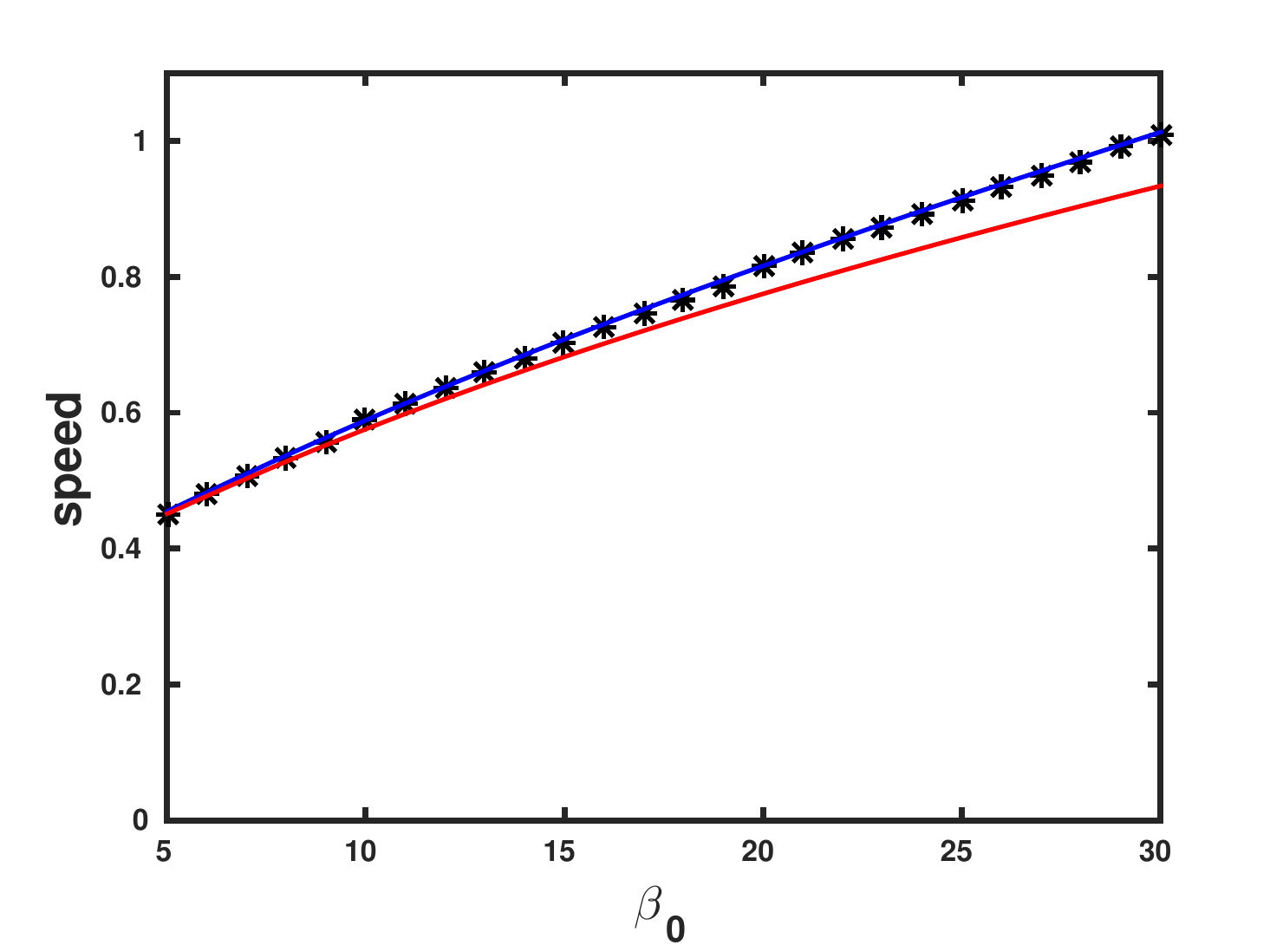}}
\caption{Comparison of numerically observed spreading speeds (black) for equation (\ref{nfe}) for Model 1 (a)-(b) and Model 2 (c)-(d), compared to linear predictions generated by the amplitude equations \eqref{ampNFE} (red) and the criterion in Definition~\ref{defi:quad} (blue).  In (a)-(c), we vary $\e$ and fix all other parameters to $\sigma=1$, $\beta_0=9.25$ and $\beta_c=0.25$.  Note that all three speeds agree for small values of $\e$ while the criterion in Definition~\ref{defi:quad} remains valid for larger values of $\e$.  In (b)-(d), we vary $\beta_0$ while fixing all other parameters to $\e=0.1$, $\sigma=1$ and $\beta_c=0.25$.}
\label{fig:SpeedMod1}
\end{figure}

\begin{figure}[ht]
\centering
 \subfigure[$S_\e$ and $\beta_0=9.25$.]{\includegraphics[width=0.4\textwidth]{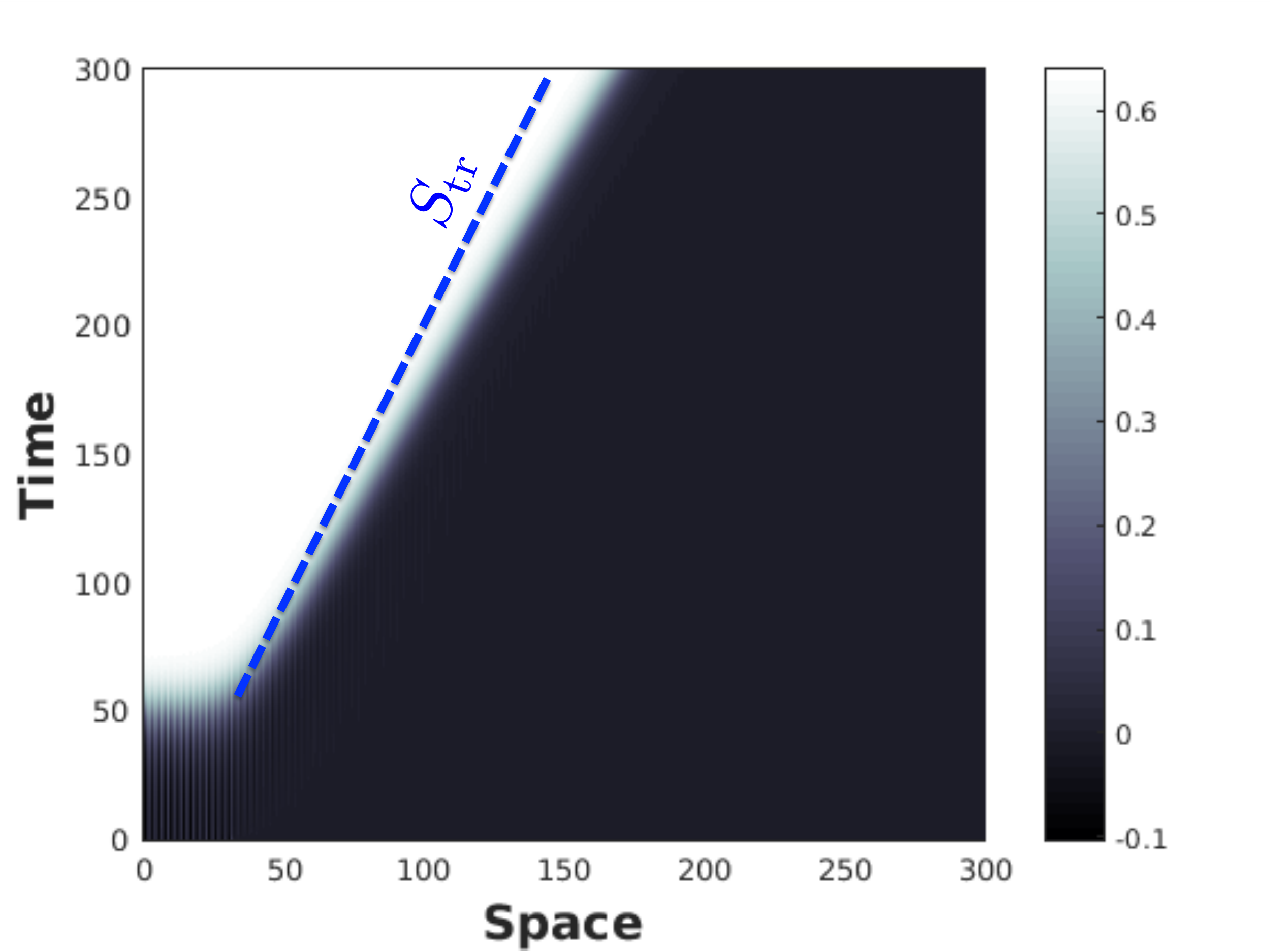}}
 \subfigure[$S_\text{tr}$ and $\beta_0=9.25$.]{\includegraphics[width=0.4\textwidth]{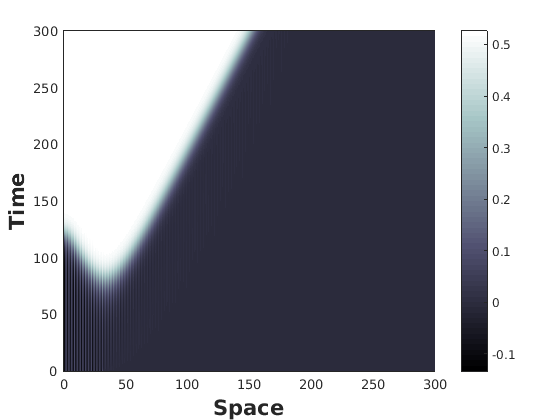}}
 \subfigure[$S_\e$ and $\beta_0=3.25$.]{\includegraphics[width=0.4\textwidth]{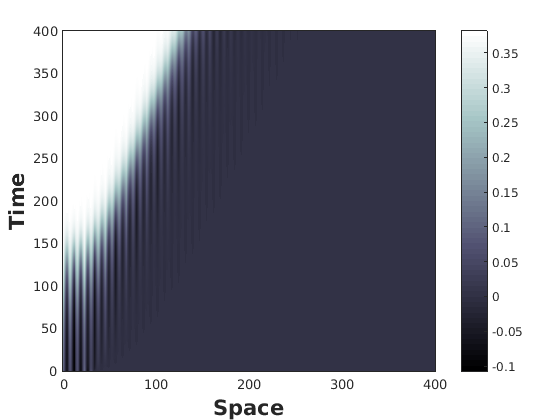}}
 \subfigure[Profiles from (a)-(b) at $t=300$.]{\includegraphics[width=0.4\textwidth]{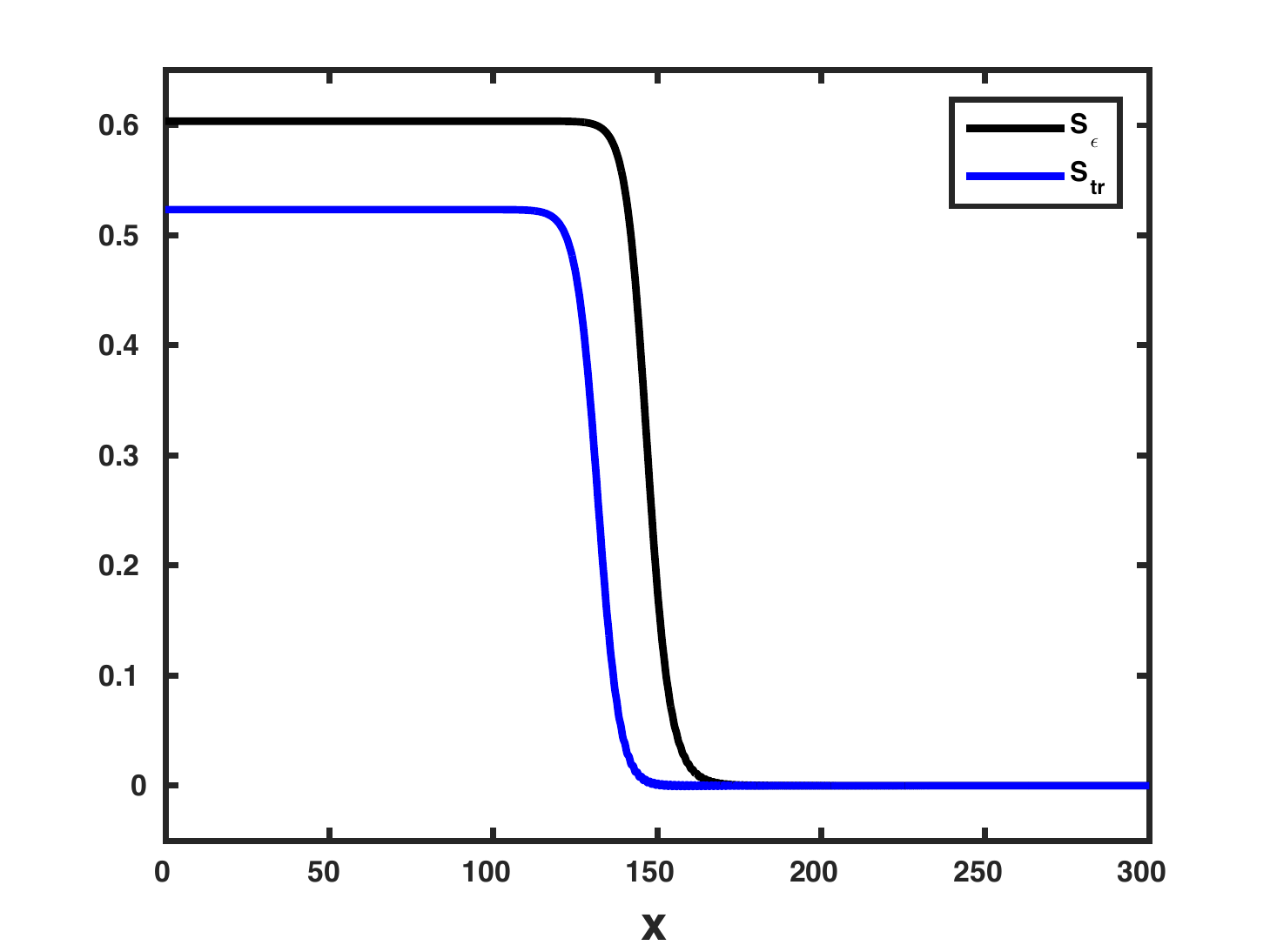}}
\caption{(a) Space-time plots of the solution of equation \eqref{nfe} with $S_\e$ and $\beta_0=9.25$. (b) Space-time plots of the solution of equation \eqref{nfe} with $S_\text{tr}$ and $\beta_0=9.25$. (c) Space-time plots of the solution of equation \eqref{nfe} with $S_\e$ and $\beta_0=3.25$. (d) Plot of the profiles at $t=300$ of the solutions taken from space-time plots (a) and (b). The connectivity function is set to Model 2 and  values of the parameters are fixed to $\e = 0.1$, $\sigma = 1$ and $\beta_c = 0.25$.}
\label{fig:STplotNFE}
\end{figure}

Typical results are plotted in Figure~\ref{fig:SpeedMod1} for Model 1 and Model 2, with space-time plots reported in Figure~\ref{fig:STplotNFE}. Once again, we find that the prediction from the linear criterion ($s_\mathrm{quad}$) matches well with the speed from the amplitude equation \eqref{ampNFE} when $\e$ is small.  For larger values of $\e$, the two speeds deviate and the spreading speeds for (\ref{nfe}) observed in direct numerical simulations match closely with the speed $s_\mathrm{quad}$ in this regime.  

\begin{figure}[ht]
\centering
\subfigure[]{
\label{fig:nfewoquad}
\includegraphics[width=0.4\textwidth]{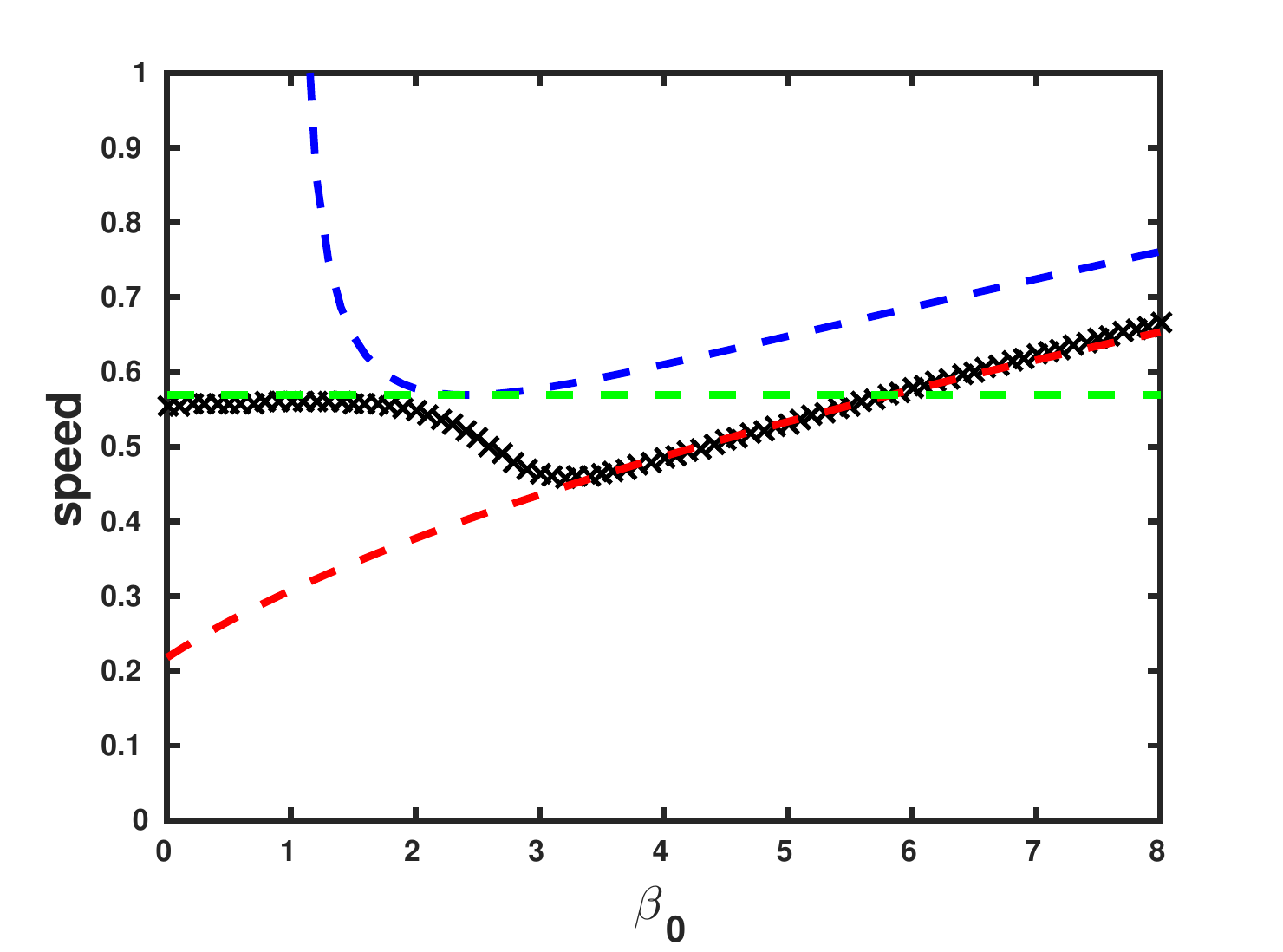}}
\hspace{.3in}
\subfigure[]{
\label{fig:SHKPPstable}
\includegraphics[width=0.4\textwidth]{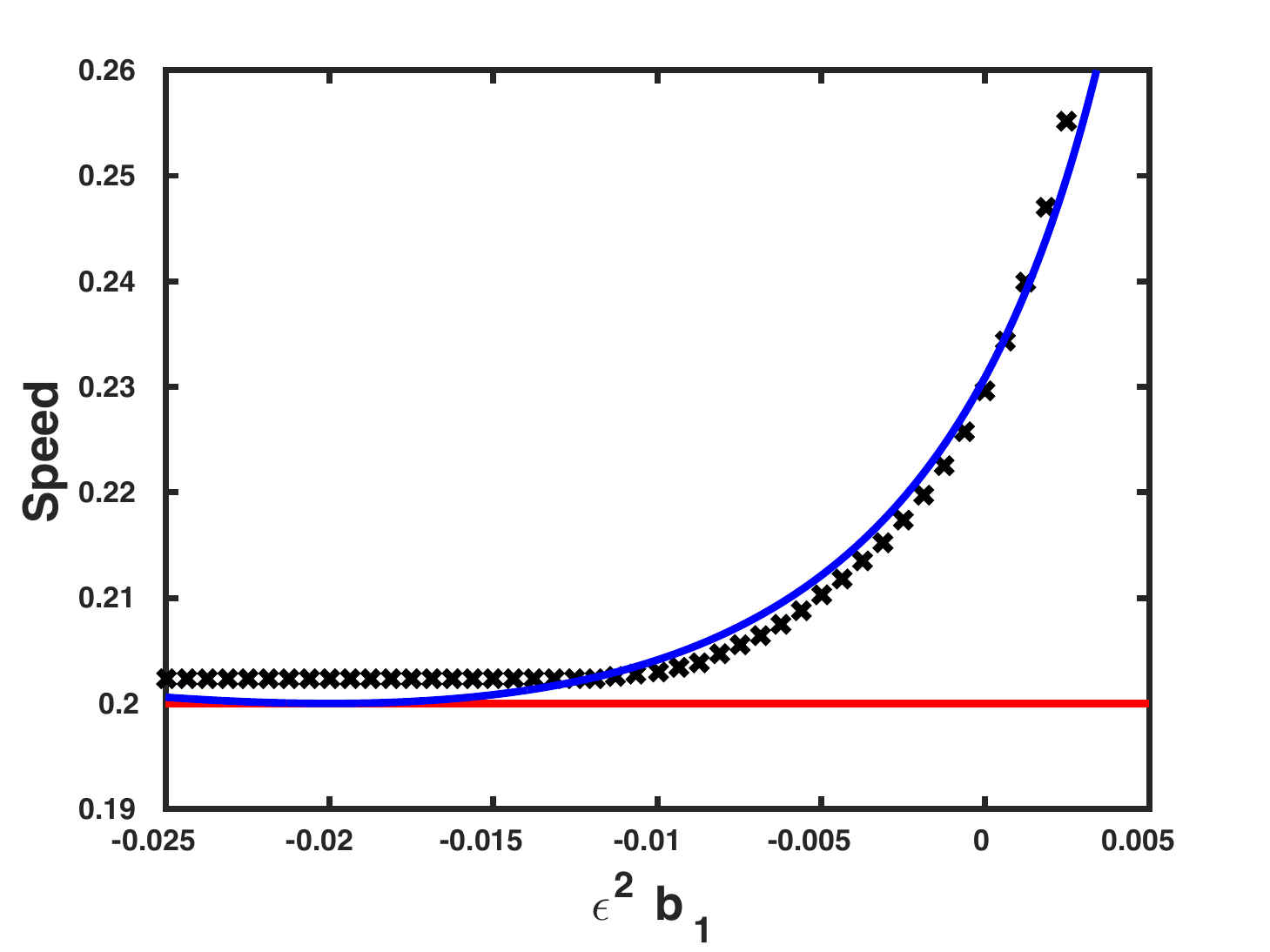}}
\caption{(a) Numerically observed spreading speeds (black) for equation (\ref{nfe}) and Model 1 as $\beta_0$ is varied from $0$ to $8$ in the case where the nonlinearity is given by equation \eqref{sigtrunc}. Linear predictions generated by the amplitude equations \eqref{ampNFE} for the A-component in isolation (red) and for the U-component in isolation (green) together with the criterion in Definition~\ref{defi:quad} (blue). (b) Numerically observed spreading speeds (black) for a modified version of equation (\ref{eq:SHKPPdiscussion}) for varying values of the instability parameter $b_1$.  All other parameters are held fixed at $\e=0.05$, $\alpha=8.0$, $d=0.5$.  The red line is the speed of the zero mode in isolation $2\e\sqrt{d\alpha}$ while the green curve is the $2:1$ resonant spreading speed predicted by the resonance criterion.}
\end{figure}

We also numerically investigated the influence of the quadratic interactions coming from the nonlinearity by studying the neural field equation \eqref{nfe} with a truncated nonlinearity of the form
\bqq
\label{sigtrunc}
S_\text{tr}(u):=\left\{
\begin{array}{lcl}
\frac{4-\e^2}{12}\sqrt{\frac{4-\e^2}{1-\e^2}} & \text{ for } & u \geq u_\e^+, \\
\left(1-\e^2/4 \right)u+(\e^2-1)u^3/3 & \text{ for } & u\in[u_\e^-,u_\e^+],\\
-\frac{4-\e^2}{12}\sqrt{\frac{4-\e^2}{1-\e^2}} & \text{ for } & u \leq u_\e^-,
\end{array}
\right.
\eqq
where
\bqs
u_\e^\pm:=\pm \frac{1}{2}\sqrt{\frac{4-\e^2}{1-\e^2}}.
\eqs
Typical results are plotted in Figure~\ref{fig:nfewoquad} for Model 1 with nonlinearity \eqref{sigtrunc}.We remark that the selected wave speed is either the linear wave speed associated to the A-component of the amplitude equation \eqref{ampNFE} in isolation or  the linear wave speed associated to the U-component of the amplitude equation \eqref{ampNFE} in isolation with a transition in wave speed for values of $\beta_0$ between $2$ and $3$. As expected, our linear criterion from Definition~\ref{defi:quad} fails to predict the selected wave speed in that case.

\section{Discussion}\label{sec:discussion}

In this paper, we identified resonances, in particular $2:1$-resonances as a mechanism for instability that determines spreading speeds into unstable states. The common pinched double root criterion that is used in order to determine spreading speeds could be interpreted as identifying a strong $1:1$-resonance in the leading edge. The higher resonances that we discuss here, as well as more general $1:1$-resonances that are associated with double double roots \cite{holzsch} require the presence of a coupling term in the equation. For $2:1$-resonances, this coupling term is quadratic and its mere presence enables the propagation at the $2:1$-resonant spreading speed. Unlike the case of pushed fronts, the speed is independent of the strength of the quadratic term. Beyond heuristics for the derivation, we show rigorously that the $2:1$-resonant speed is the spreading speed in a simple, almost scalar example. The main motivation for our work stems from mode coupling induced by the simultaneous presence of weakly stable or unstable Turing \emph{and} homogeneous modes. In fact, our simple example is a special case of amplitude equations that describe such mode interactions near onset of instability.  We support our Definition further by comparing spreading speeds from direct simulations with the theoretical prediction according to our definition, in more general coupled amplitude equations, in Swift-Hohenberg model coupled to a scalar reaction-diffusion system, and in a nonlocal neural field model. The last example illustrates in particular that the phenomenon is not reliant on an artificial decoupling of modes.  We also show how the resonance leads to \emph{pointwise} linear instabilities of traveling-wave profiles that propagate with speeds less than the resonant spreading speed. In particular, we show that the linear resonances give accurate predictions although invasion profiles create large-amplitude states, since resonances are relevant for determining speeds in the leading edge of the profile where amplitudes are small. 

It is important to notice that the resonant mechanism does not require an unstable Turing mode. Key to the instability is a large effective diffusivity in the Turing mode which generates slow spatial decay in the homogeneous mode through the resonant coupling. As an example, consider
\begin{subequations}
\begin{align}
u_t &= du_{xx}+\e^2\alpha u-u^3+\e v^2  \\
v_t &= -(\partial_x^2+1)^2v +\e^2b_1 v-v^3, 
\end{align}
\label{eq:SHKPPdiscussion}
\end{subequations}
where the instability parameter of the Turing mode is allowed to vary.  When $b_1<0$, the zero state for the component is stable.  Nonetheless, quadratic mode interactions persist and can lead to faster invasion speeds in an entirely analogous manner; see Figure~\ref{fig:SHKPPstable}.  

Unlike the usual pinched double root criterion, our definition is not symmetric in the modes involved: the necessity of a nonlinear coupling term introduces a preferred direction of resonance, in our case from Turing to homogeneous modes. This is extends the observations in \cite{LV,holzsch} for ``relevant'' and ``irrelevant'' double roots, where relevance relies on a coupling term respecting the asymmetric pinching condition between the two modes. 

Of course, the criterion readily generalizes to higher resonances, such as $3:1$-resonances, involving cubic coupling terms between linearly independent homogeneous modes, for instance. One notices very quickly that,  in general, it may be difficult to identify the most relevant resonance mechanism relevant for spreading. Interesting examples in this direction arise when coupling Hopf modes with Turing or homogeneous modes. 

Intriguing questions arise in connection with the min-max construction, already in the case of simple pinched double roots. One usually assumes that the effective diffusivity, $\mathcal{D}=-\partial_\nu^2 D/\partial_\lambda D$ at a pinched double root has negative real part. As a consequence, pinched double roots are the most unstable points of absolute spectra \cite{sandstede00,rademacher07}. One may suspects that unstable absolute spectra, or $\Re \mathcal{D} <0$, leads to resonant unstable spectrum. From a different point of view, one can ask if the min-max characterization of resonances can be formulated as a global variational problem for the dispersion relation. 

More subtly, we suspect that the coupling condition in Definition \ref{defi:quad} is not necessary. We verified in direct simulations that almost coupling through terms of the form $A(\partial_x-\nu_2)A$, which vanishes at the resonance but couples any near-resonant modes does generate the resonant speed. It is conceivable, that the convergence of the speed towards the resonant speed would be altered, however. 

Lastly, we notice that beyond the determination of spreading speeds, the resonance criterion introduced here can of course be used in order to determine transitions from convective to absolute instabilities. Somewhat surprisingly, the possibility of such transitions has not received much attention in the literature.

\bibliographystyle{abbrv}

\end{document}